\documentclass{biophys-new}
\usepackage[utf8]{inputenc}
\usepackage{graphicx}
\usepackage[colorlinks,allcolors=cyan!70!black]{hyperref}
\usepackage{subfig}
\usepackage{enumerate}

\usepackage{lipsum}

\newcommand{\bn}{{\boldsymbol{n}}}
\newcommand{\be}{\begin{equation}}
\newcommand{\ee}{\end{equation}}
\newcommand{\ph}{\varphi}

\newcommand{\e}{\varepsilon}
\newcommand{\he}{H_\varepsilon}
\newcommand{\de}{\delta_\varepsilon}

\newcommand{\Om}{\Omega}
\newcommand{\Omt}{{\Om_t}}
\newcommand{\pot}{{C_t}}

\newcommand{\RR}{\mathbb{R}}

\newcommand{\bu}{{\boldsymbol{u}}}

\newcommand{\bS}{{\boldsymbol{S}}}

\newcommand{\bbc}{{\boldsymbol{c}}}

\newcommand{\hot}{\chi_{{}_\Omt}}

\newcommand{\bI}{{\boldsymbol{I}}}
\newcommand{\bo}{{\boldsymbol{0}}}
\newcommand{\bx}{{\boldsymbol{x}}}
\newcommand{\bv}{{\boldsymbol{v}}}
\newcommand{\bb}{{\boldsymbol{b}}}

\newcommand{\po}{{\partial\Omega}}
\newcommand{\diiv}{\nabla\!\cdot }

\newcommand{\bbx}{{\bar \bx}}
\newcommand{\lamc}{lamellipodium, }
\newcommand{\lamp}{lamellipodium. }
\newcommand{\lam}{lamellipodium {}}

\newcommand{\bbv}{{{\bar\bv}}}

\def\XXint#1#2#3{{\setbox0=\hbox{$#1{#2#3}{\int}$}
 \vcenter{\hbox{$#2#3$}}\kern-.5\wd0}}

\title{A Minimal Mechanosensing Model Predicts Keratocyte Evolution on Flexible Substrates}
\runningtitle{Mechanosensing  Keratocytes} 

\author[1]{Zhiwen Zhang}
\author[2,3,*]{Phoebus Rosakis}
\author[4]{Thomas Y. Hou}
\author[5]{Guruswami  Ravichandran}
\runningauthor{Zhang, Rosakis, Hou and Ravichandran}
\affil[1]{Department of Mathematics, The University of Hong Kong, Pokfulam Road, Hong Kong SAR. Email: zhangzw@hku.hk}
\affil[2]{Department of Mathematics and Applied Mathematics, University of Crete, Heraklion 70013 Crete, Greece. Email: rosakis@uoc.gr}
\affil[3]{Institute of Applied and Computational Mathematics, Foundation for Research and Technology-Hellas, Voutes 70013 Crete, Greece}
\affil[4]{Computing and Mathematical Sciences, California Institute of Technology, Pasadena, CA 91125, USA. Email:hou@cms.caltech.edu}
\affil[5]{Division of Engineering and Applied Science, California Institute of Technology, Pasadena, CA 91125, USA. Email: ravi@caltech.edu}

\corrauthor[*]{rosakis@uoc.gr}

\papertype{Article}

\begin{document}

\begin{frontmatter}

\begin{abstract}
A \textcolor{black}{mathematical} model is proposed for shape evolution and locomotion of fish epidermal keratocytes on elastic substrates.
 The model is based on mechanosensing concepts: cells apply contractile forces onto the elastic substrate, while cell shape evolution depends locally on the substrate stress generated by themselves or external mechanical stimuli acting on the substrate. We use the level set method to study the behavior of the model numerically, and predict a number of distinct phenomena observed in experiments, such as (i) symmetry breaking from the stationary centrosymmetric to the well-known steadily propagating crescent shape, \textcolor{black}{(ii) asymmetric bipedal oscillations and traveling waves in the  \lam leading edge} (iii) response to mechanical stress externally applied to the substrate (tensotaxis),
 (iv) changing direction of motion towards an interface with a rigid substrate (durotaxis) and (v) the configuration of substrate wrinkles induced by contractile forces applied by the keratocyte.
 \end{abstract}

\end{frontmatter}

\section*{Introduction}

It has long been known that various types of biological cells exert forces that substantially deform their surroundings, such as the elastic substrate they crawl on, or the extracellular matrix they are embedded in \cite{harris,cellsdeformsubstrate,wrinkle2,wrinkleforce}.
It is also recognised that cells sense deformations or stresses that they themselves generate \cite{tensoduro}, or that are caused by external factors, and that they also sense the stiffness of the substrate \cite{rigiditysensing}. These activities are known as mechanosensing, and they facilitate some important modes of cell migration or evolution: tensotaxis \cite{tensotaxis}, the movement or protrusion towards regions of higher tensile stress, and durotaxis \cite{demboneedle}, the tendency to move towards
regions of higher stiffness. These processes play a key role in wound healing, fibrosis and tumor formation \cite{medical}.

The cells whose mechanosensing behavior has been studied the most are fibroblasts \cite{harris,demboneedle,hardsoft}. More recently it was determined that fish epidermal keratocytes also exert strong contractile forces on their elastic surroundings, to the extent that they can cause a sufficiently compliant elastic substrate to wrinkle \cite{wrinkle2}. Keratocytes are well known for their persistent, high-speed, steady locomotion while maintaining a characteristic crescent-like shape that is quite different from their stationary round configuration, e.g., \cite{needle2,polarvel}. Because of this, they have served as a model system for the study of cell locomotion on substrates of various types, through experiments \cite{wrinkle2,friction,polarvel,mushape} and theoretical modeling \cite{shapemogli,actinspeed1,mogilnervariousmodels,coupling,modeltraction,review}.

Theoretical models have largely focused on the detailed biophysical and biochemical processes within the cell \cite{mogilnervariousmodels,coupling}, but have rarely considered mechanosensing \cite{modeltraction,review,durohan}.

Here we adopt an alternative approach: we propose a mathematical model for the evolution of keratocytes on elastic substrates that is entirely based on hypotheses of active mechanosensing. The model is intentionally minimal in describing the cell, focusing instead on purely mechanical interaction of the \lam with the substrate, through active force generation, passive stress detection, and active response to stress sensing via local shape evolution. The proposed mechanism of cell evolution is a feedback loop: the \lam applies tractions onto the elastic substrate; the resulting stress field in the substrate depends on the instantaneous shape of the cell, while the evolution of the cell shape depends on the substrate stress, closing the feedback loop. The shape of the cell evolves according to a local evolution law: at each point on the \lam boundary, the normal boundary velocity is determined by the local stress state of the substrate, in a way that favors local protrusion under tension and retraction under compression.

We model the substrate as a 2D linear elastic isotropic medium, such as a thin sheet in plane stress, as in experiments on compliant silicone sheets \cite{wrinkle2} that facilitate the visualization of substrate deformation caused by keratocyte-applied tractions.

We assume that there is a centripetal retrograde velocity field in the \lam (representing actin flow) proportional to the traction the \lam applies to the substrate.  While appropriate for static keratocytes \cite{polarvel}, which are round in shape, and for fibroblasts of arbitrary shapes \cite{lemmon,rape} the centripetal form of the actin velocity field is less accurate for the steadily locomoting state of keratocytes \cite{polarvel}. 
In accordance with experimental observations \cite{polarvel,thermogliwaves1}, we thus include a generalization, where we assume the velocity field to be polarized in the direction of motion.

Tractions applied onto the substrate by the cell are assumed proportional to the actin velocity field relative to the substrate; they act as a body force in the elastic equilibrium of the substrate. This results in a stress field that is determined by the shape of the \lamp
The motion of the \lam boundary is determined by a competition between retrograde actin velocity and the actin polymerization speed normal to the boundary. We assume that at each boundary point, this speed is equal to a function of the component of the substrate stress normal to the \lam boundary. While we cannot point to the structural mechanism behind this, we note that actin fibers are known to act as tension sensors \cite{rigiditysensing,tensionsensor}; also cyclic variations in the assembly/disassembly rate of actin seem to be connected to traction fluctuations at focal adhesions \cite{plotnikov2}. This could point toward a link between polymerization speed and tension.

This constitutive assumption on the polymerization rate implies local tensotaxis: outward motion (protrusion) of the \lam edge is favored in regions of substrate tensile stress in the local normal direction;  retraction occurs locally if the \lam boundary normal is a direction of compression. Cells are known to move away from regions of compressive stress \cite{compresso}, in addition to favoring tensile stress.
 In the context of the model, such tension is generated by the cell exerting traction onto the substrate, but possibly also by external agents, such as microneedle manipulation of the substrate \cite{demboneedle,needle2,keratino}  in the vicinity of the cell.
As a result, given the shape of the \lamc the normal \lam boundary velocity is determined at each point. This determines the evolution of the \lam shape through the solution of a Hamilton-Jacobi equation, coupled to the elastic equilibrium equation. The resulting mathematical problem is amenable to numerical simulation via the level set method \cite{osher1988fronts,Chang} which has been applied to cell evolution study \cite{levelcell,mogilnervariousmodels}. In addition to the substrate stress field, the evolving shape of the \lam is the main output of the model.

Despite its simplicity, the model predicts different modes of locomotion behavior, owing to its rich bifurcation response.  In a computation starting from the annulus-shaped \lam typical of stationary keratocytes, a slight perturbation induces symmetry breaking and a topological change that leads to the well known steadily propagating crescent shape. This simulated sequence (Fig. \ref{figtransition}) closely resembles all stages of the observed transition from the static to the locomoting state of keratocytes reported in \cite{polarvel}; see also Fig. \ref{steady}.

\textcolor{black}{In addition, when a  parameter that controls polarization of the actin velocity field is increased, steady motion gives way to wiggly locomotion, with asymmetric bipedal oscillations of the \lamc similar to those observed \cite{theriotbipedal}. A further increase in the polarization results in the appearance of transverse traveling waves in the leading \lam edge, which were also reported in experiments \cite{thermogliwaves1}}.

Compressive stresses due to moving keratocytes in sufficiently thin silicone substrates cause the latter to wrinkle \cite{cellsdeformsubstrate,wrinkle2}; our model predicts the direction and relative magnitude of the wrinkles based on the computed substrate stress field (Fig. \ref{figwrinkle}).

Tensotaxis is the tendency of cells to move or extend protrusions toward regions of higher tensile stress, as observed with fibroblasts \cite{demboneedle}. In our simulations we start with a circular initial shape, representing a static \lam fragment as observed in \cite{needle2}. Exerting a force onto the substrate some distance from the fragment, but pointing toward it, breaks the symmetry; the fragment becomes crescent shaped, then moves steadily away from the force (Fig. \ref{Push}) in agreement with experiments \cite{needle2}. In another simulation, a fragment moves toward a force pointing away from it (Fig. \ref{Pull}). \textcolor{black}{Similar to  recent experiments on human keratinocytes \cite{keratino}, we find that a locomoting cell changes direction to move at right angles, then elongates toward a  needle pulling the substrate behind it, then gradually turns towards the needle. These are examples of tensotaxis, as the localized force creates either a compressive or tensile stress gradient (when pointing toward or away from the cell, respectively) which repels or attracts the fragment.}

On substrates with regions of different stiffness, cells similar to keratocytes lying initially on the softer region, have been observed to turn toward, and cross into, the stiffer portion of the substrate \cite{duroturning}. Under zero displacement boundary conditions, the simulation domain boundary is equivalent to an interface with a region of infinite stiffness (rigid). Simulated locomoting cells closer to one side of the boundary do not move straight; instead they follow a curving trajectory, approaching and eventually contacting the rigid boundary, simultaneously turning almost rigidly. This attraction by a rigid boundary is an instance of durotaxis \cite{duroturning}. The \lam motion (Fig. \ref{Turn}) agrees with observations of keratocytes following a curved trajectory while turning almost rigidly with little shape change, e.g., \cite{friction}.  \textcolor{black}{We also find reverse durotaxis; cells move away from a traction free boundary, as the later is equivalent to an interface with a softer material in the zero stiffness limit.}

 \section*{Methods}

We model fish epidermal keratocytes crawling on a thin deformable substrate, represented by a 2D medium that occupies the entire plane. It is composed of linear elastic homogeneous isotropic material undergoing small in-plane deformations. The linear theory of elasticity is used; out-of-plane displacements are neglected. The time dependent displacement vector field is ${\bu=\bu(\bx,t)}$, where $\bx$ is position vector in the plane and $t$ is time. The stress tensor is related to the displacement gradient
\be\label{sss}\bS=\lambda(\nabla\cdot \bu)\bI+\mu(\nabla \bu+\nabla\bu^T).\ee
 in the isotropic case considered here, where $\lambda>0$ and $\mu>0$ are the Lam\'e constants and $\bI$ the identity tensor.

The cell is modeled as a time-dependent region $\Omt$ in the plane.
The cell interacts with the substrate by exerting forces on it. This occurs mostly in the \lamc while the part of the cell body around the nucleus need not even be in contact with the substrate \cite{mushape}. Accordingly, $\Omt$ represents the lamellipodium only. The forces exerted by the lamellipodium onto the substrate are assumed to be in-plane; they are due to retrograde actin flow within the cell caused by myosin contraction pulling at radial actin fibers; see e.g., \cite{polarvel}. The actin exerts a force onto the substrate through drag and/or adherence to focal adhesions that are attached to it. For stationary cells, there is evidence \cite{polarvel,lemmon,rape} that the actin network within the cell arranges itself radially from the centroid of the cell and exerts centripetal tractions onto the substrate \cite{wrinkle2}. For fibroblasts on elastic substrates this occurs independently of shape \cite{rape}. Stationary keratocytes assume a disk shape; the \lam is approximately an annulus surrounding the nucleus. The direction of the actin flow velocity is radially inward toward the cell center \cite{polarvel} and the magnitude increases with distance from the centroid. 

We generalize this for moving cells.   We assume that the actin velocity relative to the substrate is radially inward towards a point $\bx_0(t)$ traveling with the cell and its magnitude increases linearly with distance from $\bbc$. Thus the actin velocity in the substrate frame is
$$\bv_s(\bx,t)=-\gamma(\bx-\bx_0(t))$$
for $\bx$ in $\Omt$  with the actin velocity coefficient $\gamma>0$ a constant. Further, we suppose that the traction exerted onto the substrate by the keratocyte \lam is $\bb=\eta\bv_s$ where $\eta>0$ is a viscosity coefficient. As a result we have
\be\label{b0}\bb(\bx,t)=-K\chi_\Omt(\bx)(\bx-\bx_0(t)),\ee
where $ K=\gamma\eta$ and $\chi_\Omt(\bx)=1$ for $\bx$ in $\Omt$ and $0$ outside $\Omt$ is the characteristic function of $\Omt$. The total external force per unit area acting on the cell is $-\bb(\bx,t)$, the reaction exerted by the substrate. Since the process is quasistatic, the cell must be self-equilibrated, namely, 
\be\label{ceq}\int_{\Omt} \bb(\bx,t) d\bx=\bo.\ee
In view of  Eq.~\eqref{b0}, this dictates $\bx_0=\bbx$, the cell centroid, given by
\be\label{cc}\bbx=\bbx(t)=\frac{\int_\Omt \bx d\bx}{\int_\Omt d\bx }.\ee
This dictates
\be\label{vs1}\bv_s(\bx,t)=-\gamma(\bx-\bbx),\ee
so that
\be\label{b00}\bb(\bx,t)=-K\chi_\Omt(\bx)(\bx-\bbx(t))={\eta}\bv_s(\bx,t).\ee
The substrate experiences an in-plane body force (per unit substrate area) equal to $\bb(\bx,t)$, representing tractions on a 2D substrate exerted by another 2D body (the cell) in contact with it. Quasistatic equilibrium for the substrate reads
\be\label{equil}\diiv \bS(\bx,t)+\bb(\bx,t)=\bo.\ee
Here $\bS$ is the stress in the substrate, related to the substrate displacement via Eq.~\eqref{sss}, while $\bb$ is exerted by the cell onto the substrate.

A central ingredient of our model is the evolution law that governs the motion of the cell boundary curve $\pot$. It is based on the notion that cells can detect stress in the substrate (mechanosensing)  \cite{tensionsensor} and make local adjustments to their shape accordingly.

 In order to characterize the moving curve $\pot$, it suffices to specify its normal velocity $V_n(\bx,t)$ at each $\bx\in\pot$ and time $t$. To begin with, we follow previous models in assuming
\be\label{vn}V_n=\bv_s\cdot\bn+v_p\quad\hbox{on } \pot\ee
\cite{actinspeed1,actinspeed2}. Actin filaments polymerize at the boundary with outward normal speed $v_p$ but also flow inwards with velocity $\bv_s$ whose normal component is $\bv_s\cdot\bn$. Thus the net normal boundary velocity $V_n$ is the excess of the polymerization speed $v_p$ over the retrograde inward actin flow speed in the direction normal to the cell boundary. It remains to characterize the polymerization speed $v_p$. A point of departure from other models of keratocyte evolution \cite{shapemogli,actinspeed1,mogilnervariousmodels,coupling,modeltraction,review} is the incorporation of mechanosensing in a constitutive relation for $v_p$.

 In contrast with \cite{actinspeed2}, we do not take $v_p$ to be constant.
We include two contributions:
\be\label{vp}v_p=G(\bn\cdot\bS\bn)+\Lambda\left(1-A(t)/A(0)\right)\quad\hbox{on } \pot.\ee
 The second term in Eq.~\eqref{vp}  is a penalty term that tends to maintain the area $A(t)$ of $\Omt$ constant ($\Lambda=$const.$>0$.) \textcolor{black}{The  rationale behind the first term is as follows. We make a mechanosensing hypothesis, which we refer to as local tensotaxis: the lamellipodium boundary tends to protrude locally in areas of tension and recede in areas of compression. This is motivated by a global tensotaxis behavior: cells are known to move away from regions of compressive stress \cite{compresso}, in addition to favoring tensile stress \cite{demboneedle}. Since stress $\bS$, being a tensor, can be both compressive and tensile at the same point (in different directions) we clarify the precise meaning of tension and compression.  On an isotropic substrate there are no other special directions, except the lamellipodium boundary unit normal $\bn$. It is reasonable to choose  the component of stress in this normal direction, $\bn\cdot\bS\bn$, as the one related to the polymerization rate. An obvious choice would be a linear relation between normal polymerization velocity  and normal tension, however, we require the velocity to remain bounded as cells seem to move with bounded speeds on substrates, rarely exceeding a few microns per second, so it is reasonable to assume instead a relation that saturates for large values of tension.
 Thus in Eq.~\eqref{vp} we choose
 \be\label{gz}G(z)=\beta\frac{z}{\sigma_0+|z|},\ee}
 which is an odd, increasing function that remains bounded for large values of its argument, with $\beta$ a positive mobility coefficient and $\sigma_0$ a constant with dimensions of stress.  Accordingly, apart from the first term in Eq.~\eqref{vp}, $v_p$ changes signs depending on whether the normal stress component $\bn\cdot\bS\bn$ is tensile or compressive.
While we cannot point to the structural mechanism behind this, we note that actin fibers are known to act as tension sensors \cite{rigiditysensing,tensionsensor}; also cyclic variations in the assembly/disassembly rate of actin seem to be connected to traction fluctuations at focal adhesions \cite{plotnikov2}. This could point toward a link between polymerization speed and tension.

\textcolor{black}{ A  generalization of our model is motivated by observations \cite{polarvel,thermogliwaves1} of the actin velocity field of locomoting keratocytes, which loses radial symmetry and becomes polarized in the direction of cell motion \cite{polarvel}. 
Our approach is to model this variation of actin velocity in a phenomenological yet minimal form.} Accordingly, we assume that at a given distance from the centroid, the actin velocity in the cell frame is more pronounced in the direction of motion than in the perpendicular direction, depending on the cell centroid velocity $\bbv=\dot\bbx$. We still assume that $\bv_s$ is linear in $\bx-\bx_0$, but with magnitude that is larger in the direction $\bbv$ of cell motion:
\be\label{v2}\bv_s=-\gamma(\bI+e \bbv\otimes\bbv)(\bx-\bx_0),\ee
where the actin velocity coefficient $\gamma>0$ and polarization coefficient $ e \ge 0$ are constants. In a basis with vectors along, and normal to, the direction of cell motion, the matrix
$$\bI+e \bbv\otimes\bbv=
\left(
\begin{array}{ccc}
1+e|\bbv|^2 & 0 \\
 0& 1 \\
\end{array}
\right).
$$
Thus the velocity component along the direction of cell motion is amplified by a factor $1+e|\bbv|^2$ compared to the radially symmetric actin velocity field. When $\bbv=\bo$, or for the choice $e=0$, the velocity field Eq.~\eqref{v2} reduces to the radially symmetric one, Eq.~\eqref{vs1}.
 Cell equilibrium Eq.~\eqref{ceq} with  $\bb=\eta\bv_s$ and $\bv_s$ given by Eq.~\eqref{v2} determines
\be\label{b2}\bb(\bx,t)=-\chi_\Omt(\bx)K(\bI+e \bbv\otimes\bbv)(\bx-\bbx).\ee
Once an initial lamellipodium shape $\Om_0$ at $t=0$ is specified, further evolution is governed by the normal velocity $V_n$, Eq.~\eqref{vn}, where $v_p$ is given by Eq.~\eqref{vp}, $v_s$ is determined by Eq.~\eqref{v2}, and the stress $\bS$ is obtained from the solution of Eqs \eqref{equil}, \eqref{sss}, with body force $\bb$ from Eq.~\eqref{b00}.

We use the level set method \cite{osher1988fronts,Chang} which has been successfully applied to cell evolution study, e.g., \cite{levelcell,mogilnervariousmodels} to solve for the evolution of the \lam boundary $\pot$ together with the other model equations. The \emph{level set function} $\ph(\bx,t)$ vanishes on $\pot$, is positive inside $\Omt$ and negative outside it. It evolves according to the \emph{level set equation}
\be\label{lev1}\ph_t-V_n|\nabla\ph|=0,\ee
with $V_n$ the normal velocity of $\pot$, which is determined by the equation $\ph=0$. The model thus comprises Eqs \eqref{equil}, \eqref{lev1}, with $\bb$ given by Eq.~\eqref{b2}, $V_n$ supplied by Eqs \eqref{vn}, \eqref{vp}.

\textcolor{black}{ 
\subsubsection*{Nondimensional Form and Independent Parameters}
The model involves eight constitutive parameters. The substrate is characterized by the the Lam\'e constants  $\lambda>0$ and $\mu>0$, while the cell by the kinetic coefficient $\beta$, actin velocity coefficient $\gamma$, viscosity $\eta$, velocity polarization $e$, area penalty coefficient $\Lambda$ and stress coefficient $\sigma_0$.  We define the nondimentional variables
\[\tilde \bx =(\gamma/\beta)\bx,\quad  \tilde t= t/\gamma,\quad \tilde v=v/\beta ,\quad \tilde \bS= (1/\sigma_0)\bS
\]
and the nondimentional constants 
\[\tilde \eta=(\beta^2/(\gamma\sigma_0))\eta,\quad  \tilde e=\beta^2 e, \quad\tilde \Lambda=\Lambda/\beta.\]
We then revert to the same notation (without tilde) for the nondimentional variables and constants; this is equivalent to setting $\beta=1$, $\gamma=1$, $\sigma_0=1$ in the original system. The remaining  independent parameters for the cell are $\eta$, $e$, $\Lambda$.
Since the body force field Eq.~\eqref{b00} is independent of the Lam\'e  moduli $\lambda$, $\mu$, for null displacement or traction-free boundary conditions, a theorem of linear elasticity \cite{gurtin} asserts that the stress field depends on $\lambda$, $\mu$ only through their ratio, or equivalently Poisson ratio $\nu=\frac{\lambda}{2(\lambda+\mu)}$.  Thus there is one independent nondimensional parameter $\nu$ for the substrate, or a total of 4 nondimensional model parameters. Unless otherwise specified, in our simulations we used a standard parameter set of 
\be\label{sp} \beta=2.5,\quad \gamma=0.8, \quad K=\gamma\eta=3, \quad e=2,\quad \Lambda=20,\quad \nu=1/4.\ee 
Exceptions are used for study of the effect of $K$ and $e$; these are the only two parameters that we vary.}

\subsection*{Figures}

\begin{figure}
 \centering 
\subfloat[\label{tr1}]{\includegraphics[width=1.5in]{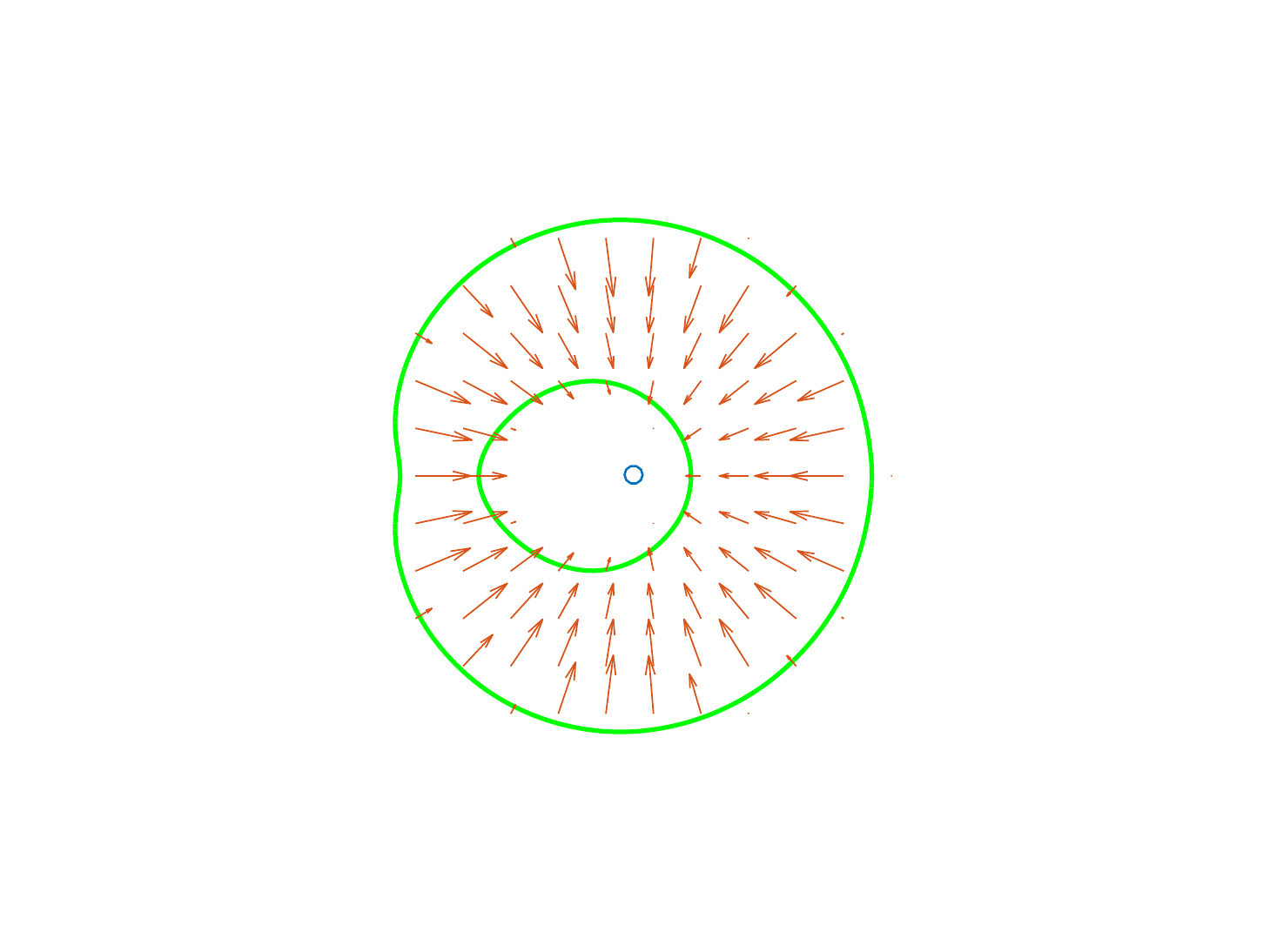}\hskip-0.5cm}
\subfloat[\label{tr2}]{\includegraphics[width=1.5in]{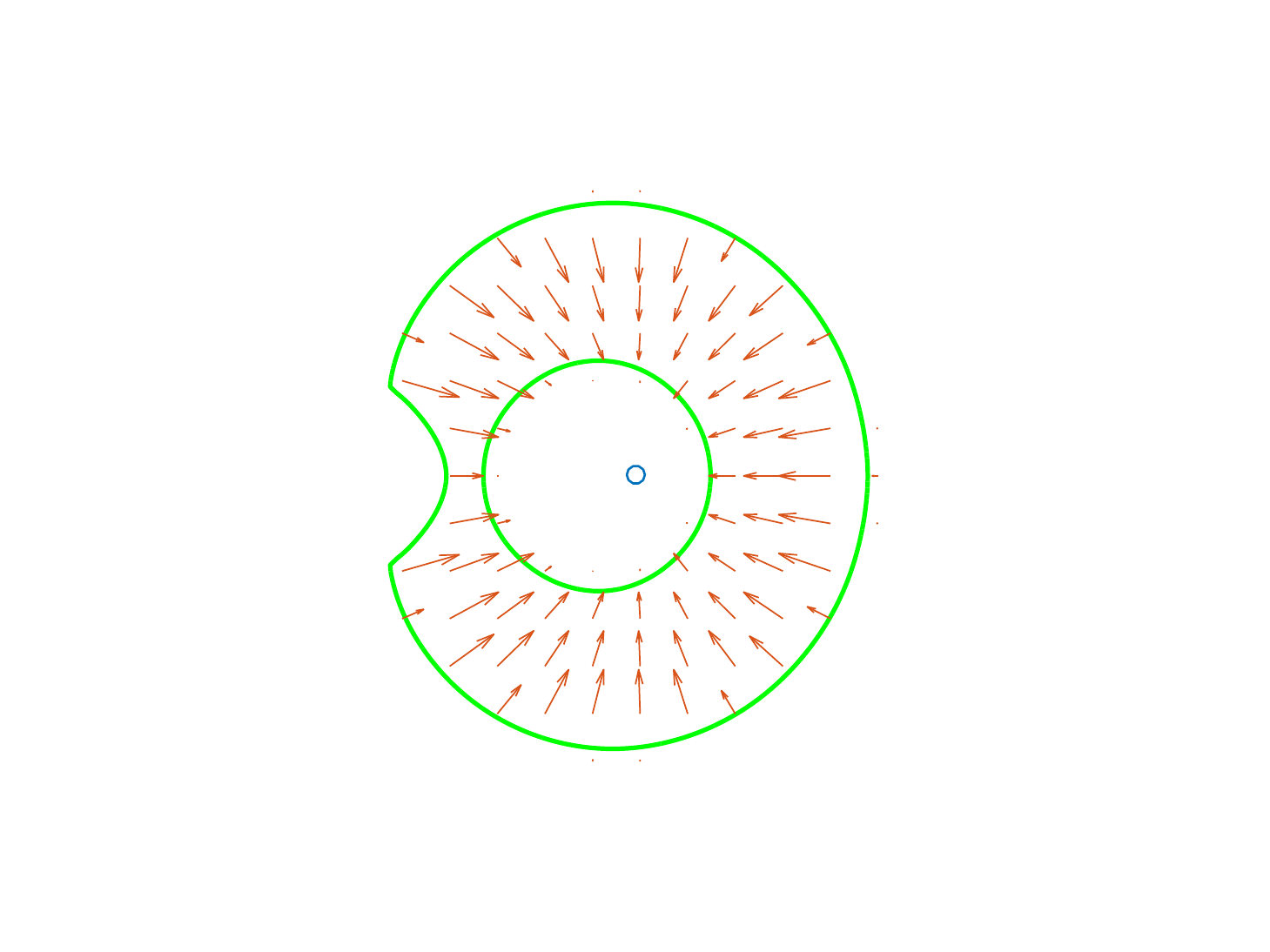}\hskip-0.5cm}
\subfloat[\label{tr3}]{\includegraphics[width=1.5in]{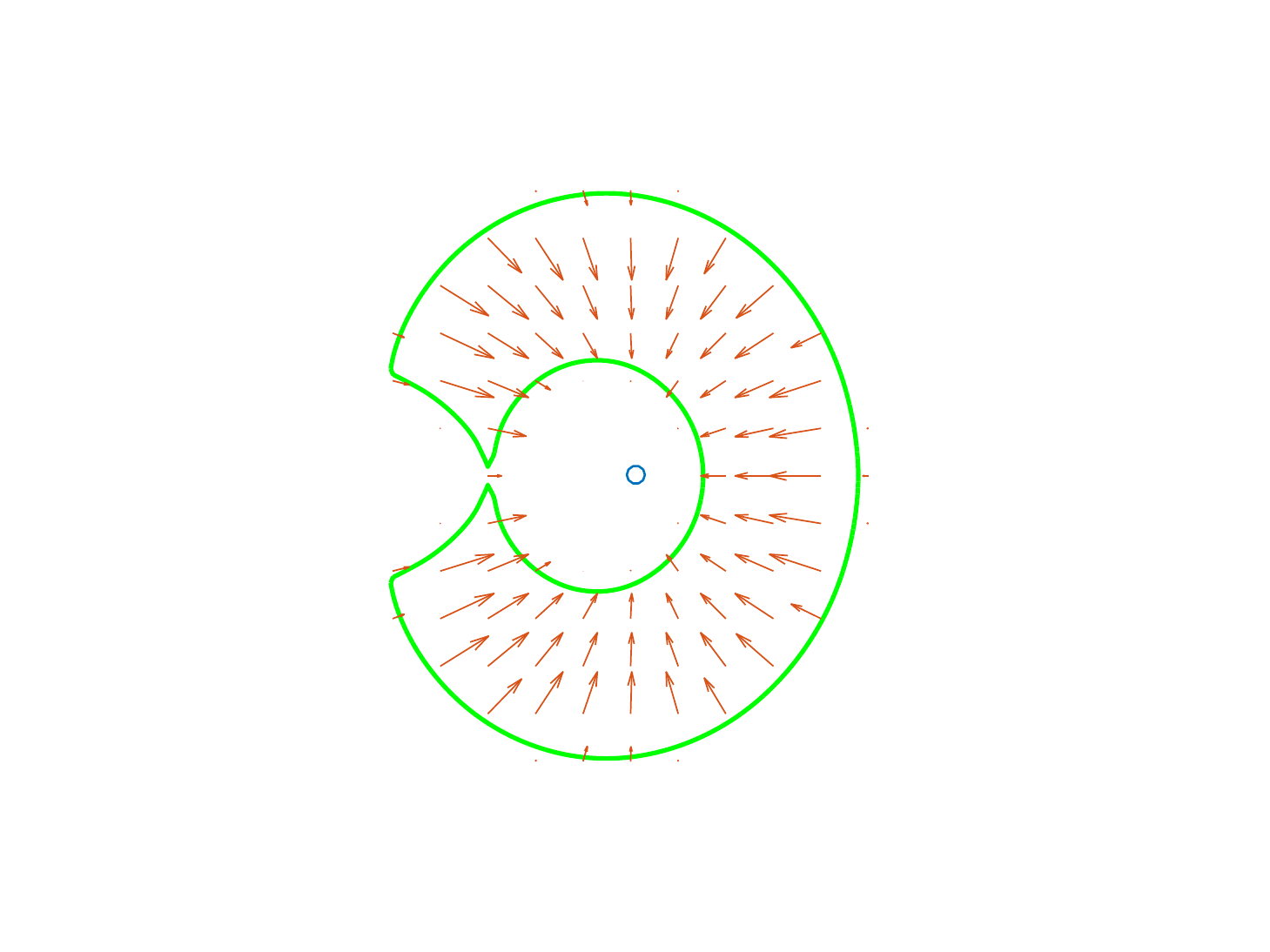}\hskip-0.5cm}
\subfloat[\label{tr4}]{\includegraphics[width=1.5in]{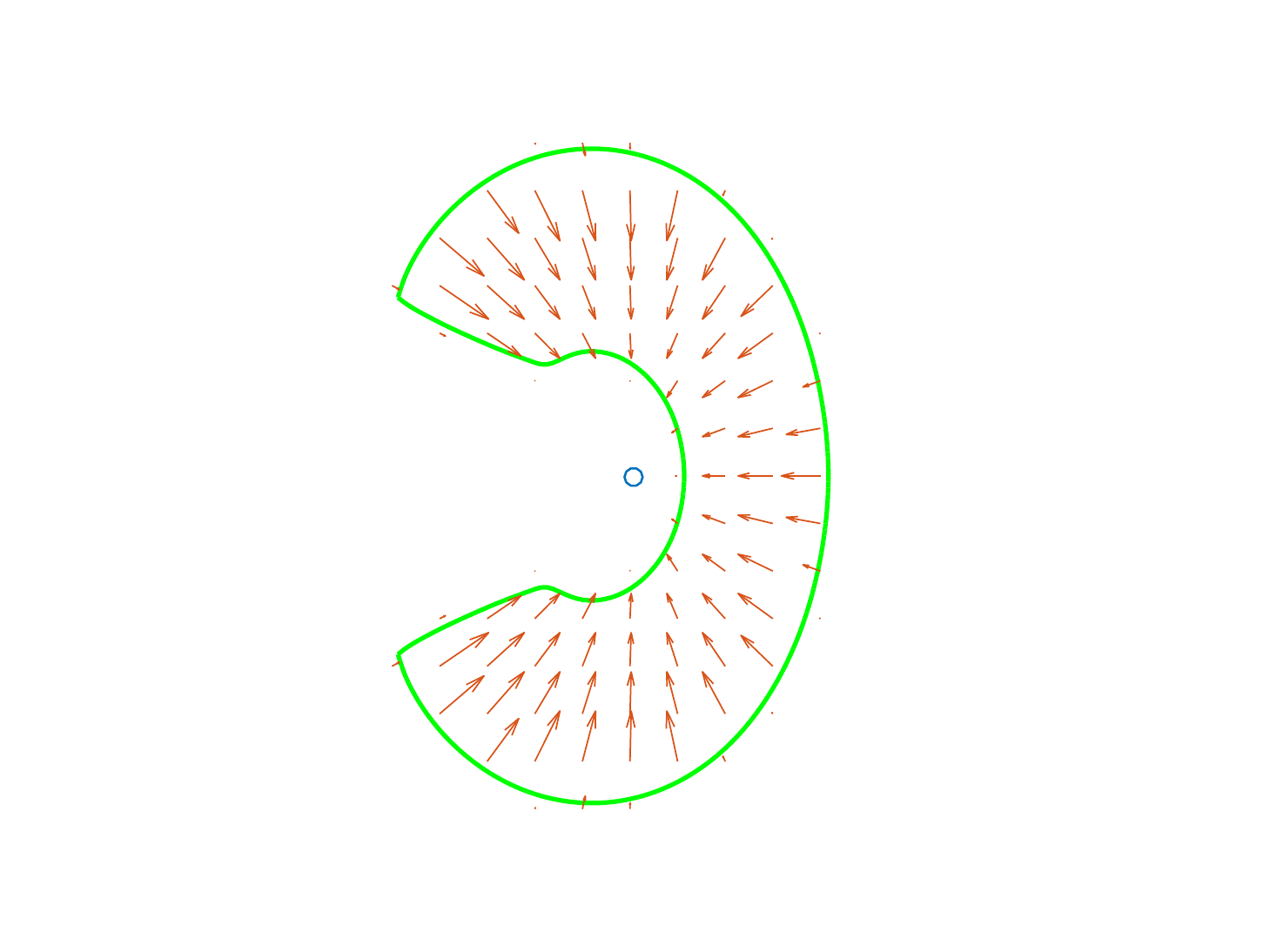}\hskip-0.5cm}
\subfloat[\label{tr4e}]{\includegraphics[width=1.5in]{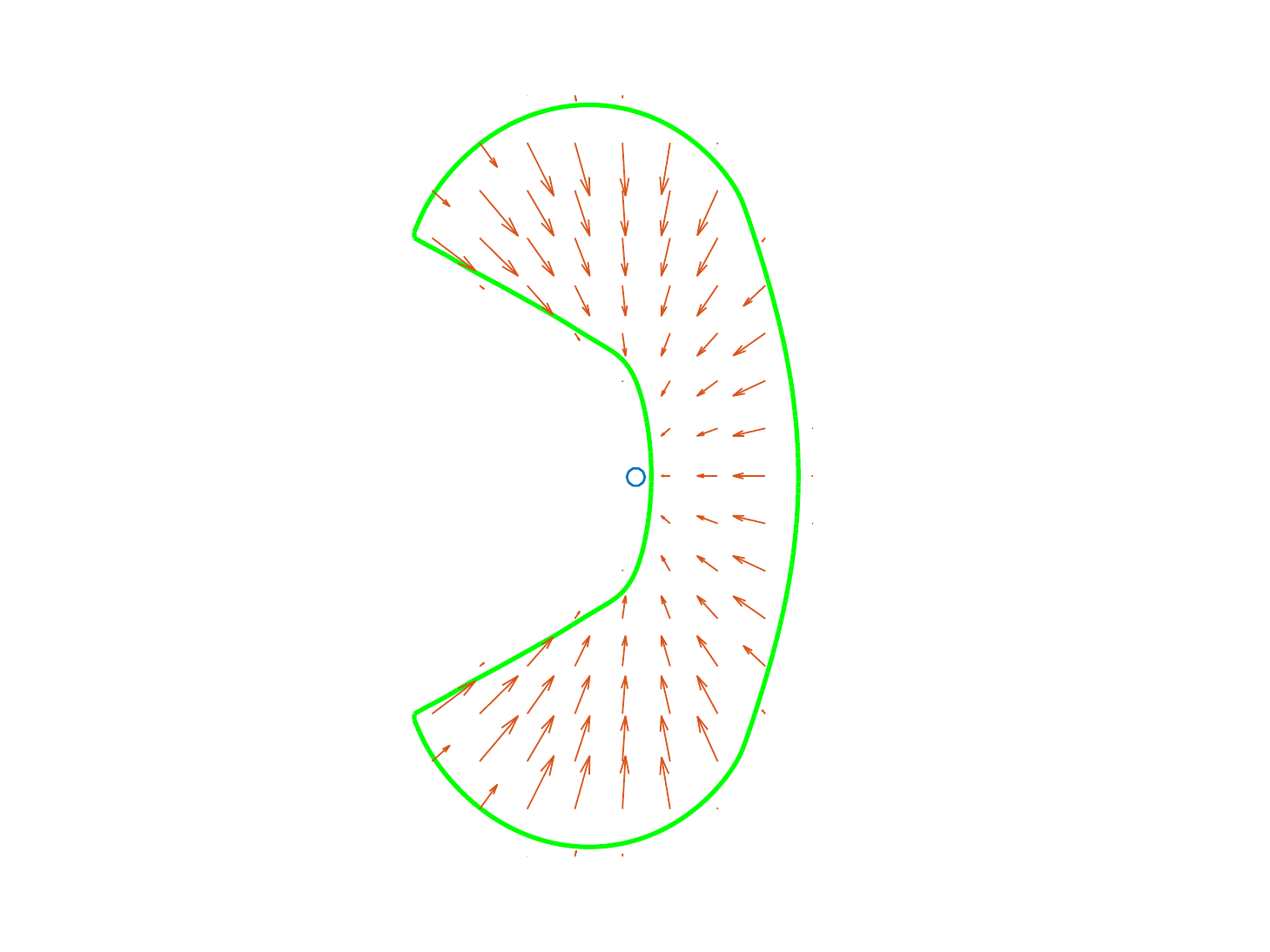}}
\\
\subfloat[\label{tr5}]{\includegraphics[width=6in]{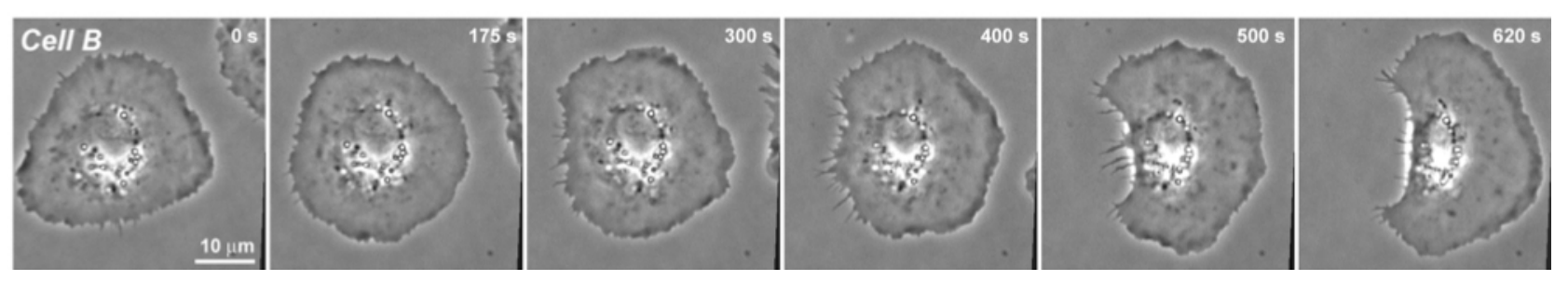}}
 \caption{Transition from the stationary annulus configuration to the locomoting crescent shape of the keratocyte \lamc from our model simulation with standard parameters Eq.~\eqref{sp}: (a) Initial condition for model simulation: stationary annular \lam with centripetal velocity field and imperfection. (b) Retraction (pinching) of the left side. (c) Topological transition. (d) Motile horseshoe shape.
 (e) Fully developed locomoting crescent shape; motion is to the right. (f)Image sequence of observed transition from \cite{polarvel}.}
 \label{figtransition}
\end{figure}

\begin{figure}
\centering
	\subfloat[\label{k3e1}]{\includegraphics[width=1.4in]{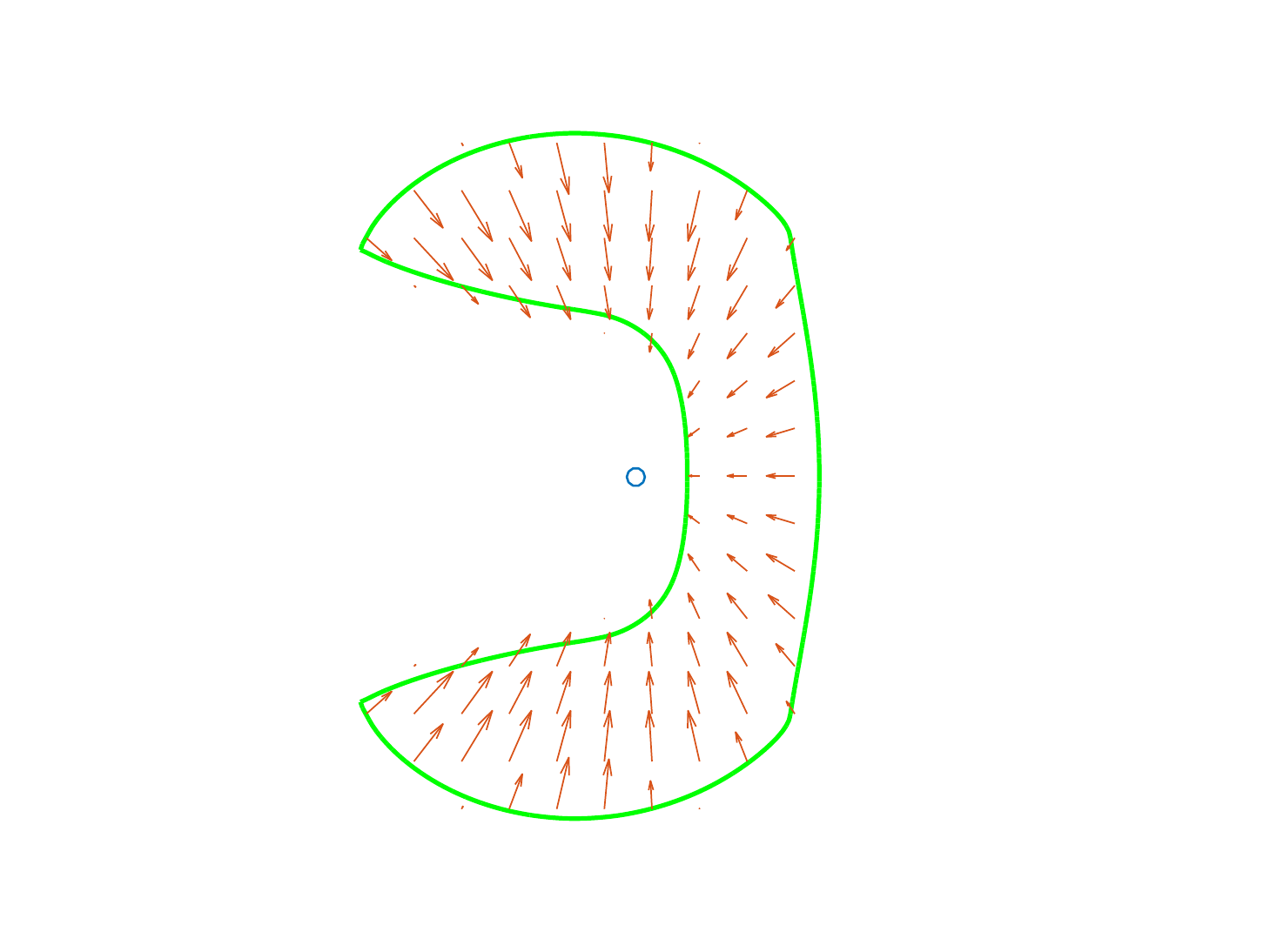}\hskip-0.2cm}
	\subfloat[\label{k3e25}]{\includegraphics[width=1.4in]{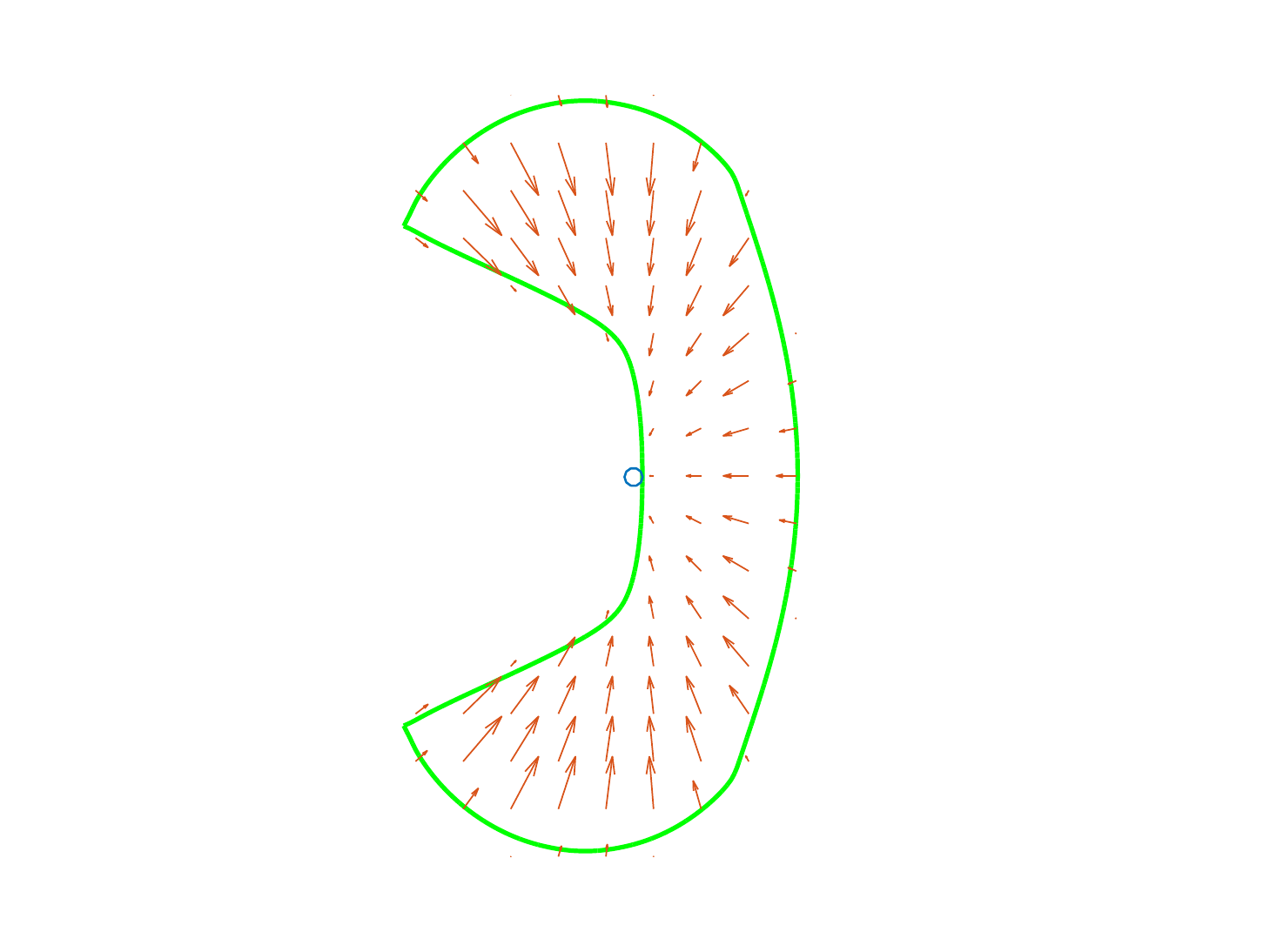}\hskip-0.2cm}
        \subfloat[\label{k3e3}]{\includegraphics[width=1.4in]{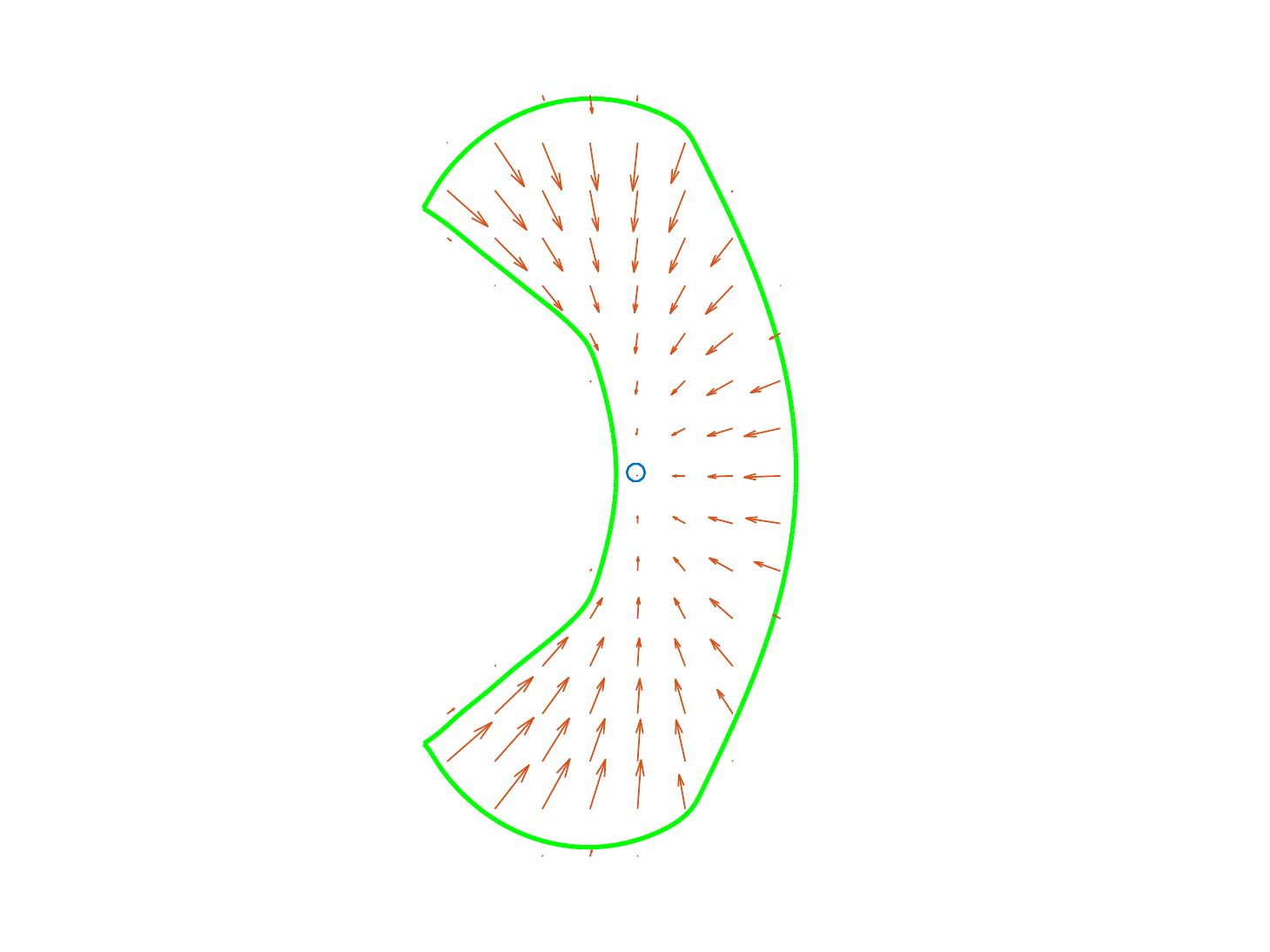}\hskip-0.2cm}	
	\subfloat[\label{k10e1}]{\includegraphics[width=1.4in]{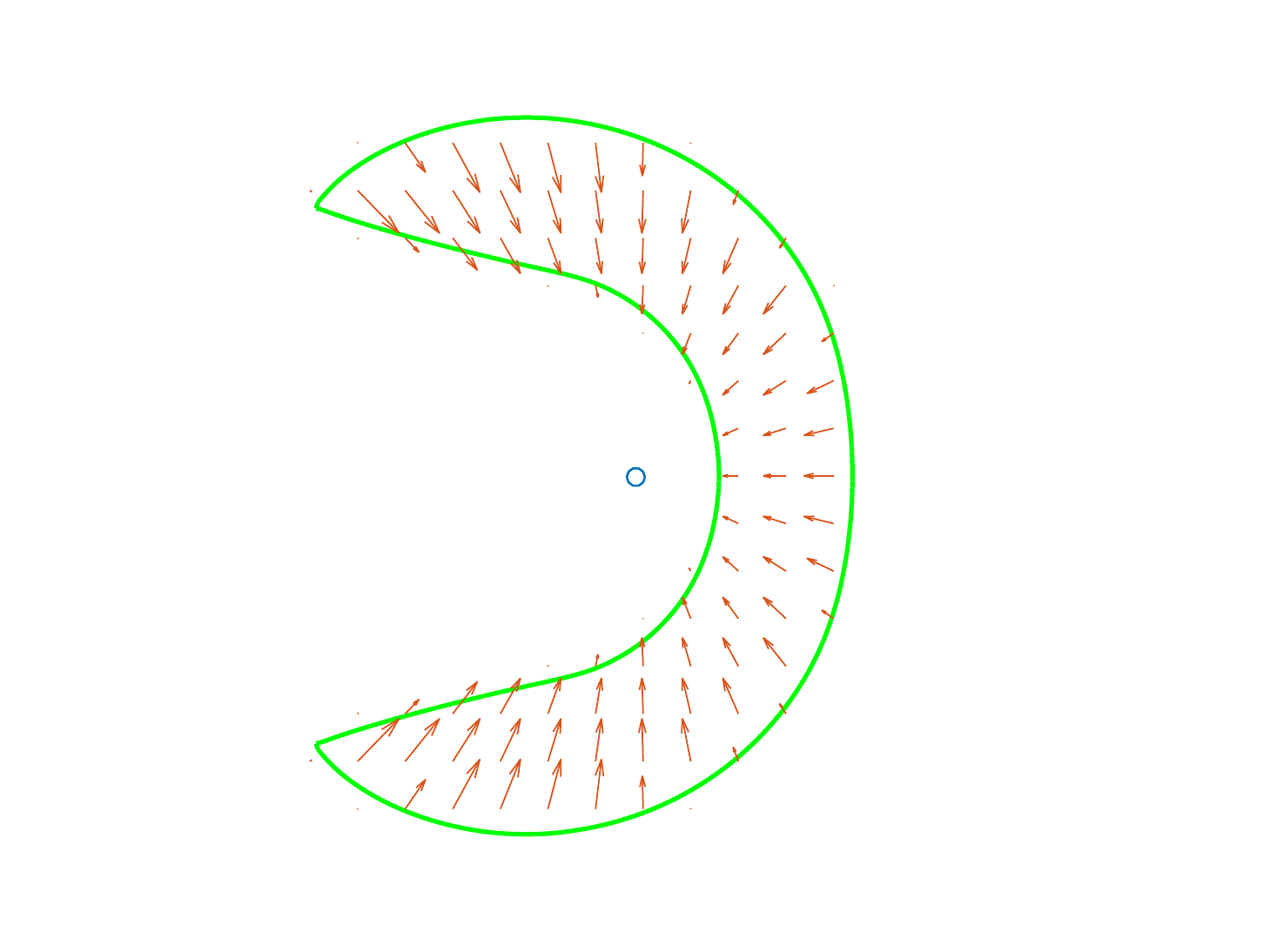}\hskip-0.2cm}
	\subfloat[\label{k10e3}]{\includegraphics[width=1.4in]{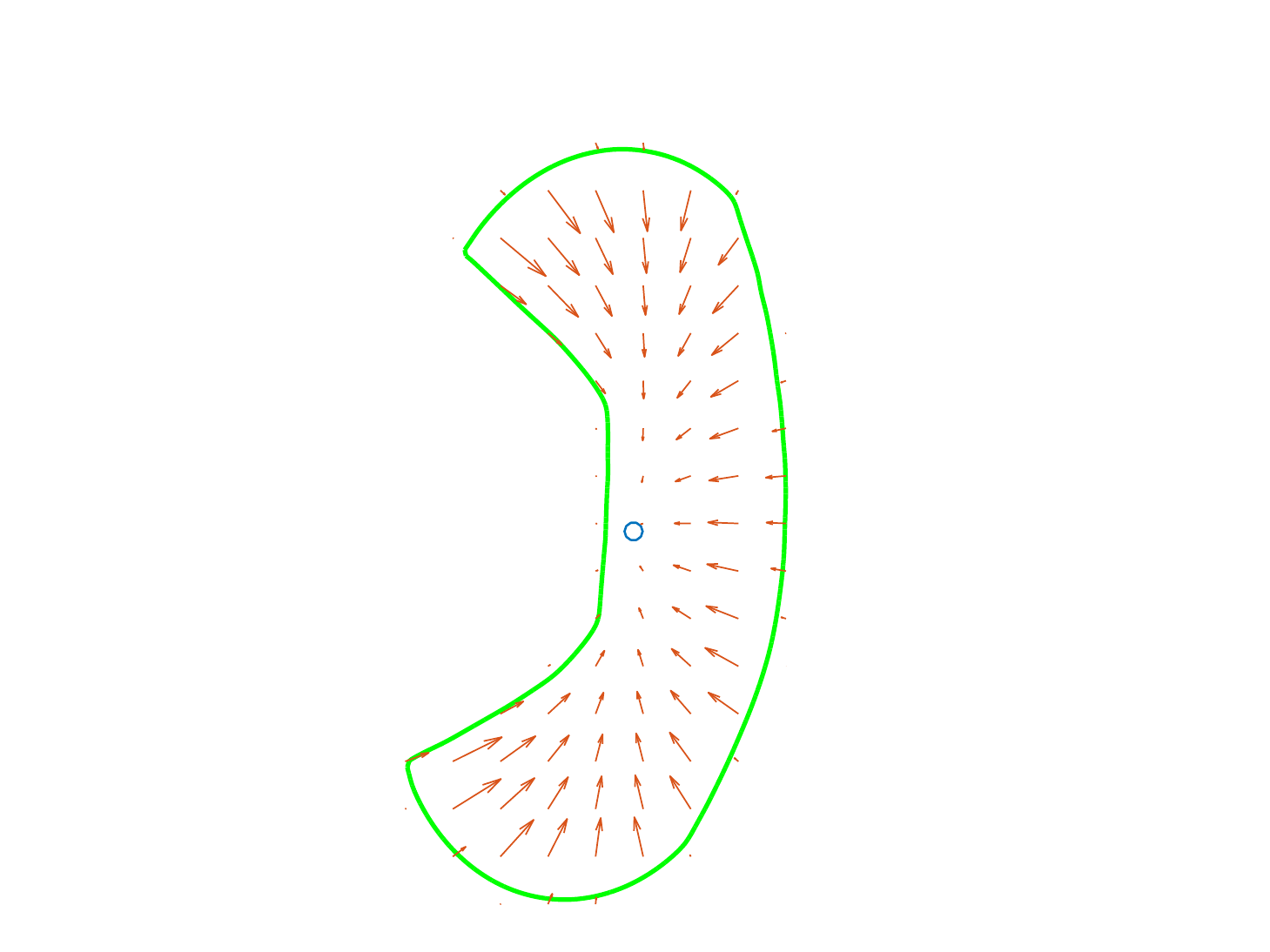}}
 \caption{Parameter dependence of fully developed locomoting crescent \lam shape in the low actin velocity polarization regime. Motile keratocyte  moving  to the right. (a)-(c)  $K=3$ and $e=0,1.5,2$, respectively. (d),(e) $K=10$ and $e=0,2$, respectively. Green curve is \lam shape, actin velocity vectors are shown red.}
 \label{123}
\end{figure}

 \begin{figure}
	\centering
	\subfloat[\label{t0}]{\includegraphics[width=3.0in]{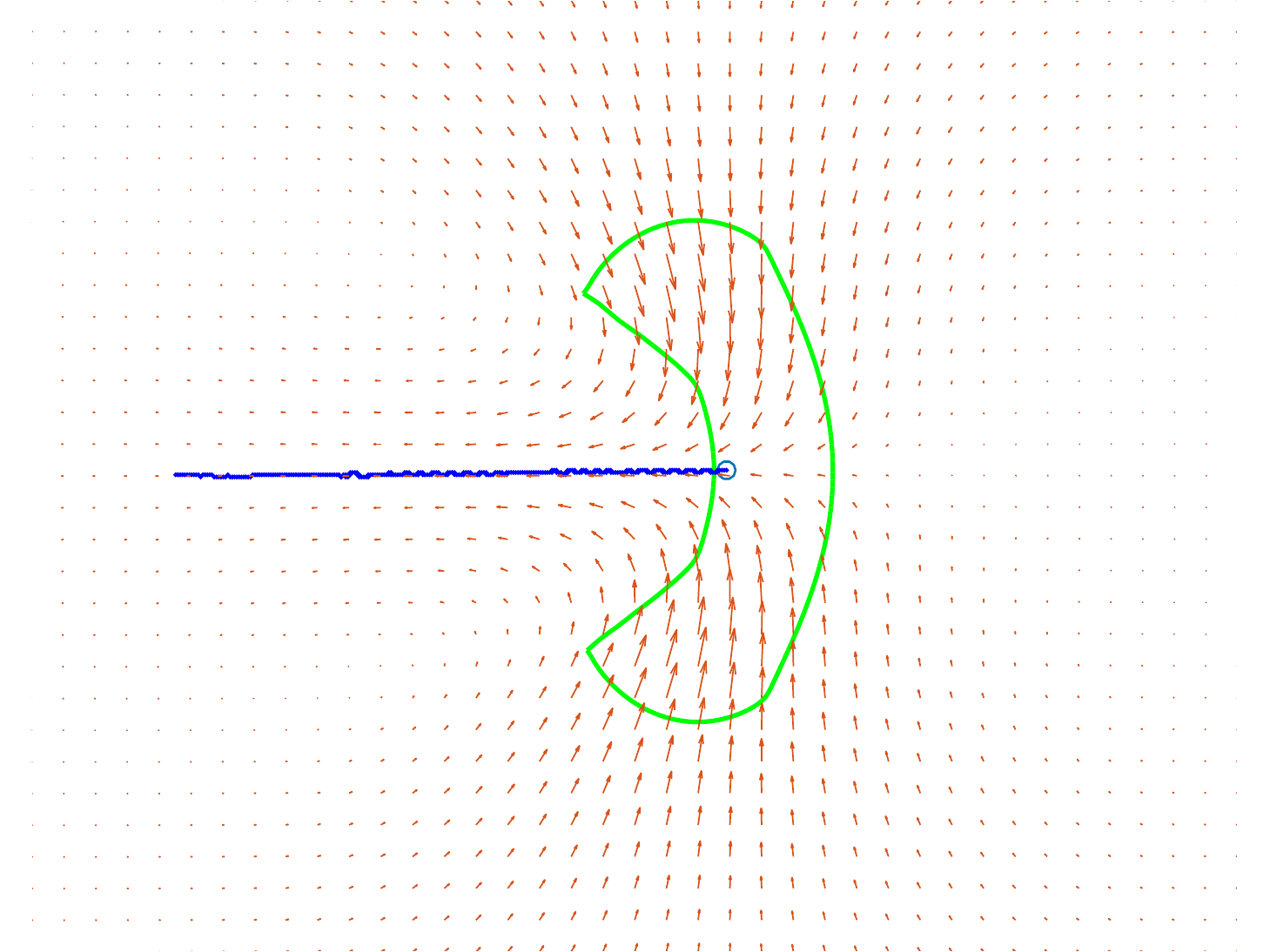}}\quad \quad
	\subfloat[\label{t1}]{\includegraphics[width=3.0in]{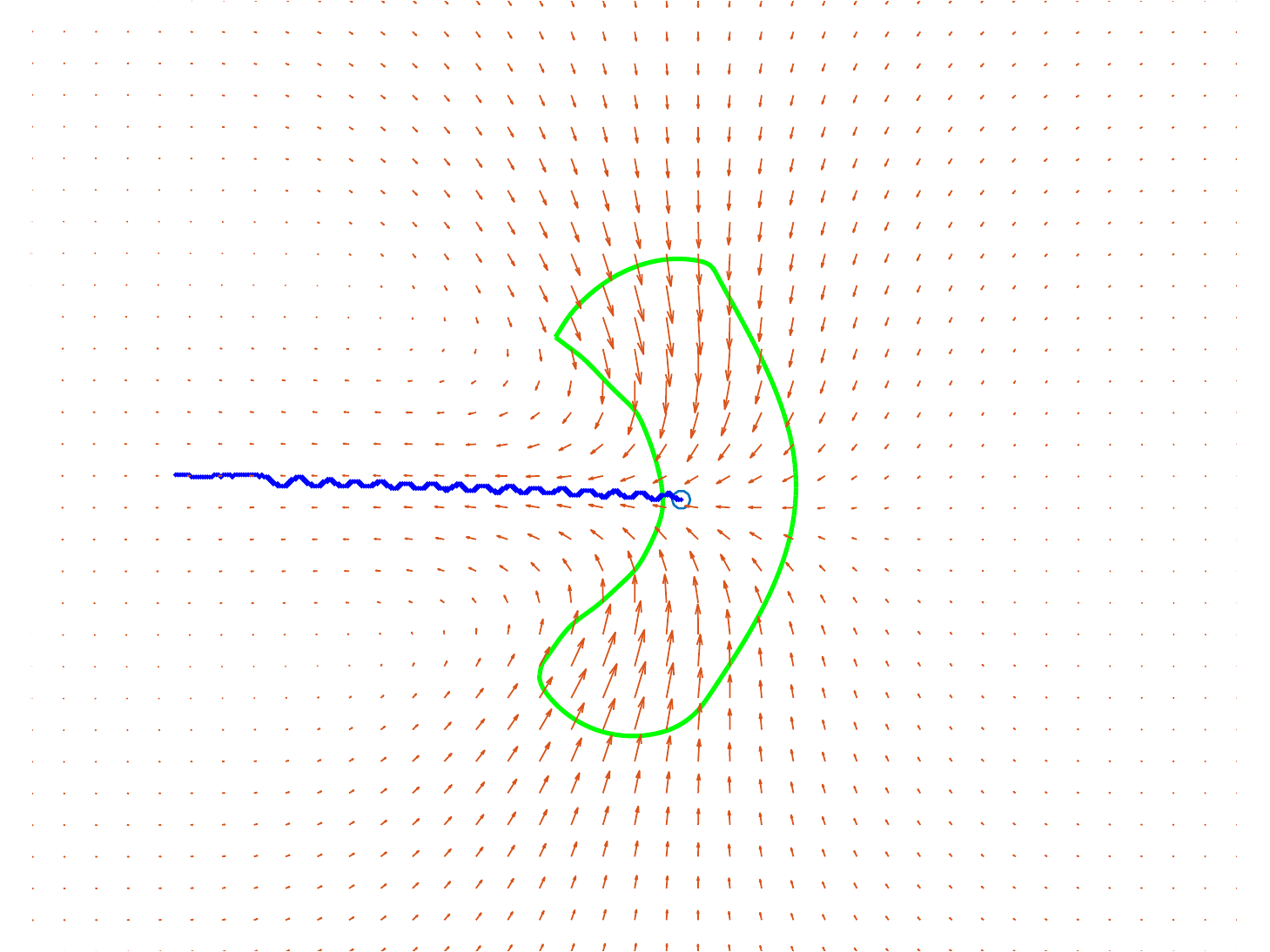}} 
	\\
	\subfloat[\label{t2}]{\includegraphics[width=3.0in]{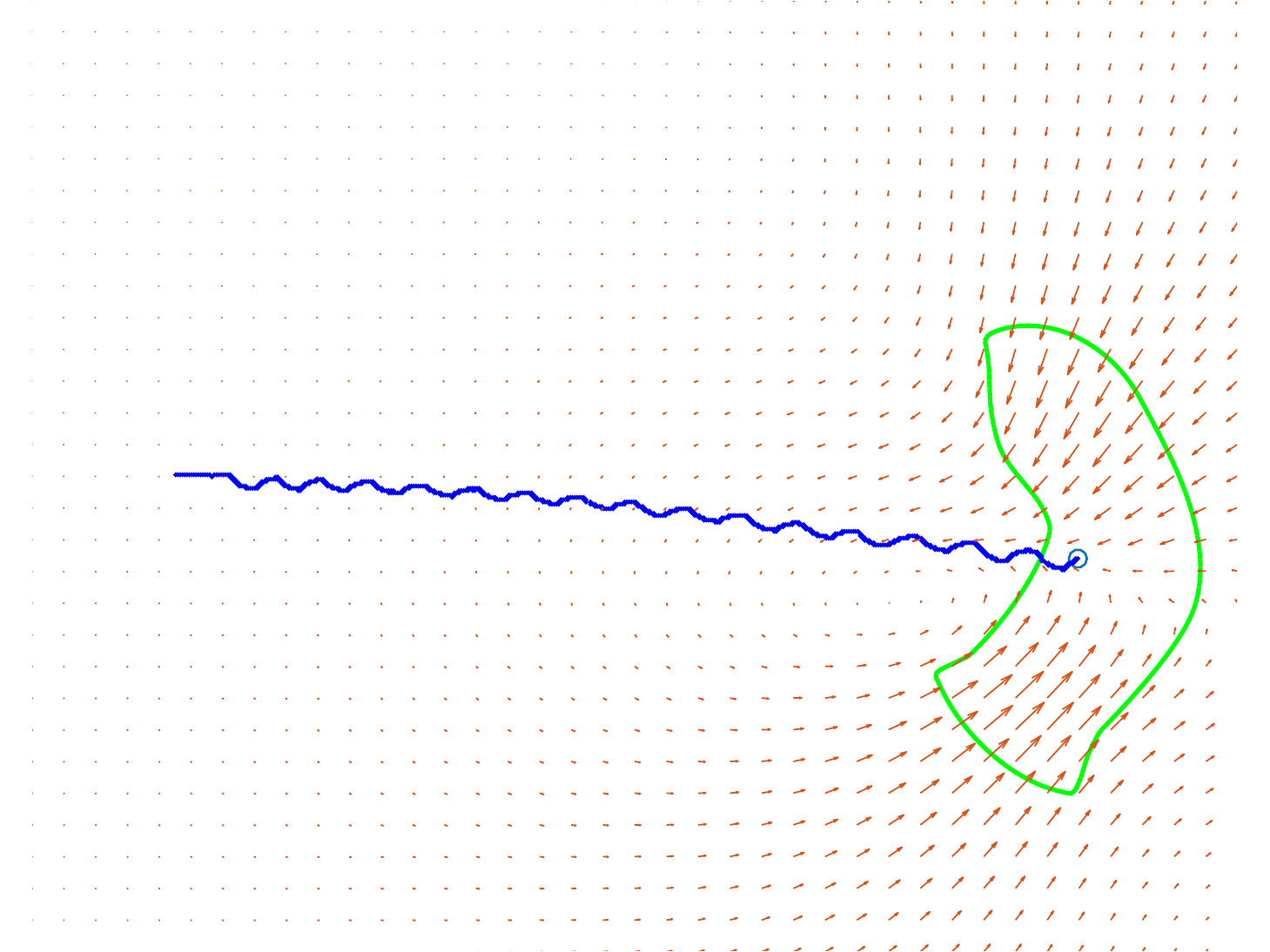}}\quad \quad
	\subfloat[\label{t3}]{\includegraphics[width=3.0in]{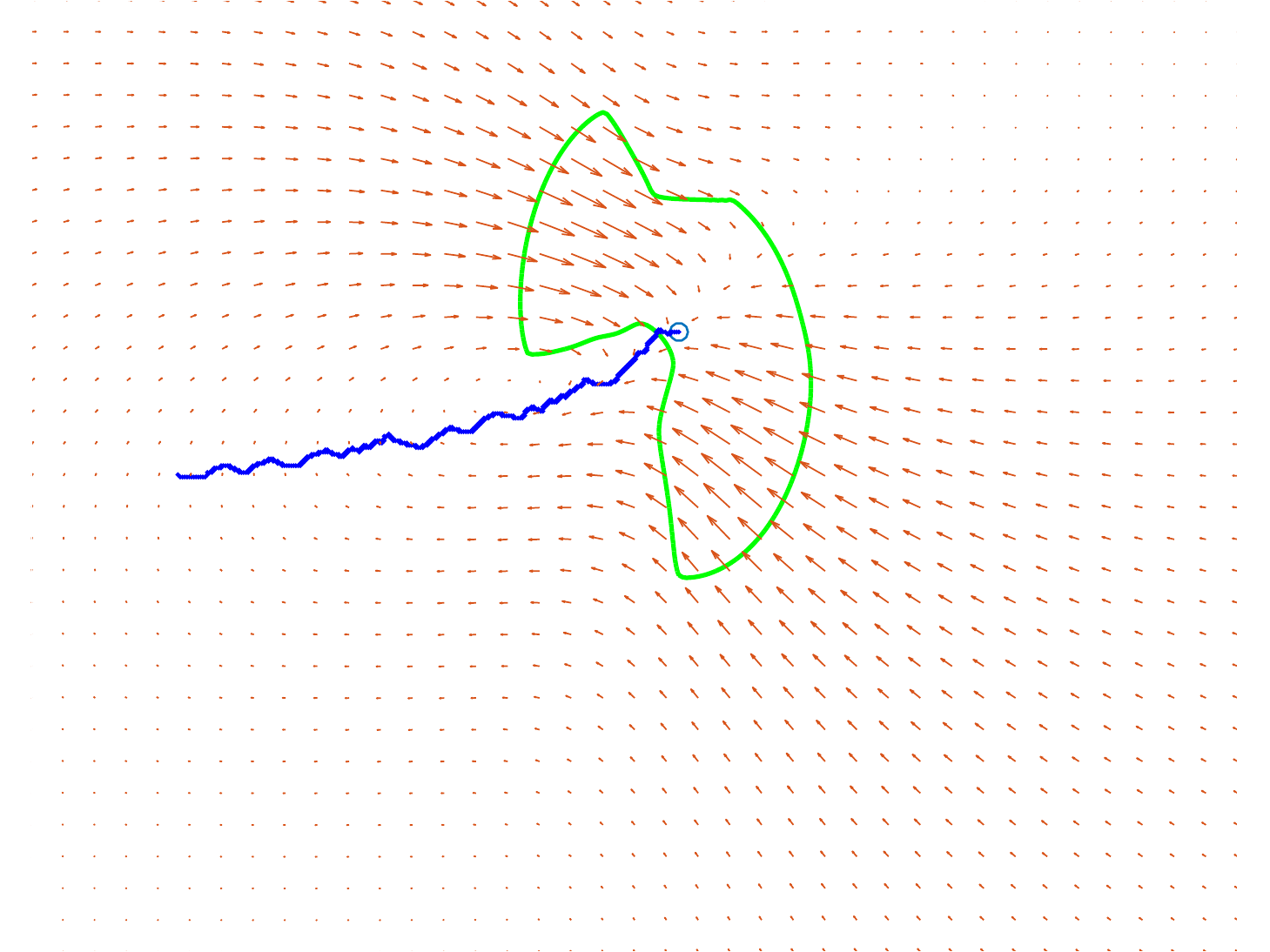}} 
	\\
	\subfloat[\label{t4}]{\includegraphics[width=2.5in]{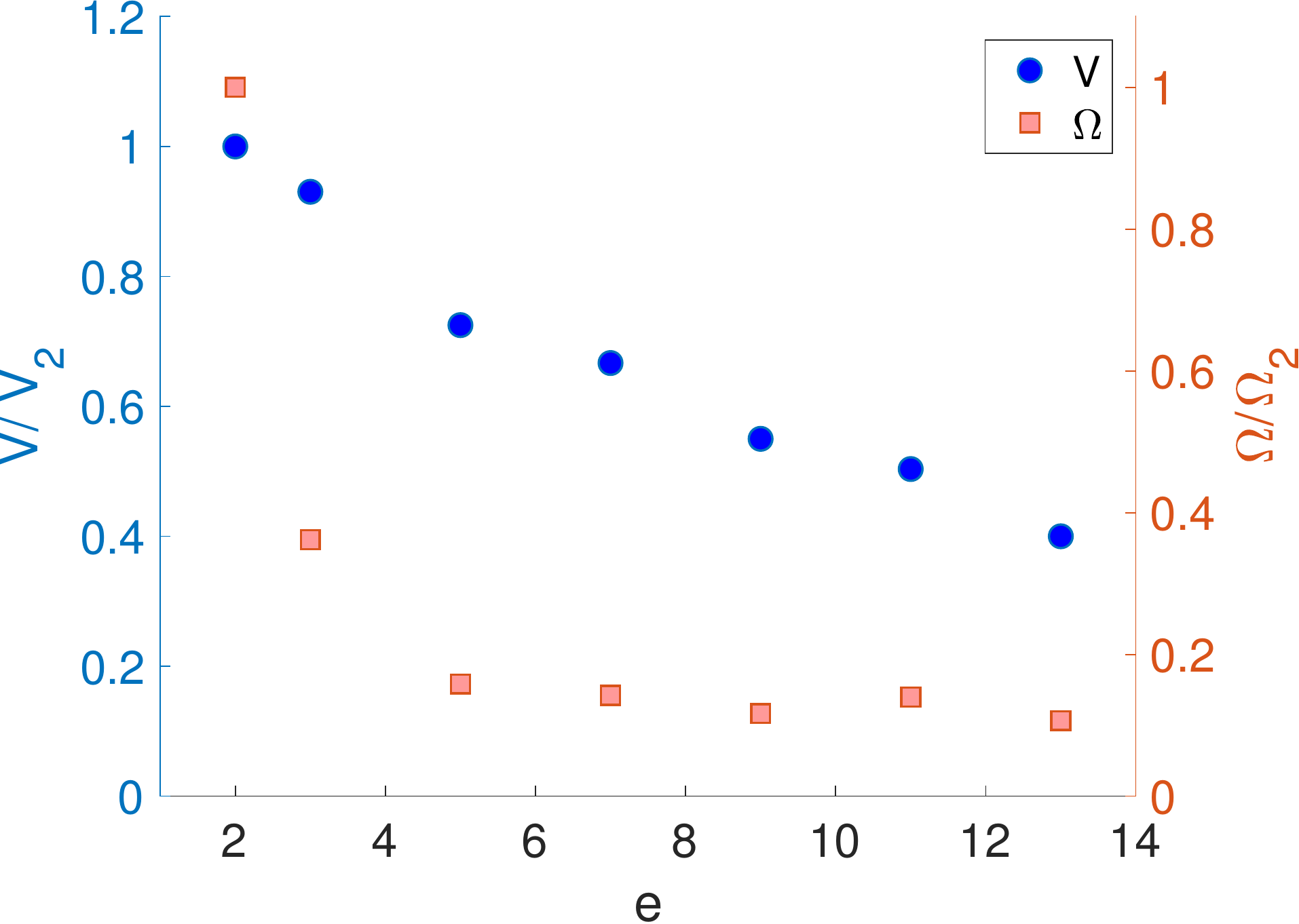}}\quad \quad
	\subfloat[\label{t5}]{\includegraphics[width=2.5in]{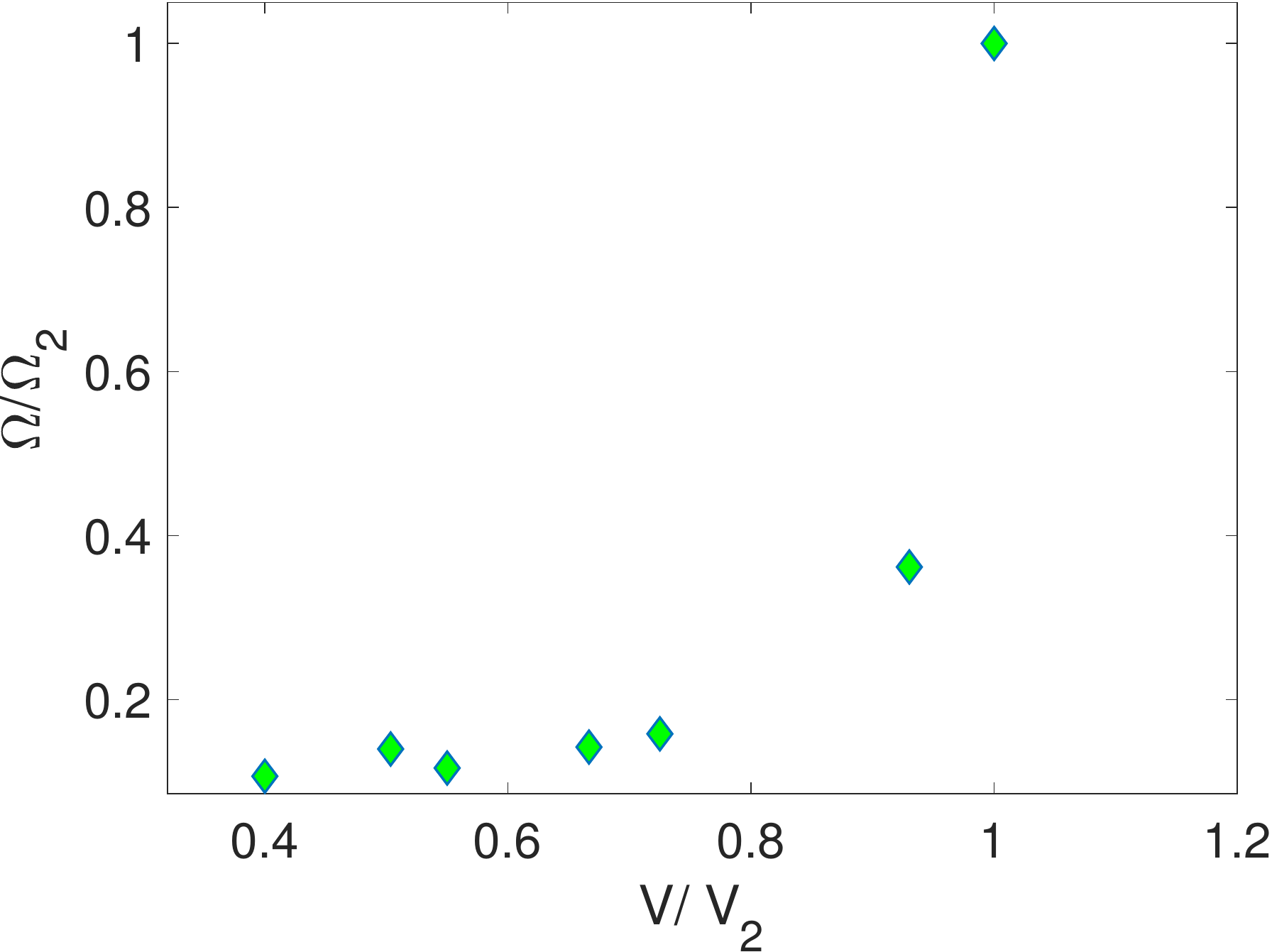}} 
	\caption{Oscillatory and unsteady motion in the intermediate and high velocity polarization regimes. Blue curve: centroid trajectory with wiggles. Note increasing wavelength, and increasing deviation from the $x$ axis, from (a) to (d). Green curve: \lam shape. Red arrows: substrate displacements. Here $K=3$ in all snapshots. (a), (b) intermediate polarization regime with bipedal oscillations and $e=2,3$, respectively. (c), (d) high polarization regime with  irregular oscillations and traveling waves (kinks) on anterior \lam boundary  and $e=5,11$, respectively.(e) average centroid speed $V$ and frequency $\Omega$ vs velocity polarization $e$ from our simulations. Both are normalized by the speed $V_2$ and frequency $\Omega_2$ from the run with  $e=2$.  
(f) Average frequency $\Omega$ correlates with  average centroid speed $V$.}
	\label{wiggly}
\end{figure}

\begin{figure}
	\centering
    \subfloat[\label{ttr1}]{\includegraphics[width=1.8in]{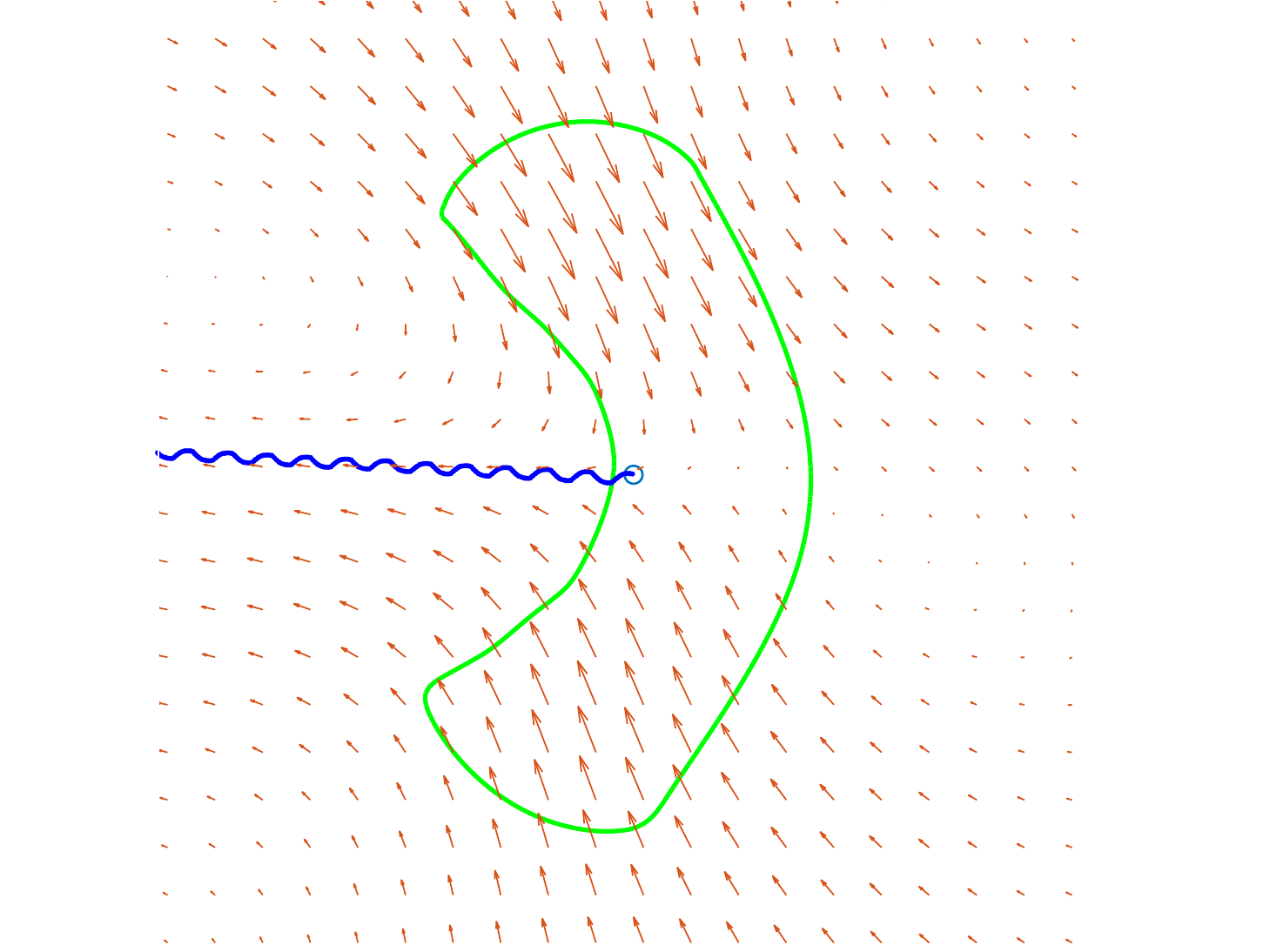}}
    \subfloat[\label{ttr2}]{\includegraphics[width=1.8in]{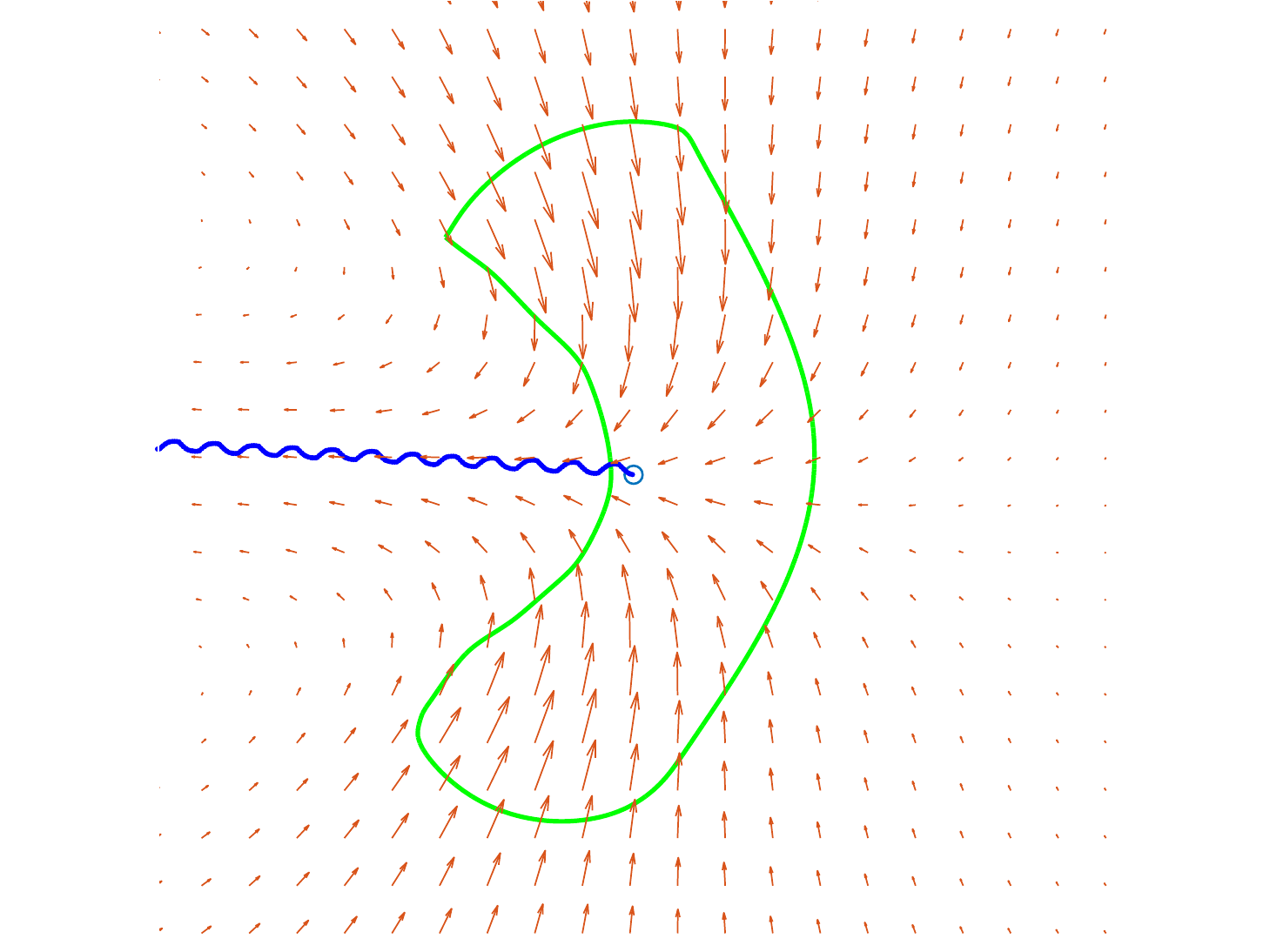}}
    \subfloat[\label{ttr3}]{\includegraphics[width=1.8in]{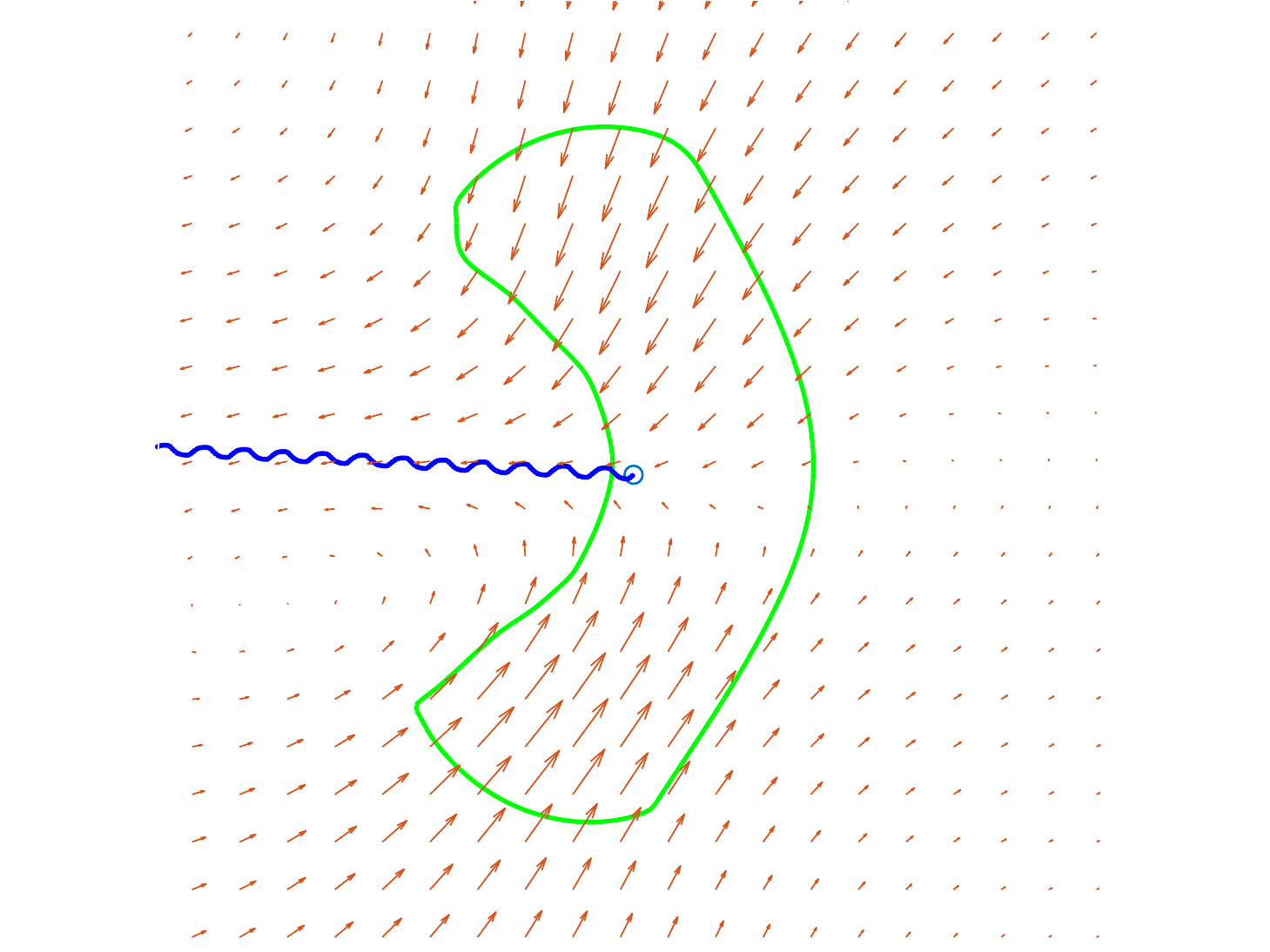}}
    \subfloat[\label{ttr4}]{\includegraphics[width=1.8in]{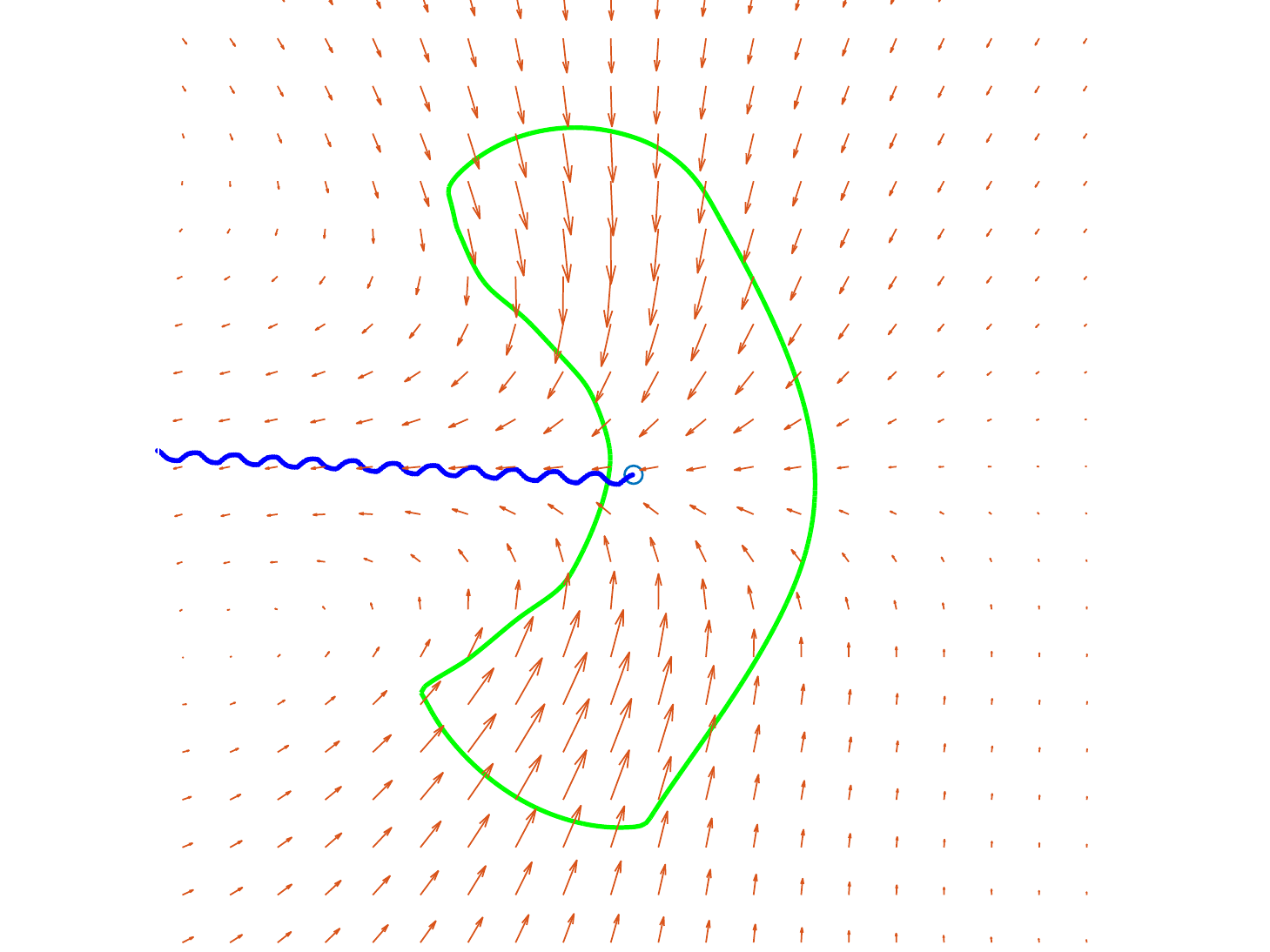}} 
	\\
    \subfloat[\label{ttr5}]{\includegraphics[width=1.8in]{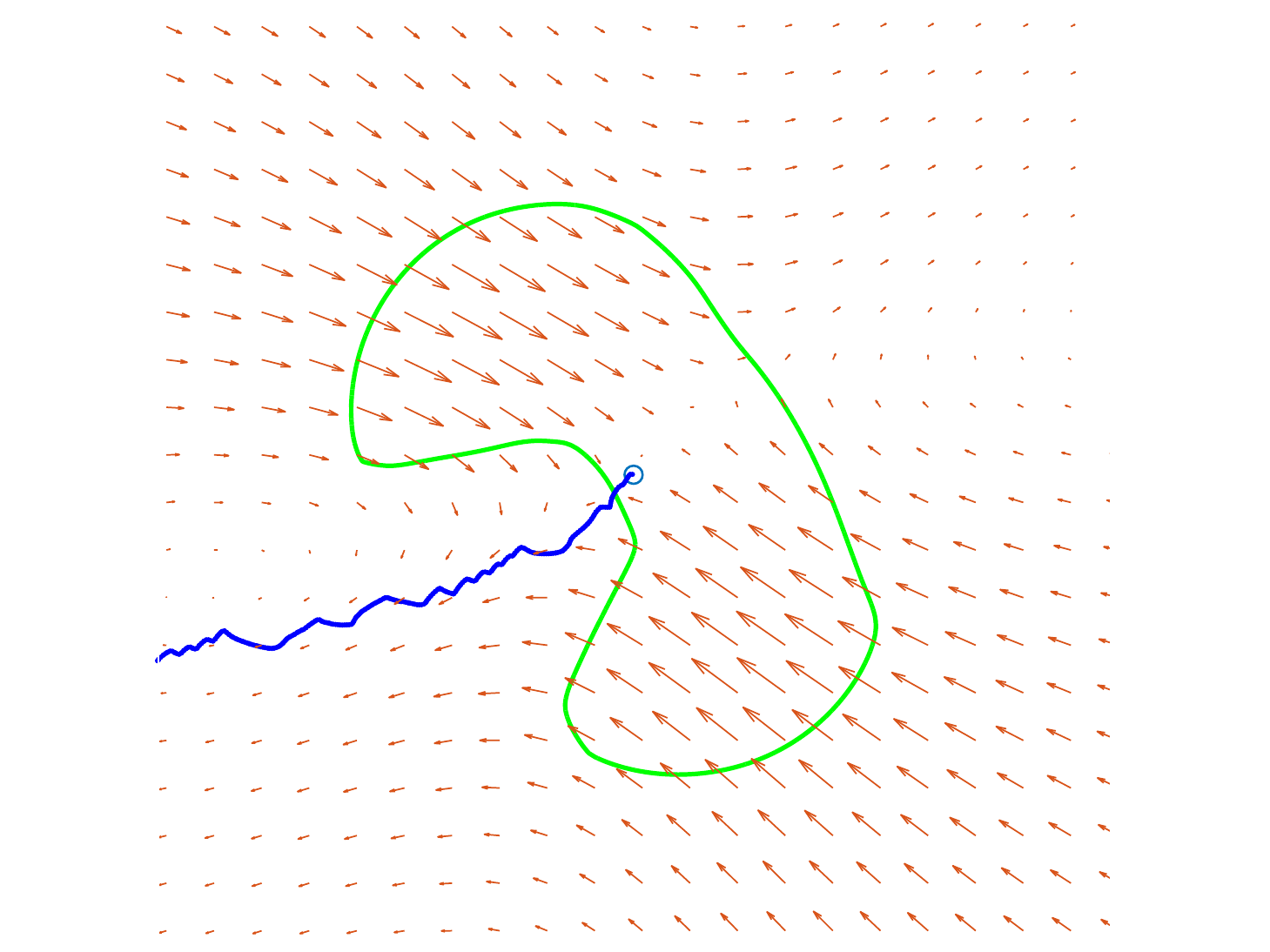}}
    \subfloat[\label{ttr6}]{\includegraphics[width=1.8in]{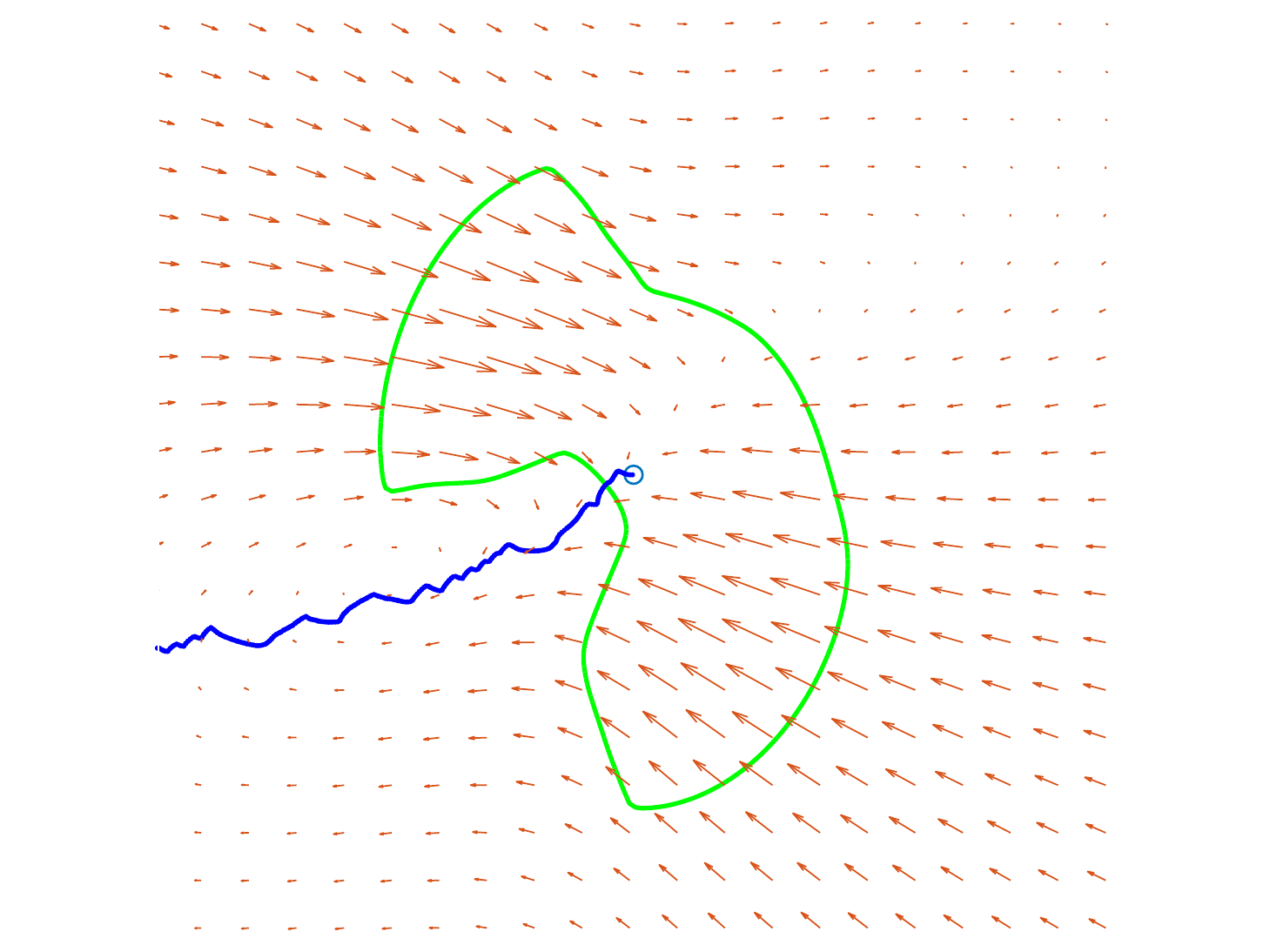}}
    \subfloat[\label{ttr7}]{\includegraphics[width=1.8in]{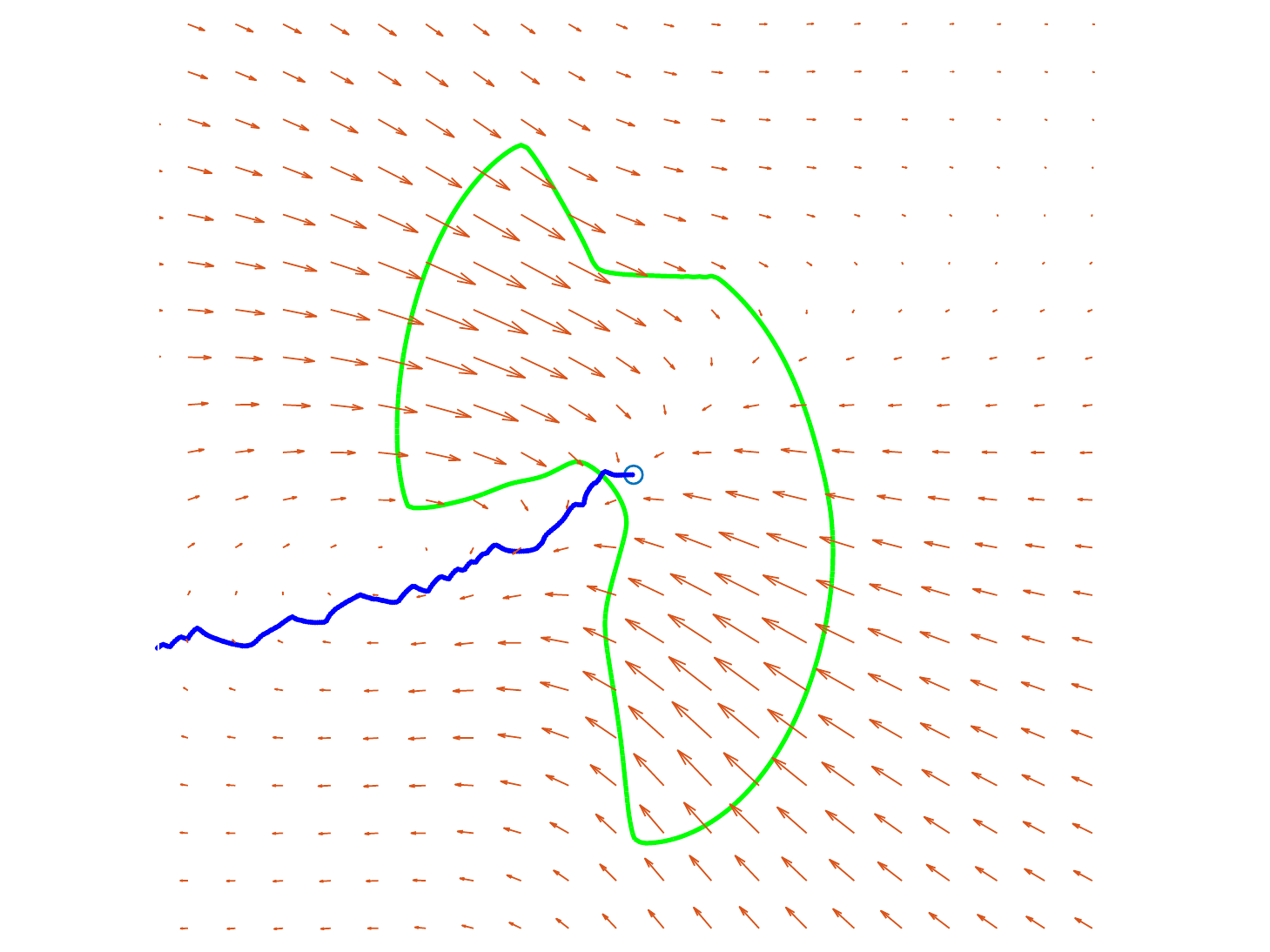}}
    \subfloat[\label{ttr8}]{\includegraphics[width=1.8in]{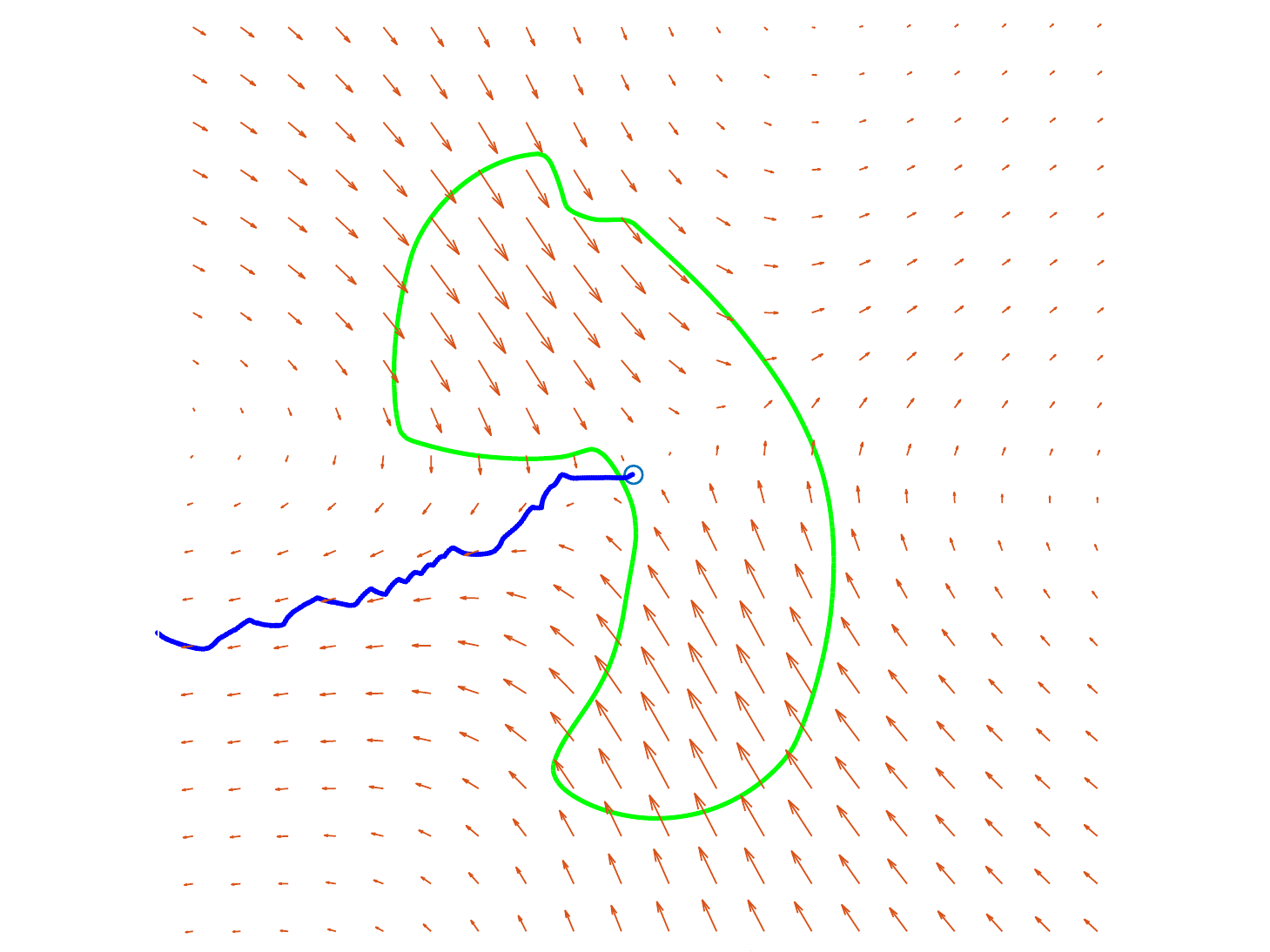}} 
	\caption{(a)-(d) Intermediate polarization regime: $K=3$, $e=3$. Four snapshots during a period of bipedal oscillation. In (a)  and (c) the substrate displacement fields (red arrows) are roughly antisymmetric about the $x$ axis and mirror images to each other. In (b) and (d) they are nearly  symmetric about the $x$ axis. In contrast,  the upper and lower trailing edges are pointed and rounded, respectively in  (b), and reversed in  (d) so locomotion is bipedal and the displacement oscillates between symmetry and antisymmetry about the $x$ axis. Note the rather regular oscillatory centroid trajectory with slight deviation from the $x$ axis. (blue curve). (e)-(h) High polarization regime: $K=3$, $e=11$. Four successive snapshots illustrating a kink (traveling wave) on the \lam front (green curve) nucleating  in (e), growing in (f), and traveling outward in (g) and (h). Note the irregular shape and curved, jagged centroid trajectory with large deviation from the $x$ axis (blue curve)}
	\label{snapshots}
\end{figure}

 \begin{figure}
\centering
	\subfloat[\label{motile}]{\includegraphics[width=2in]{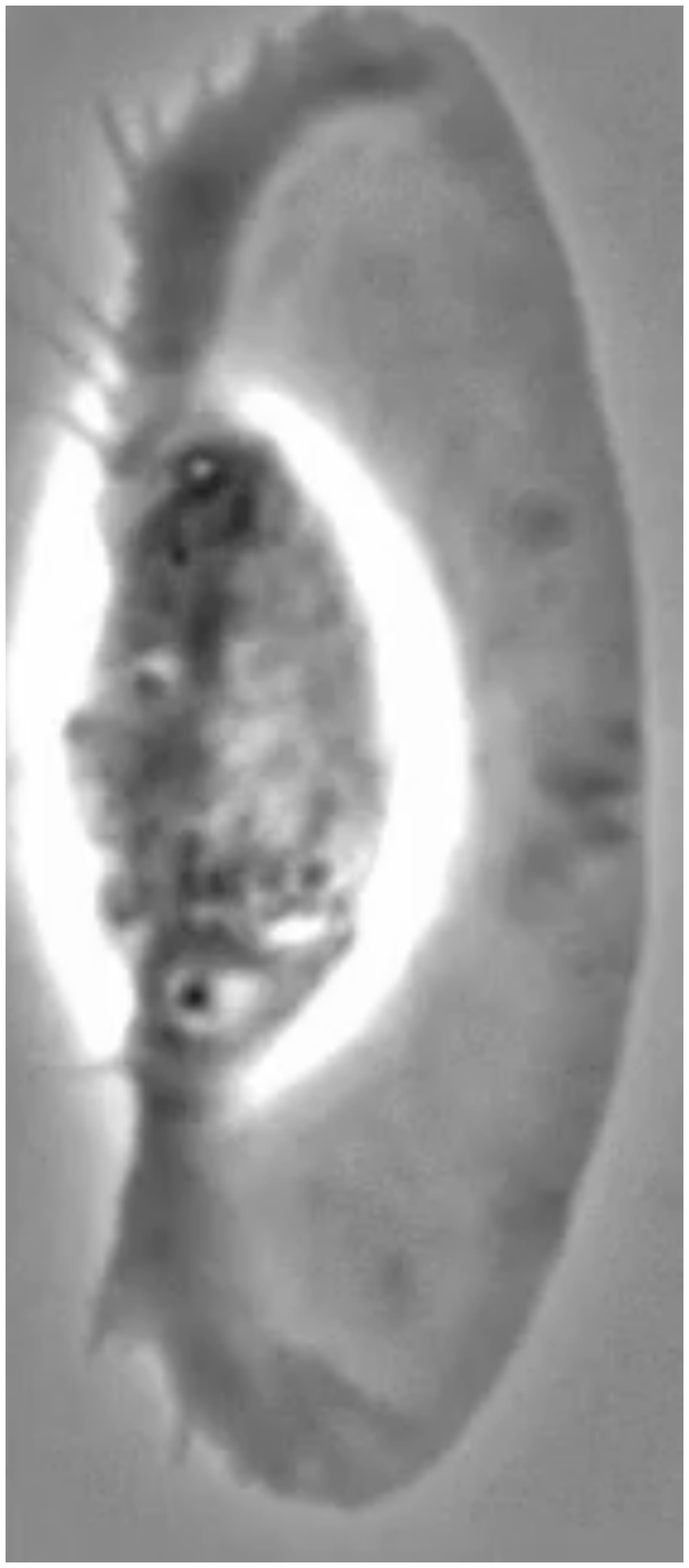}\hskip1.2cm}
	\subfloat[\label{vel}]{\includegraphics[width=1.8in]{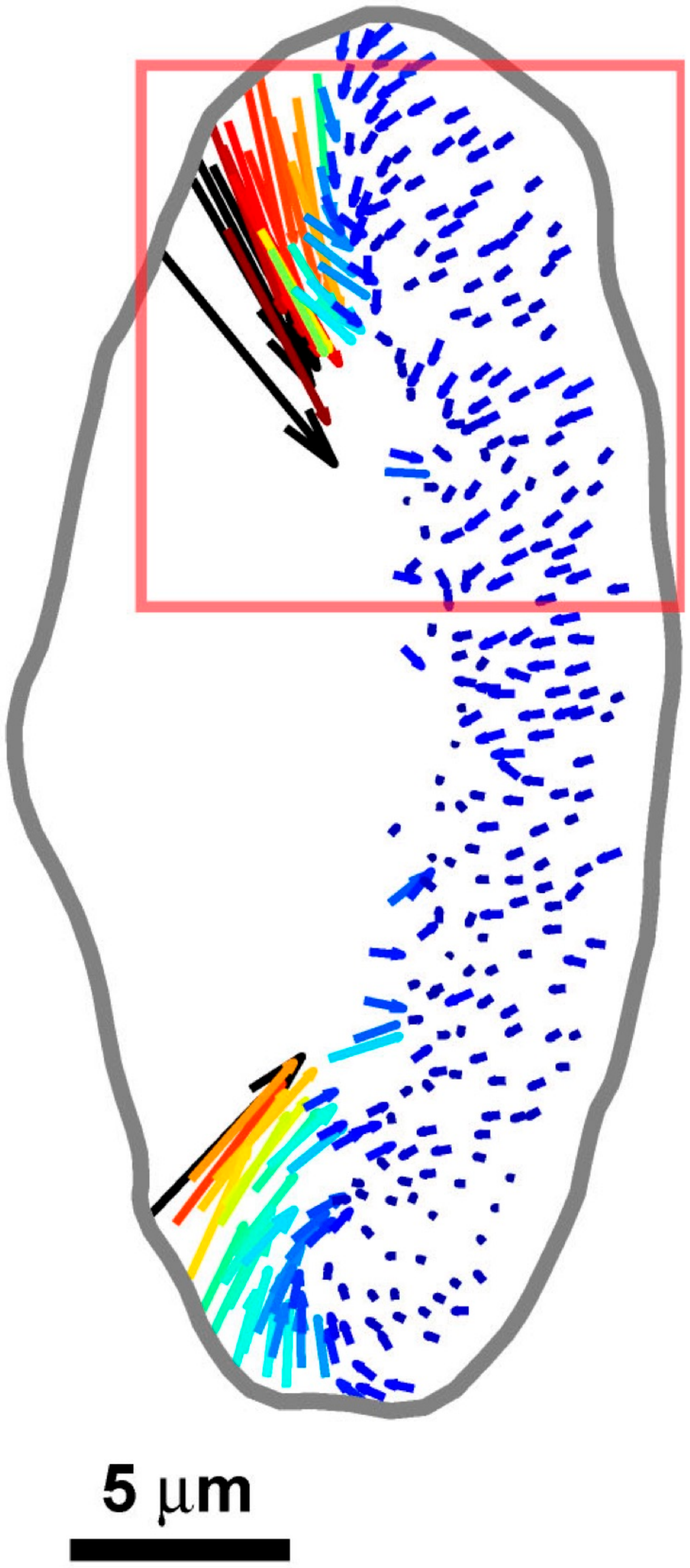}\hskip-0.5cm}
	\subfloat[\label{motsim}]{\includegraphics[width=3.5in]{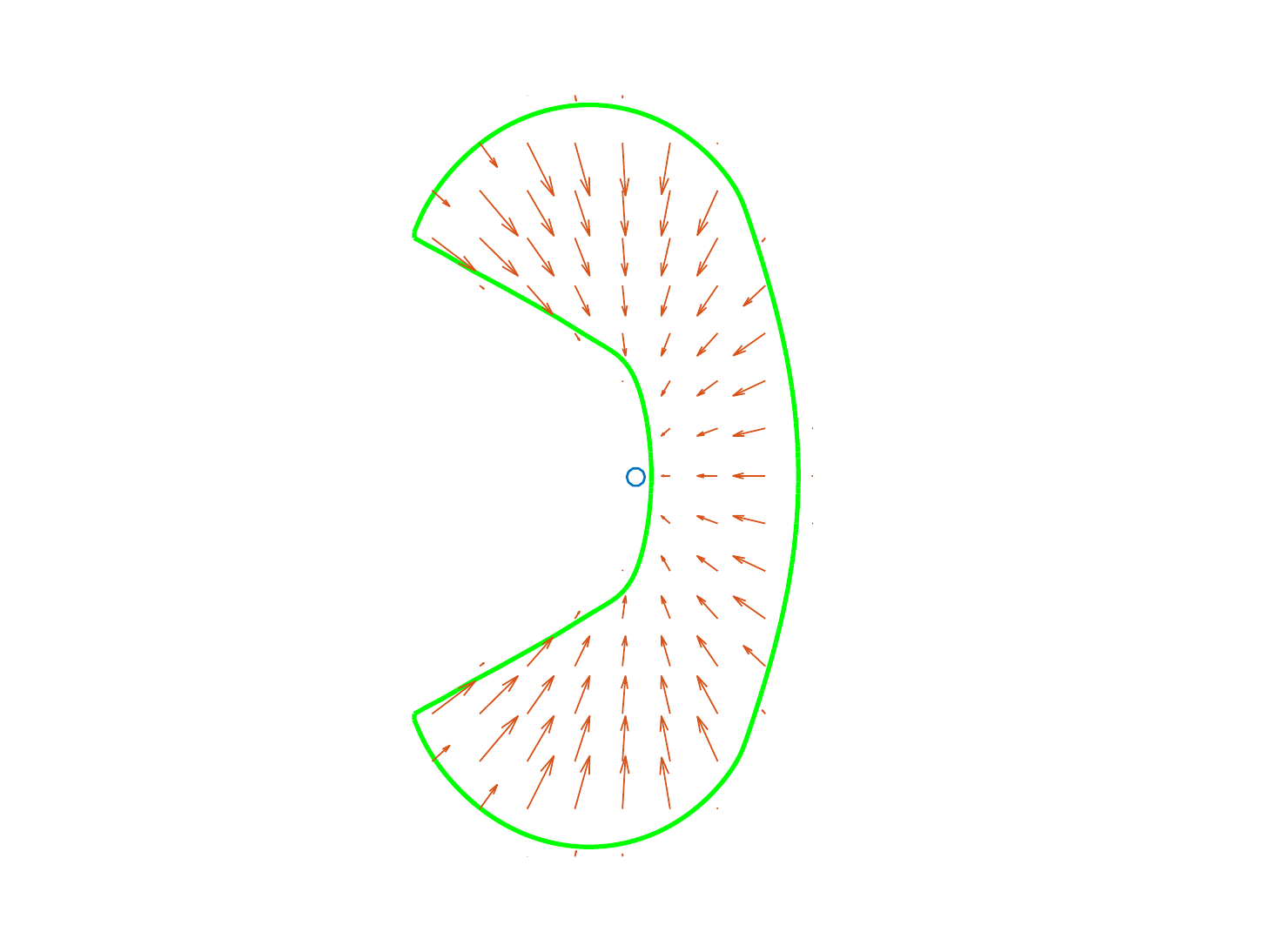}}
	\\
	\subfloat[\label{usim}]{\hskip-1cm\includegraphics[width=3.6in]{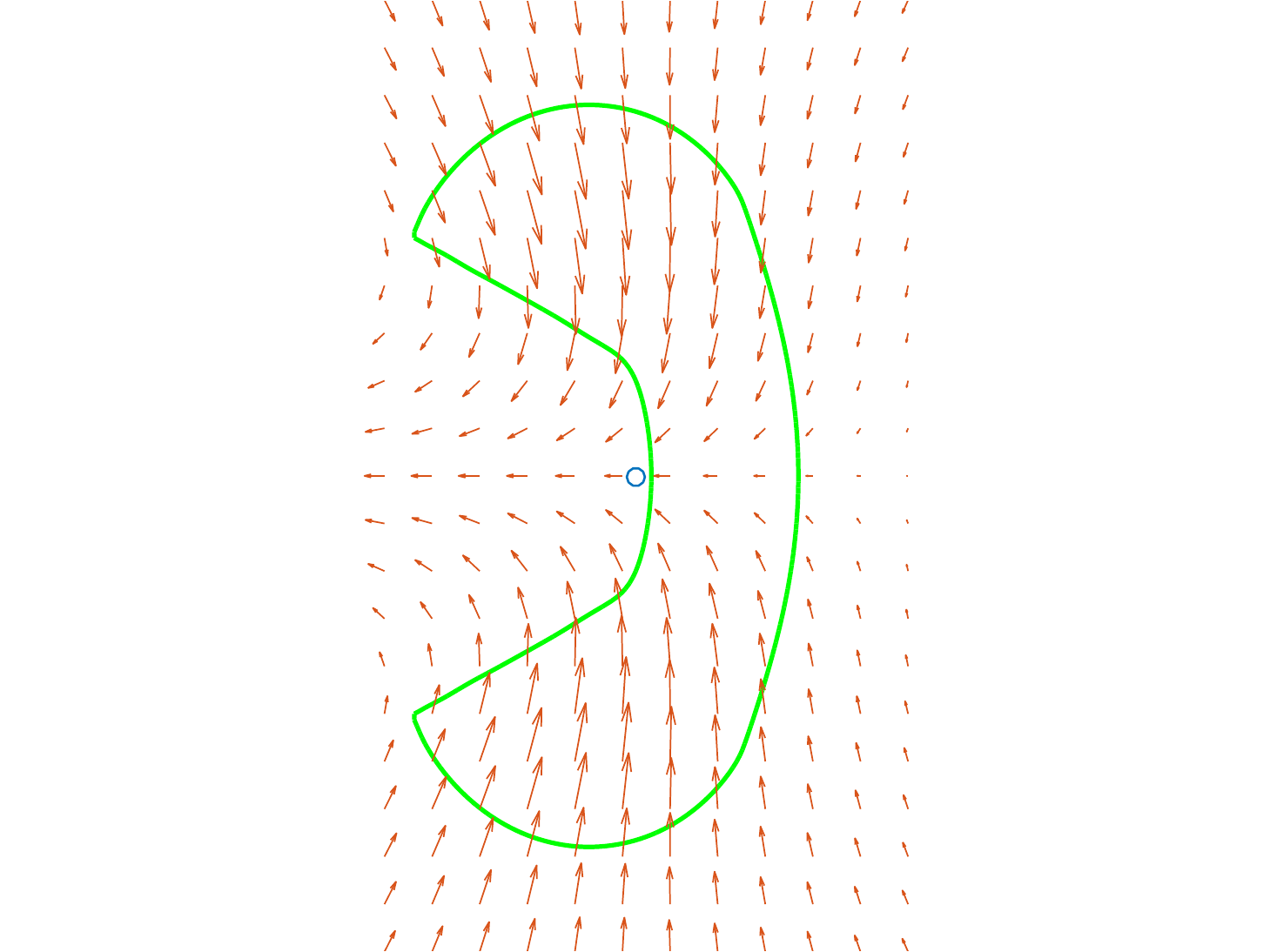}\hskip-1cm} 
	\subfloat[\label{uexp}]{\includegraphics[width=1.6in]{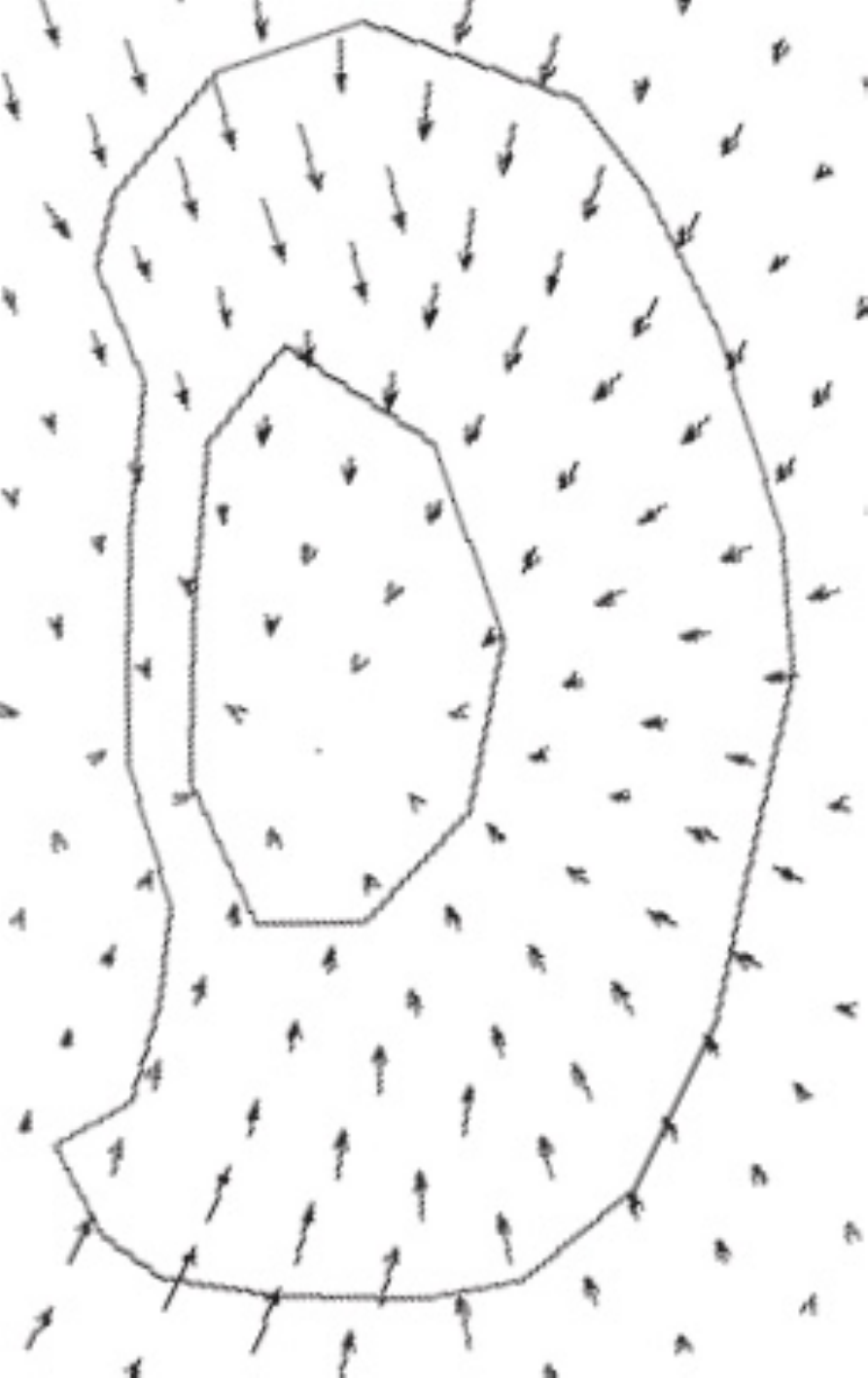}}\hskip0.56cm
	\subfloat[\label{trexp}]{\includegraphics[width=2.25in]{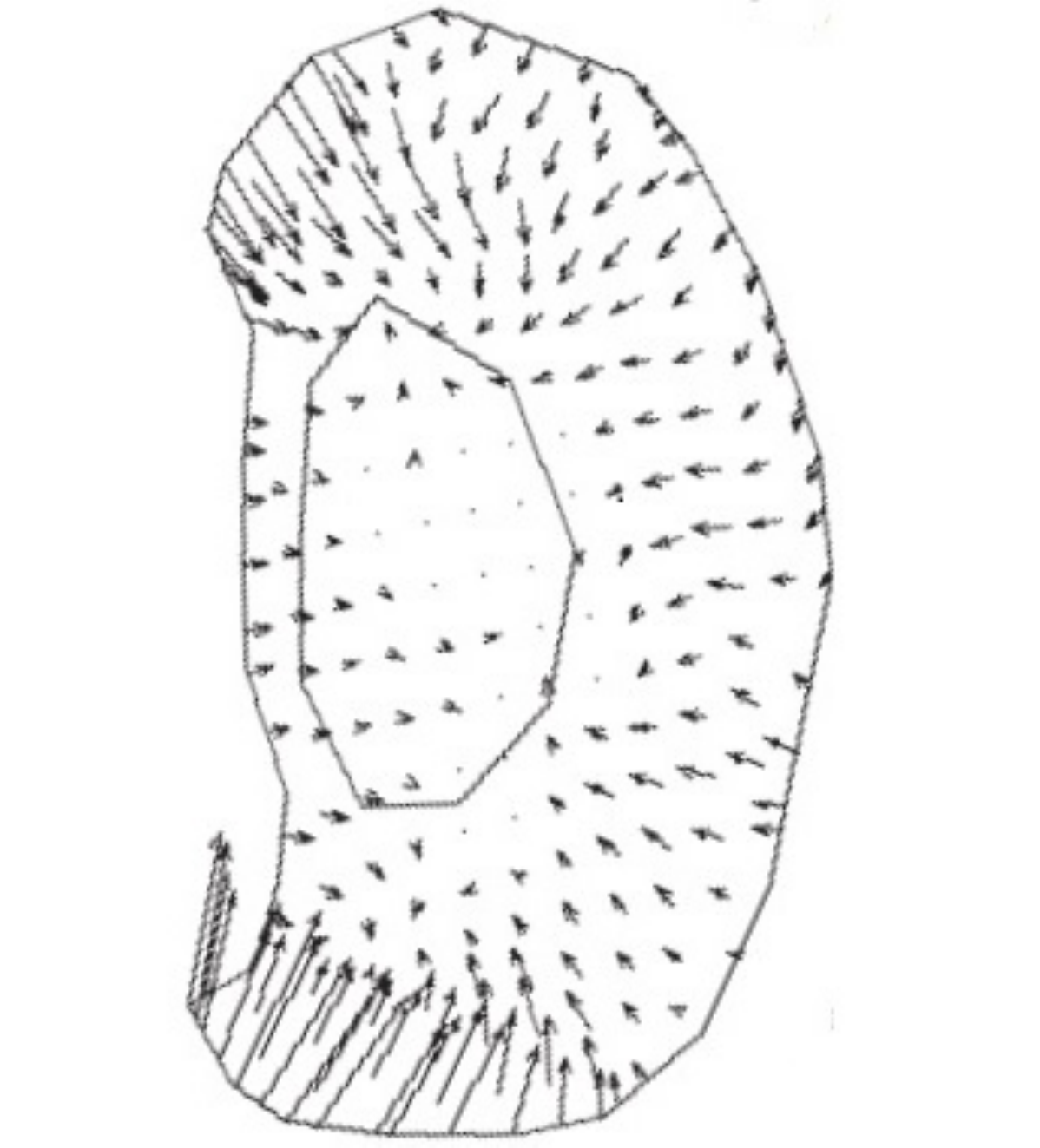}}
 \caption{Comparison of \lam shape (a) vs (c), actin velocity (b) vs (c), substrate displacement (d) vs (e) and traction (f) vs (a)  from experiments ((a), (b), (e), (f)), and our model ((c) and (d)). (a) Motile keratocyte with nearly steady shape and speed (moving to the right) from \cite{polarvel}, Fig.1E. (b) Measured actin velocity vectors in the \lam (blank region corresponds to the nucleus) \cite{polarvel}, Fig. 1F. (c) Simulation of present model predicts steady propagation of the \lam following the sequence shown in Fig. \ref{figtransition}. Green: steady \lam shape; also shown are actin velocity vectors (red); note large inward flow at the rear and smaller speeds in the front in rough qualitative agreement with (b).   (d) Same as (c) but red arrows are substrate displacements. (e) Substrate displacement and cell shape from  \cite{displ}, Fig. 2a. (f)  Substrate traction inferred from displacements shown in (e)  from  \cite{displ}, Fig. 2b. The ligament to the left of the nucleus in (e), (f) is not part of the \lamp}
 \label{steady}
\end{figure}

\begin{figure}
 \centering
\subfloat[\label{wa}]{\includegraphics[width=3in]{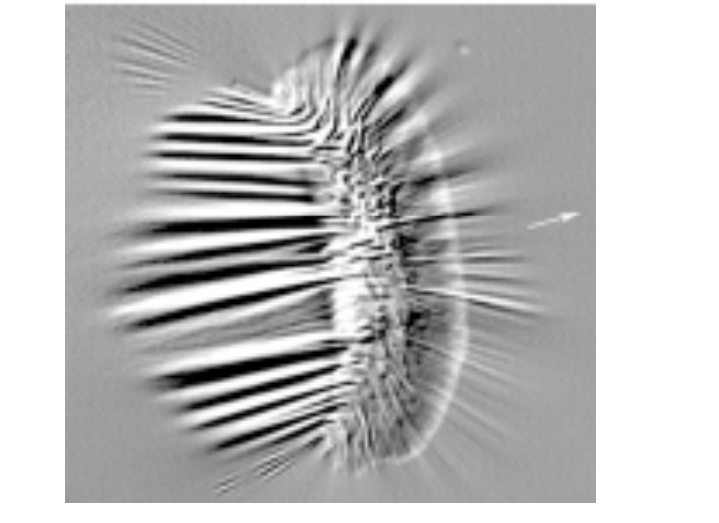}}
	\subfloat[\label{wb}]{\includegraphics[width=3.3in]{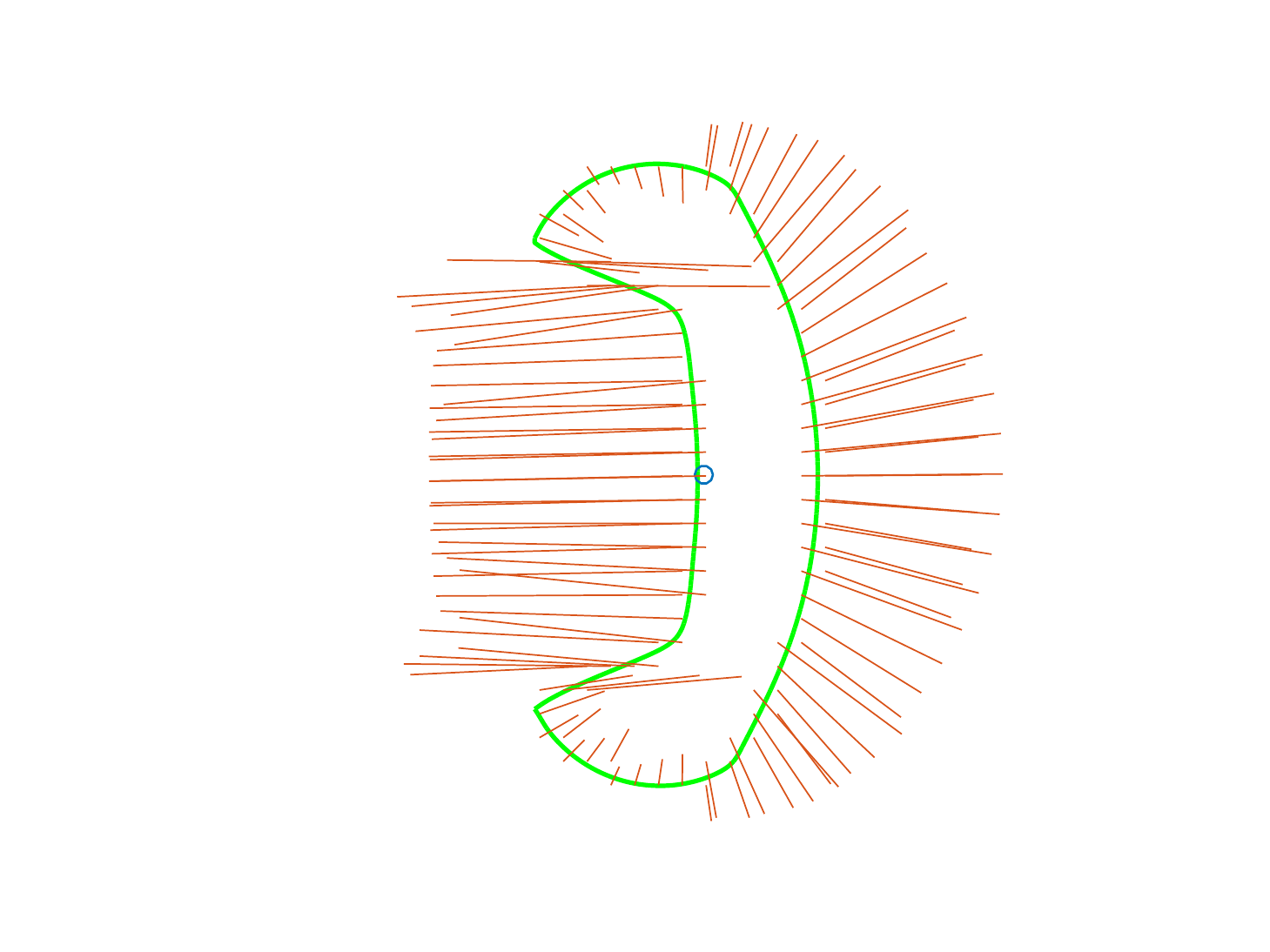}}
 \caption{(a) Motile fish keratocyte wrinkling a silicone substrate, reproduced from [\cite{cellsdeformsubstrate}, Fig. 1(a)]. (b) Simulation predictions from our model: \lam (green curve), substrate wrinkles (red lines).}
 \label{figwrinkle}
\end{figure}

 \begin{figure}
\centering
	\subfloat[\label{p0}]{\includegraphics[width=1.35in]{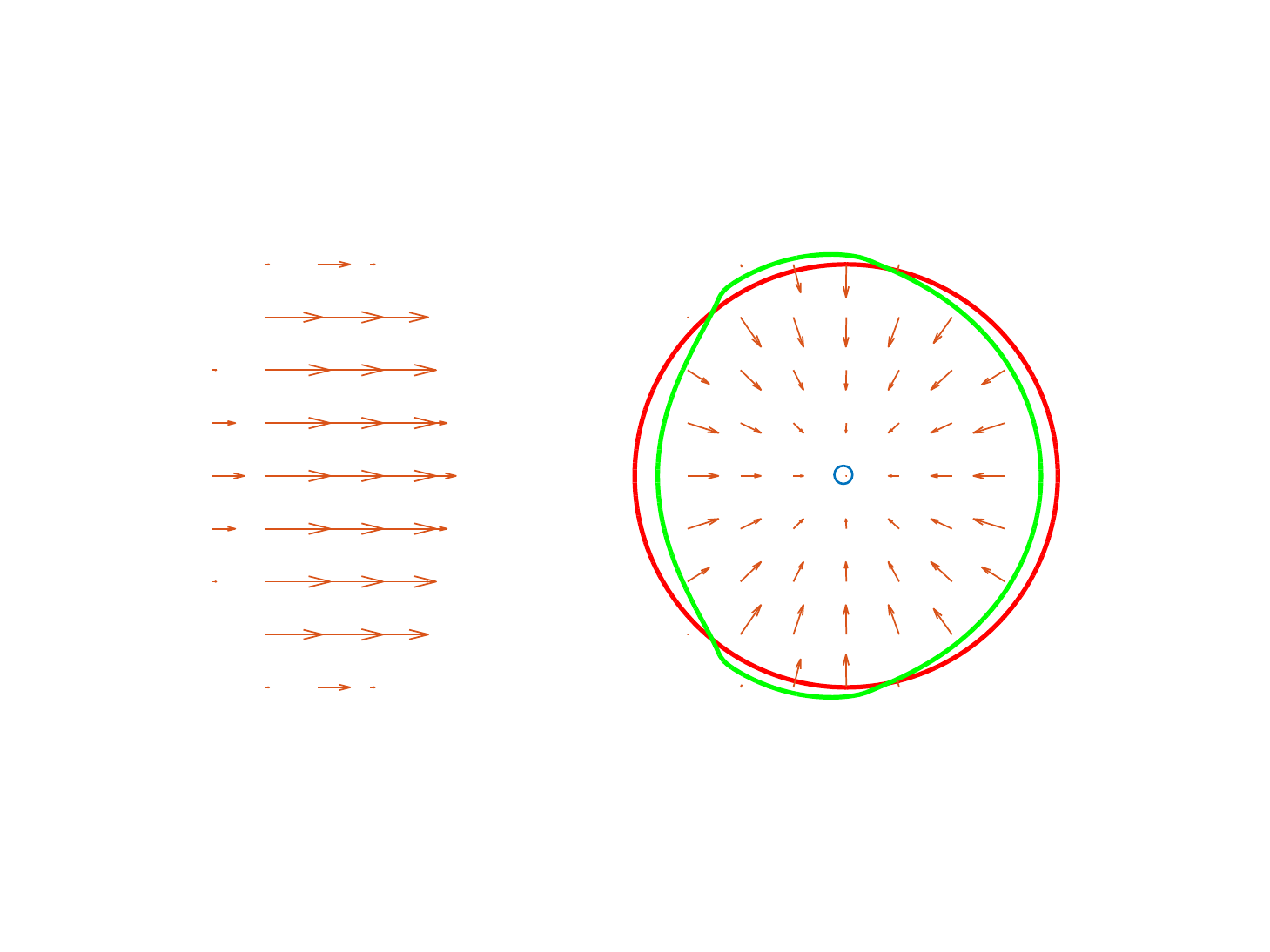}} 
	\subfloat[\label{p1}]{\includegraphics[width=1.35in]{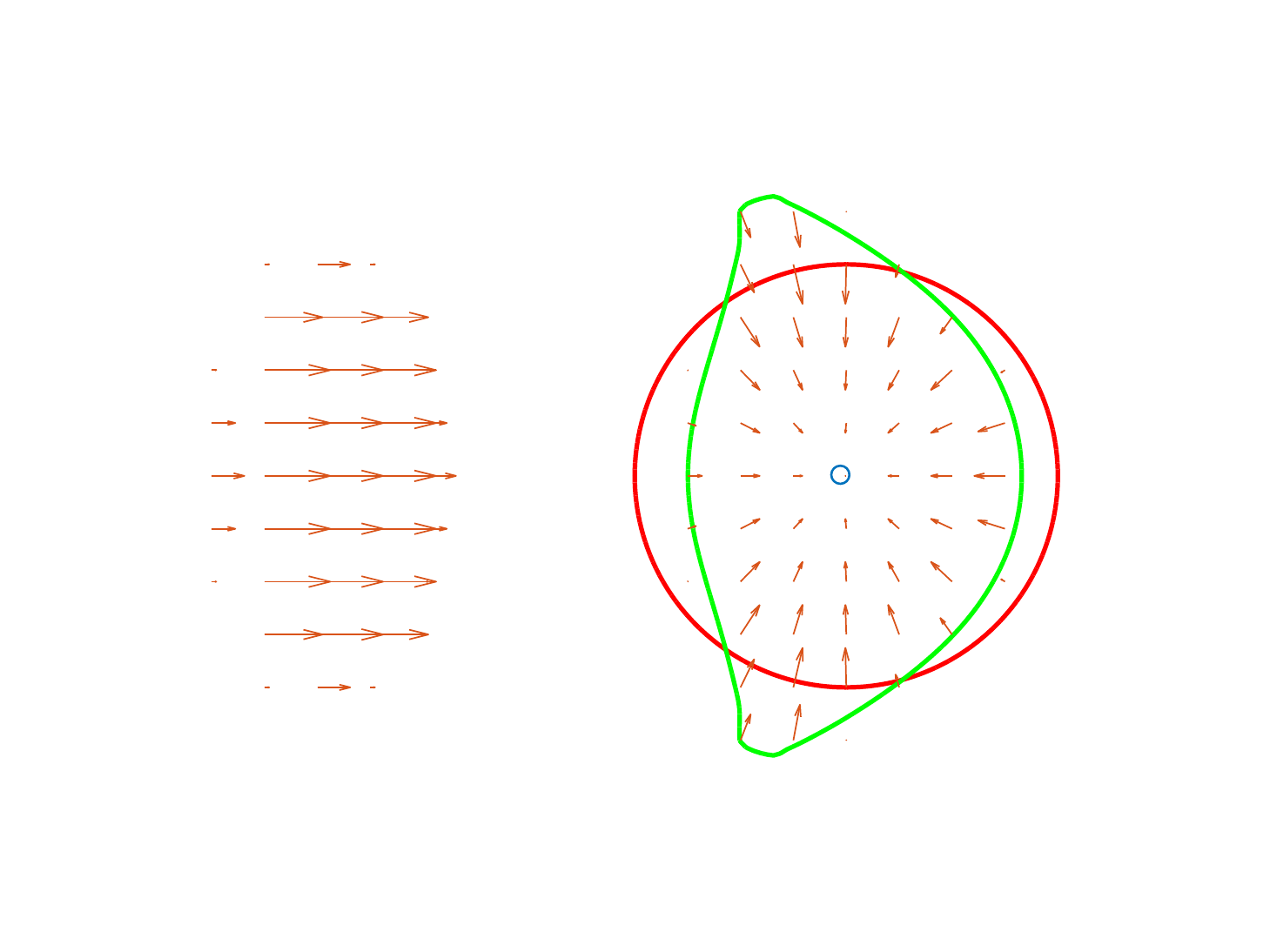}} 
	\subfloat[\label{p2}]{\includegraphics[width=1.35in]{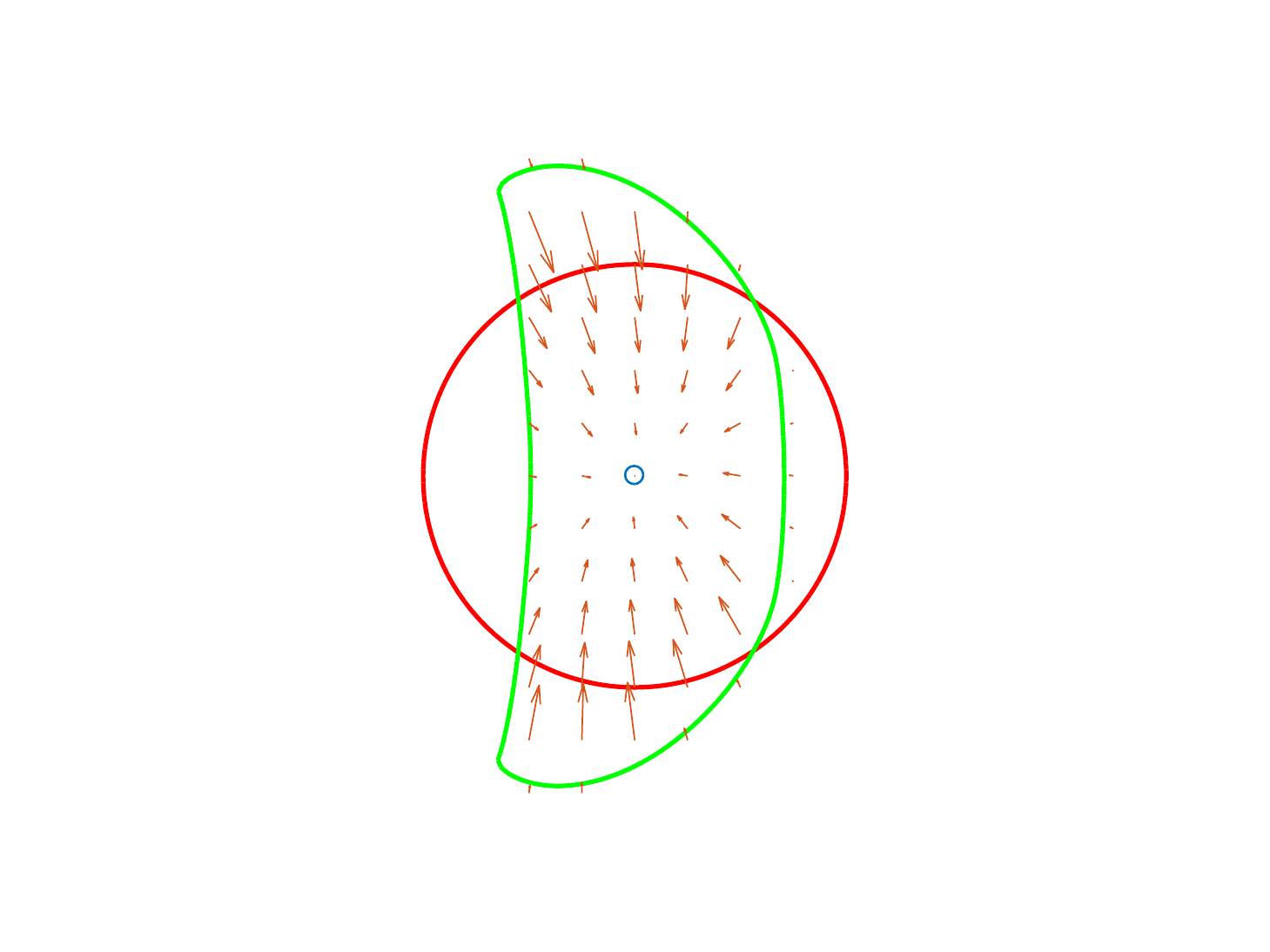}}
	\subfloat[\label{p3}]{\includegraphics[width=1.35in]{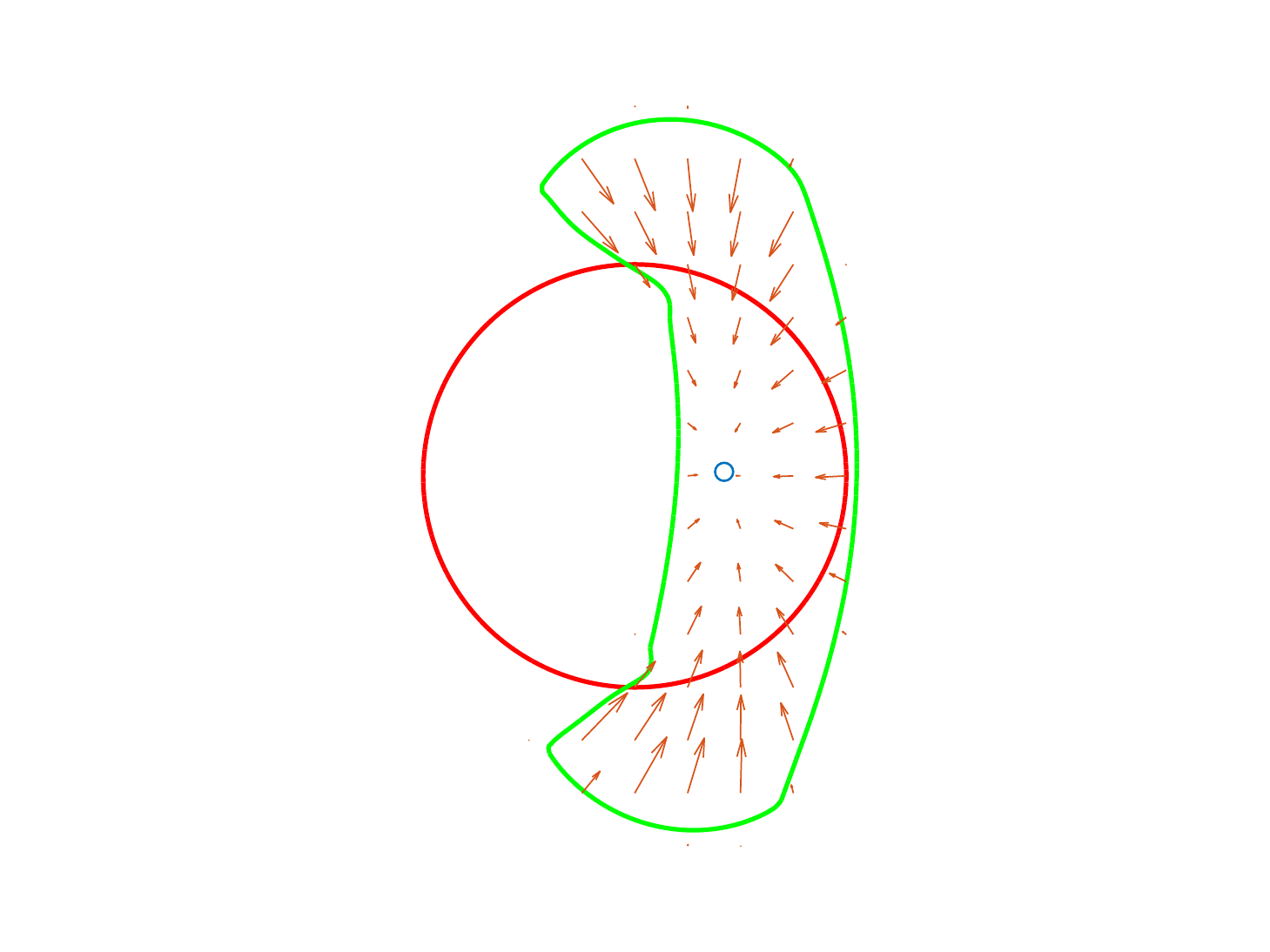}}
	\subfloat[\label{p5}]{\includegraphics[width=1.35in]{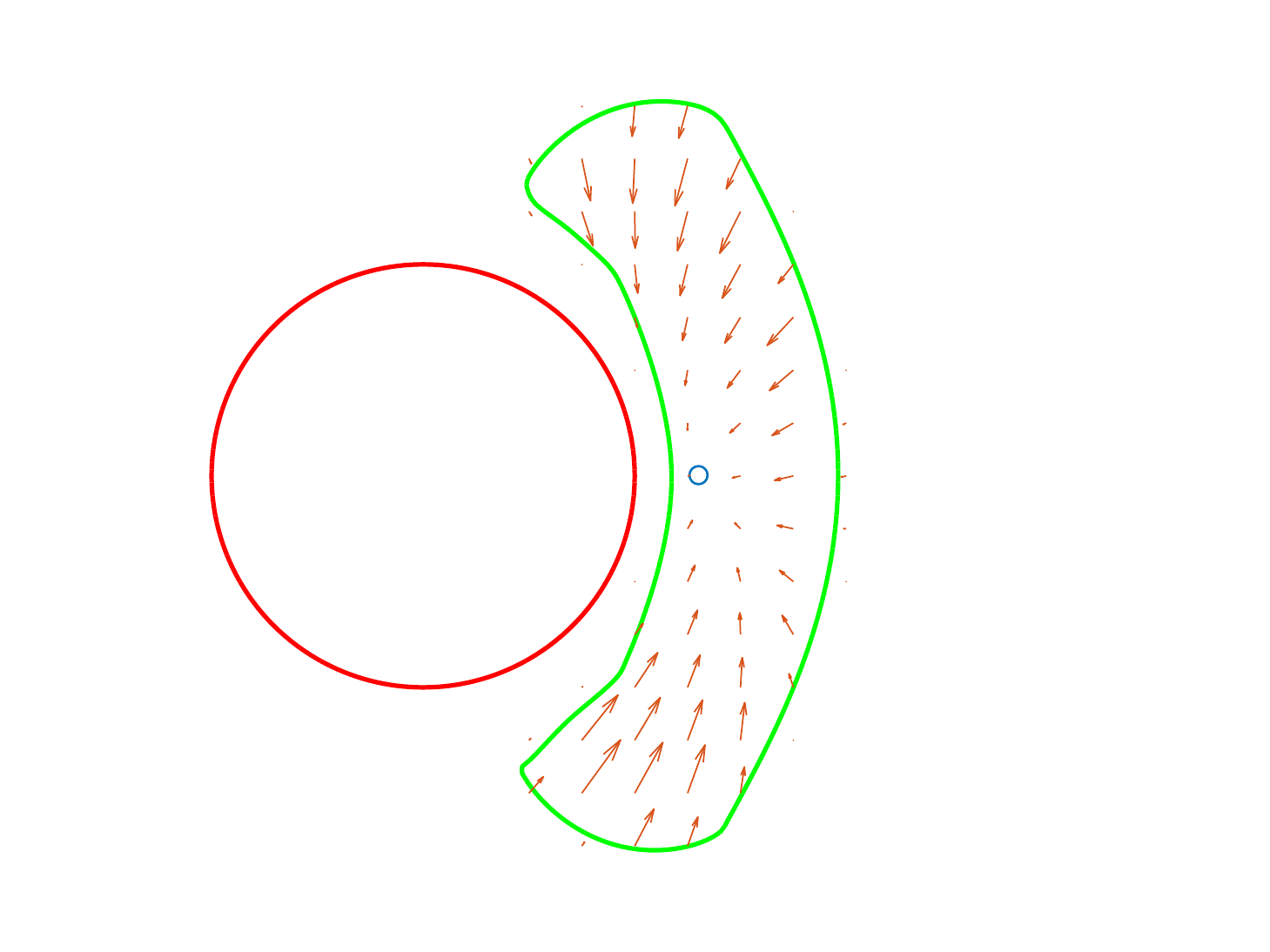}}
 \caption{ Reverse Tensotaxis: Model  simulation snapshots of a \lam fragment (red: initial fragment position, green: subsequent fragment positions). (a) External forces are exerted onto the substrate to the left of the circular fragment (purple arrows pointing to the right). (b) The fragment starts receding away from the compressive stresses induced by the forces which are about to be removed. (c), (d) The fragment becomes crescent like and starts moving to the right even after the forces are removed. (e) It assumes the usual steady shape of a crawling \lam and moves steadily to the right henceforth.}\label{Push}
\end{figure}

\begin{figure}
\centering 
	\subfloat[\label{pp0}]{\includegraphics[width=1.35in]{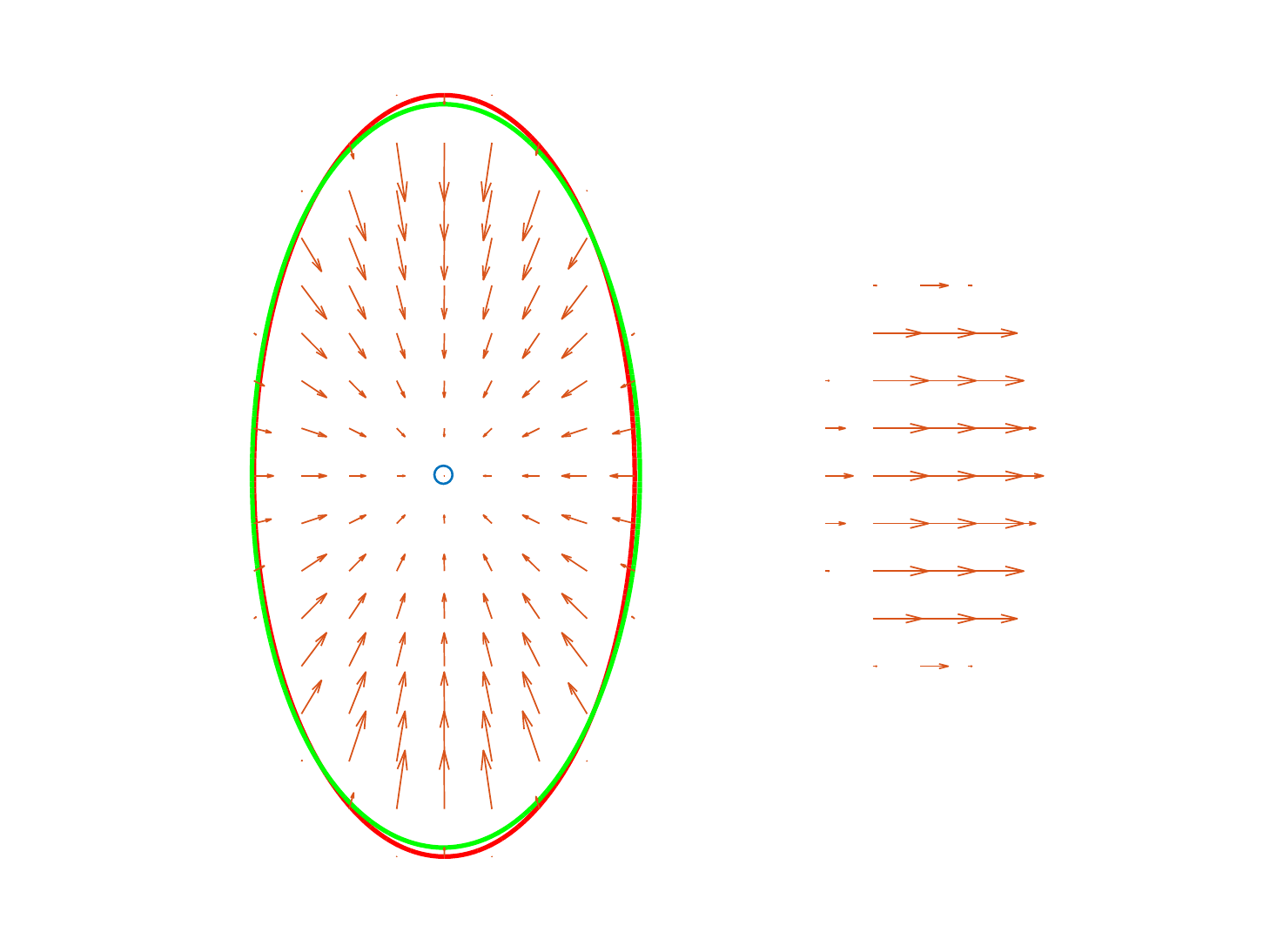}} 
	\subfloat[\label{pp1}]{\includegraphics[width=1.35in]{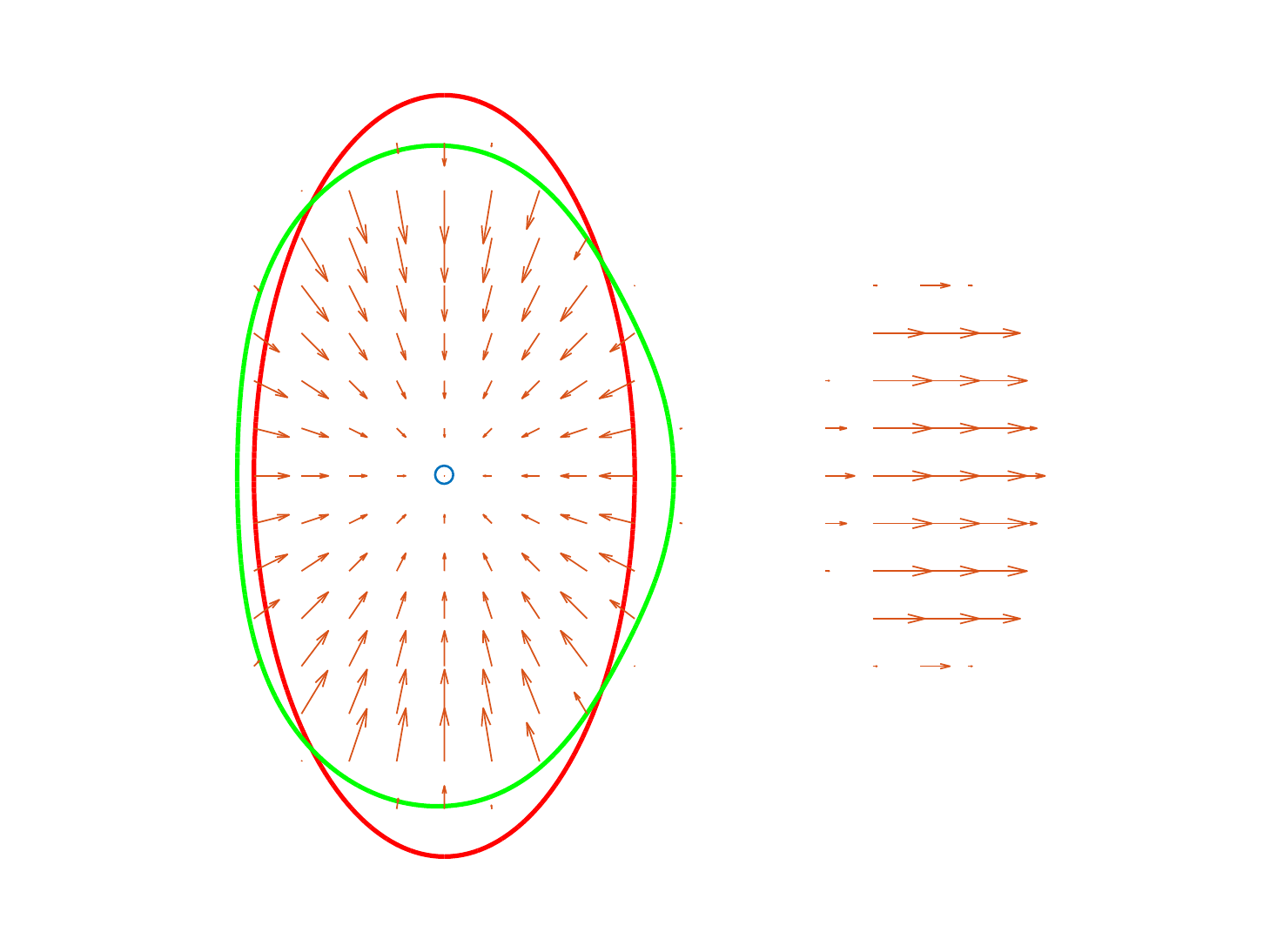}} 
	\subfloat[\label{pp2}]{\includegraphics[width=1.35in]{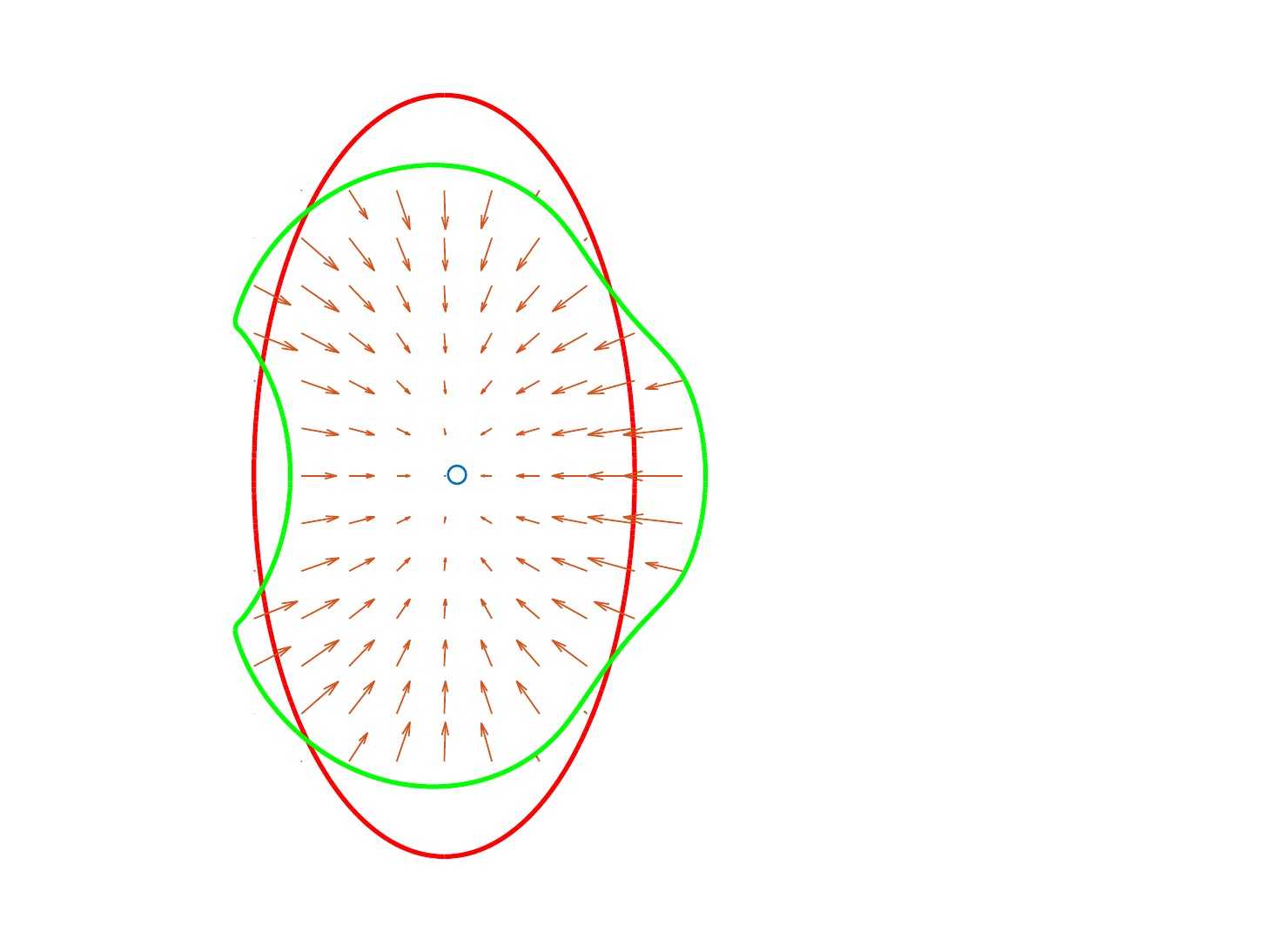}}
	\subfloat[\label{pp3}]{\includegraphics[width=1.35in]{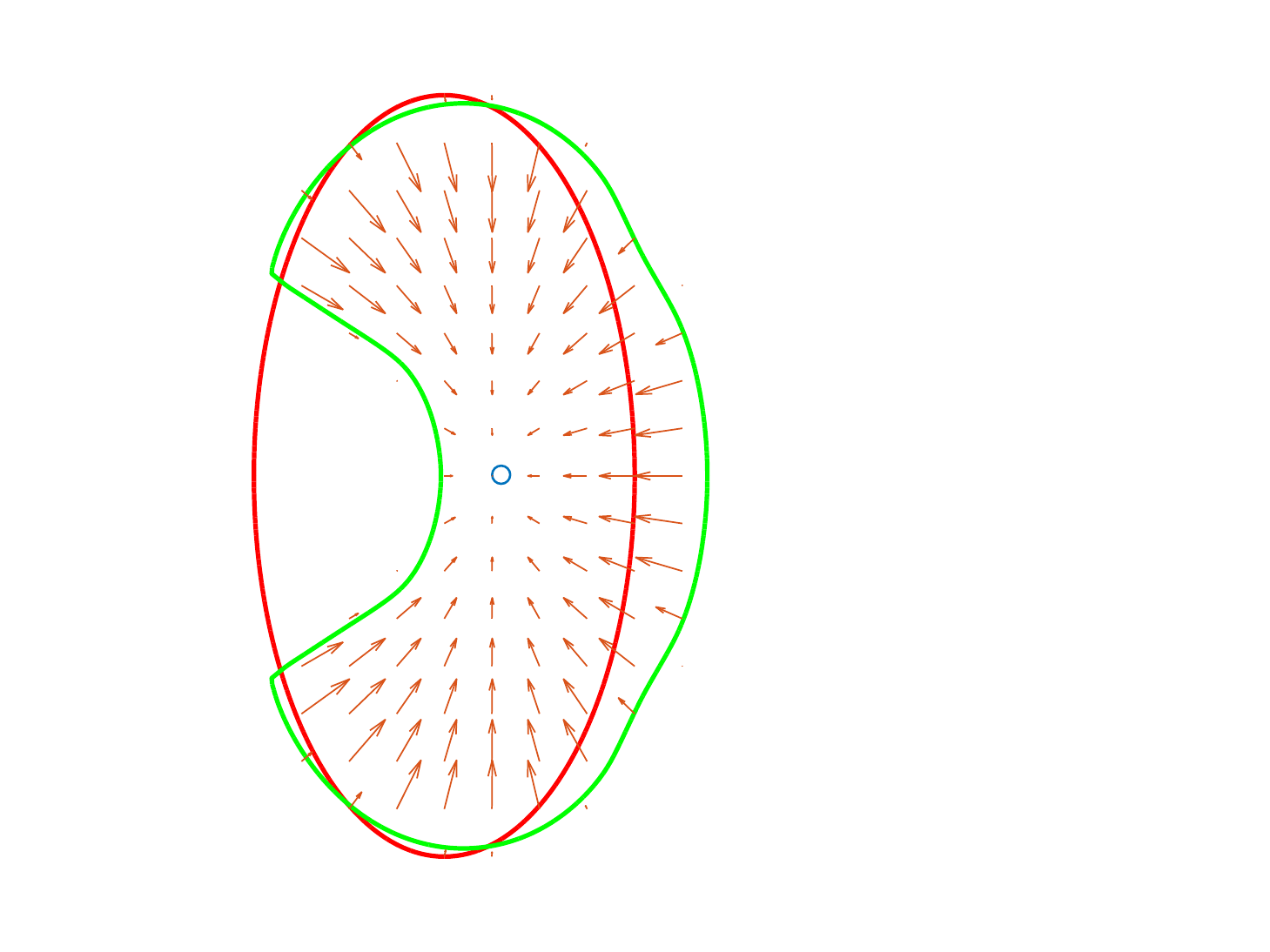}}
	\subfloat[\label{pp5}]{\includegraphics[width=1.35in]{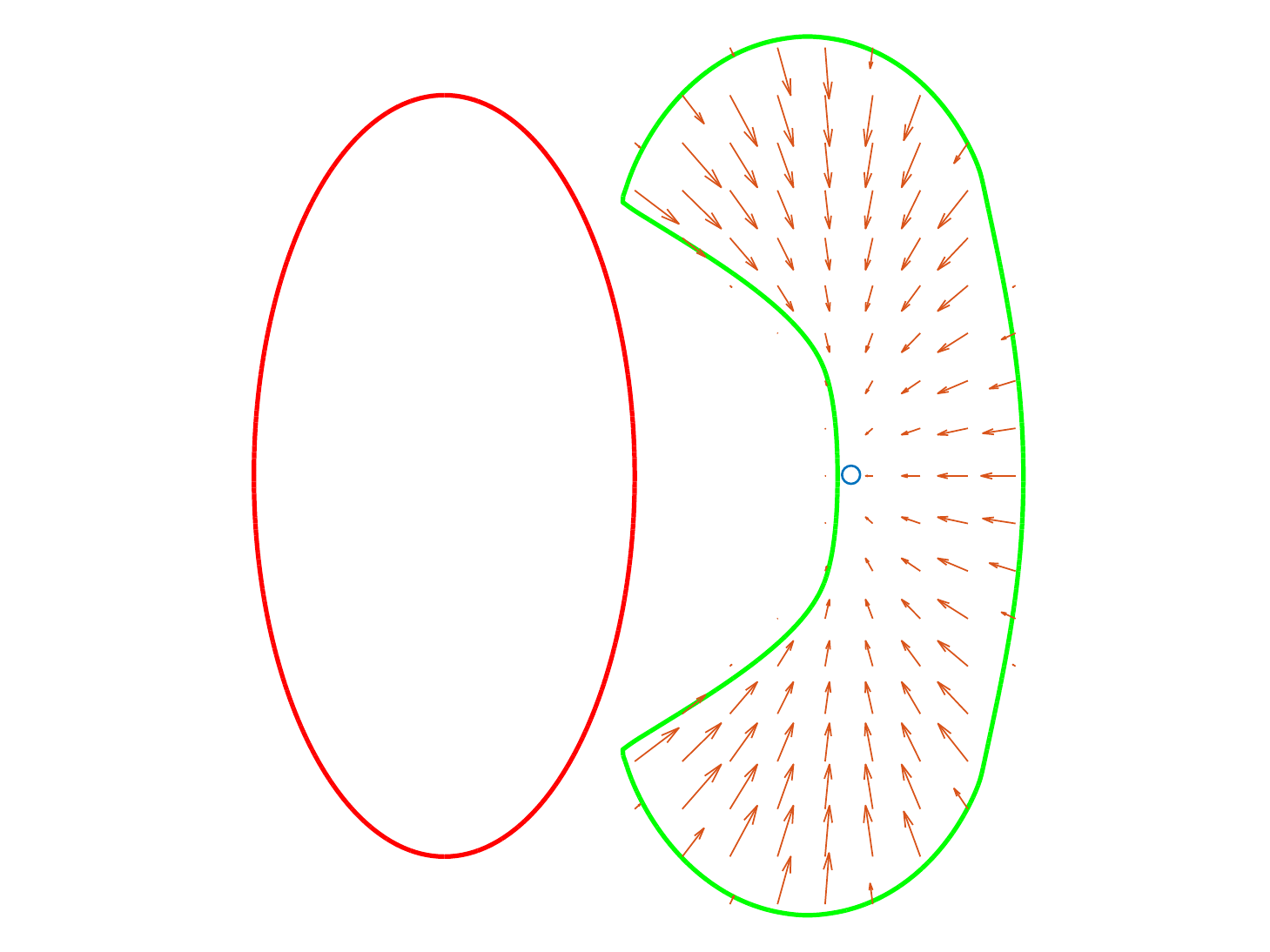}}
 \caption{Tensotaxis: Model  simulation snapshots of a \lam fragment (red: initial fragment position, green: subsequent fragment positions). (a) External forces are exerted onto the substrate to the right of the elliptical fragment (yellow arrows pointing to the right). (b) The fragment starts protruding toward the tensile stresses to its right induced by the forces (which are about to be removed). (c), (d) The fragment becomes crescent like and starts moving to the right even after the forces are removed. (e) It assumes the usual steady shape of a crawling \lam and moves steadily to the right henceforth.}\label{Pull}
\end{figure}

 \begin{figure}
\centering
	\subfloat[\label{d0}]{\includegraphics[width=1.35in]{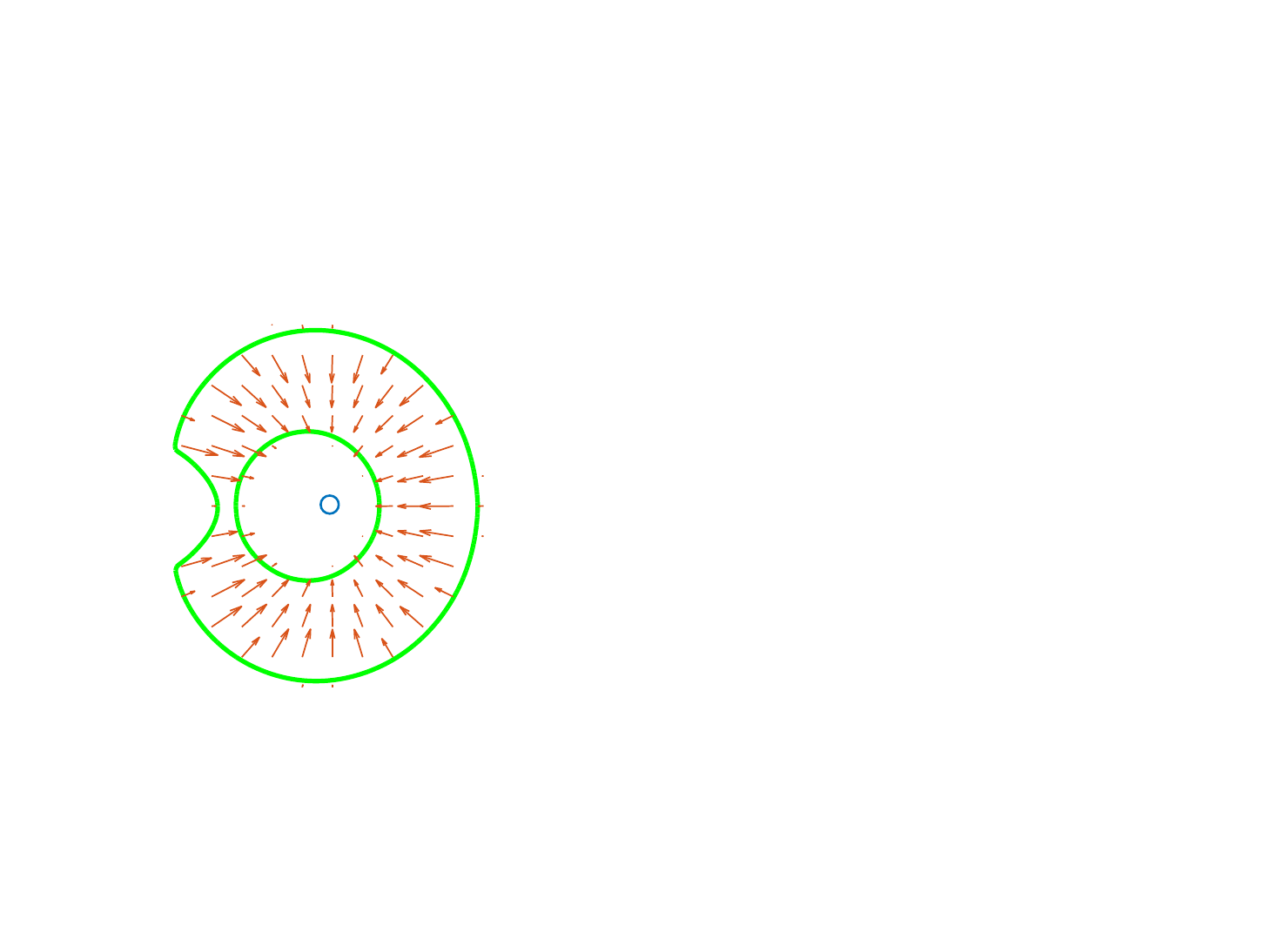}}
	\subfloat[\label{d1}]{\includegraphics[width=1.35in]{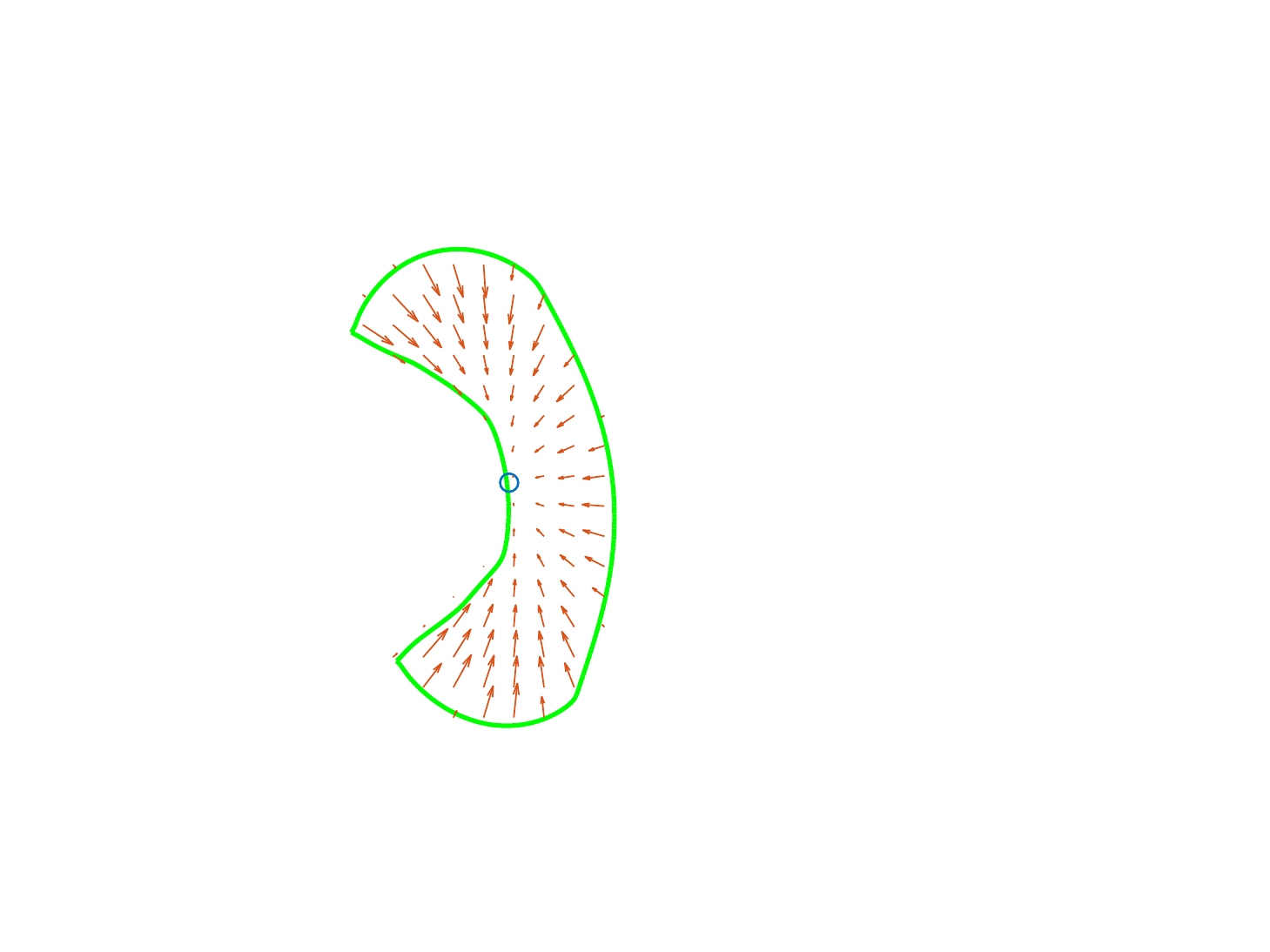}}
	\subfloat[\label{d2}]{\includegraphics[width=1.35in]{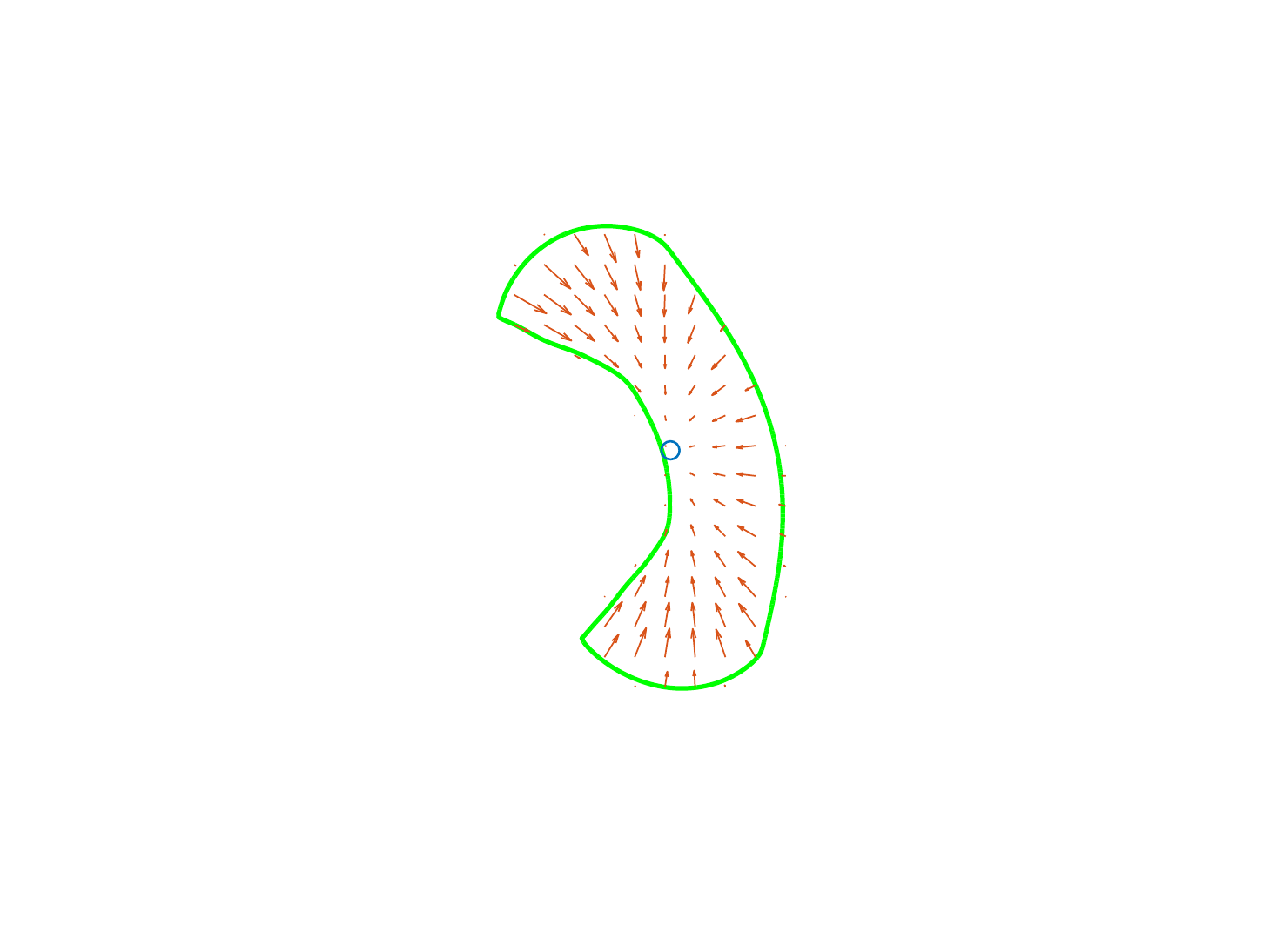}}
	\subfloat[\label{d3}]{\includegraphics[width=1.35in]{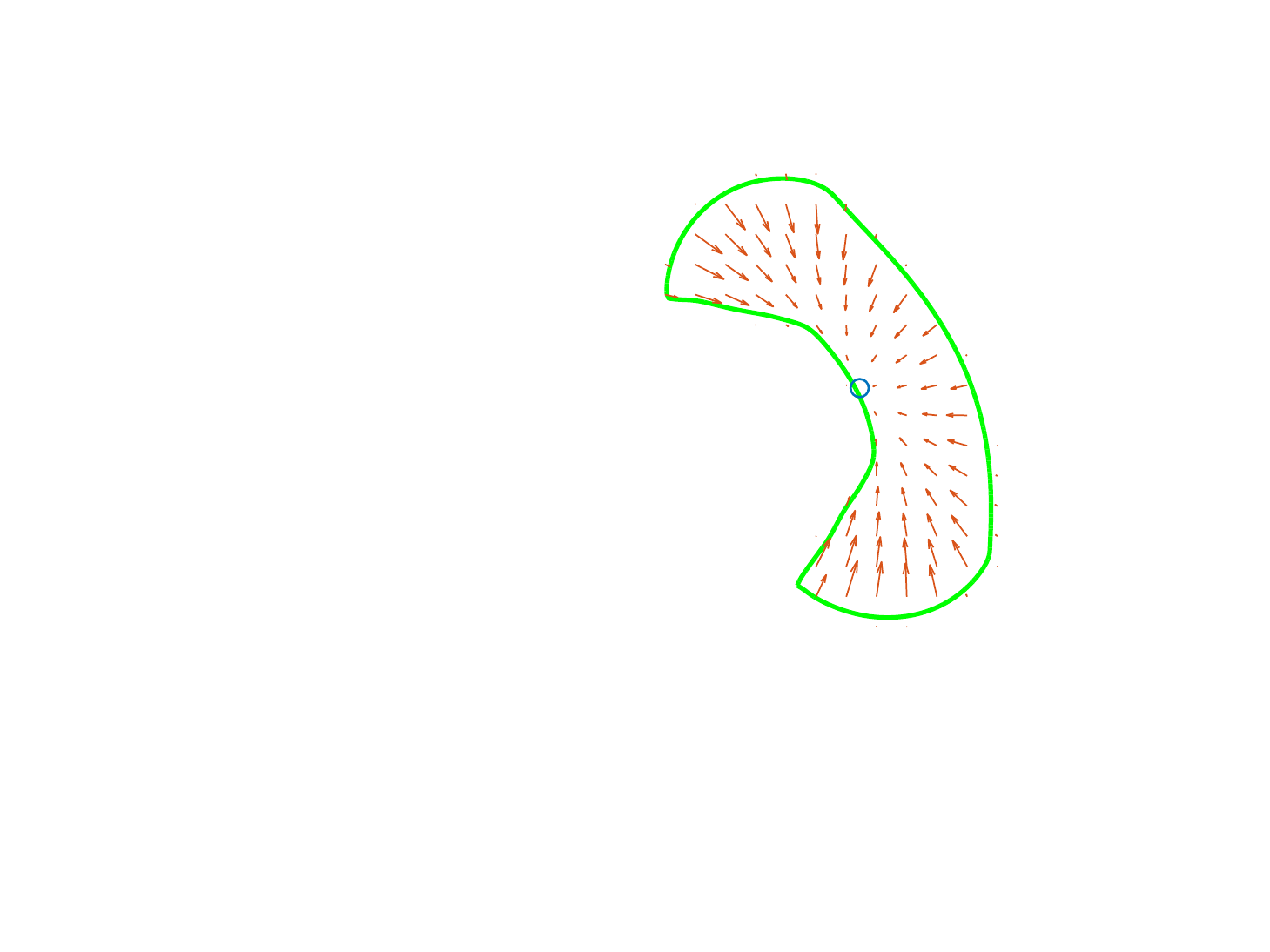}}
	\subfloat[\label{d5}]{\includegraphics[width=1.35in]{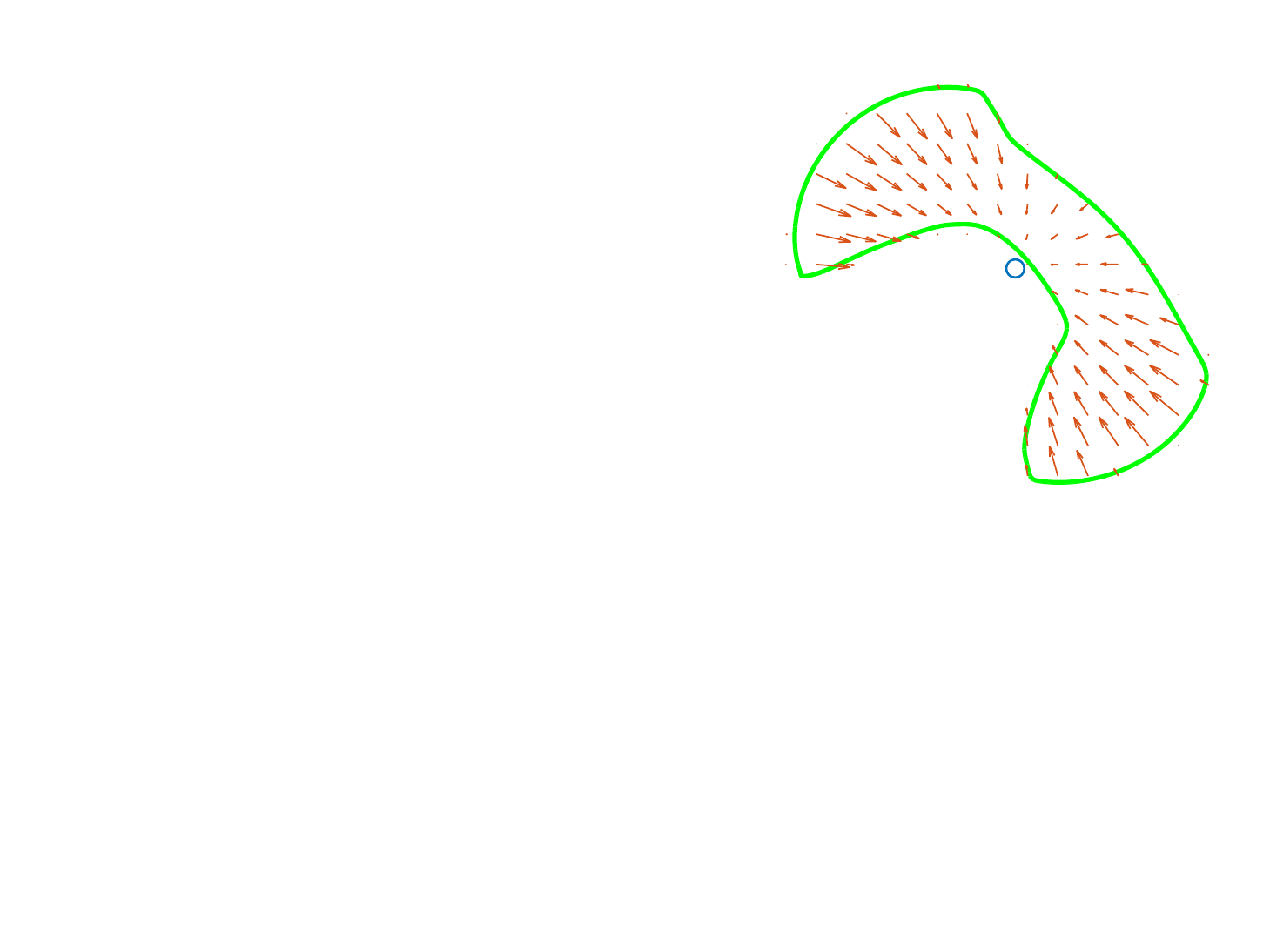}}
 \caption{ Durotaxis: Snapshots of a keratocyte (Model  simulation; green: \lamc red: actin velocity vectors) near a rigid boundary (top) starting to move to the right as in Fig. \ref{tr2}, then turning toward a rigid boundary (top of each figure). The shape is slightly distorted as the keratocyte turns, and symmetry about the instantaneous direction of motion is perturbed. Contact of the \lam with the rigid boundary occurs at (e).}
 \label{Turn}
\end{figure}

 \section*{Results and Discussion}

\subsubsection*{Symmetry Breaking and Topological Transition}
Keratocytes typically assume a roughly circular shape when stationary, with an annular \lam surrounding the nucleus \cite{polarvel}. Contact and force transmission with the substrate occurs only at the \lam and not the nucleus and organelles \cite{mushape}. Accordingly, we choose the initial \lam region $\Om_0$ to be an annulus in the center of the square domain $D$, with the nucleus excluded from description by the model.
The actin velocity field is centripetal. Next, we modify $\Om_0$ with a slight shape imperfection, in the form of a localized slight thinning at the rear of the cell (Fig. \ref{tr1}). This causes the symmetry to break and the \lam outside boundary starts to move inwards in the vicinity of the imperfection (Fig. \ref{tr2}).
The localized retraction causes further thinning until the \lam pinches off completely and a topological transition occurs (Fig. \ref{tr3}) as the annulus splits off into a simply connected, horseshoe shaped domain (Fig. \ref{tr4}). The topological change is evident as a result of excluding the nucleus from $\Omt$. Retraction of the cell rear occurs before the front starts to protrude, as reported in the experiments of \cite{polarvel}. The horseshoe flattens into a banana or crescent shape which only has symmetry about the $x$ axis. This polarized shape starts moving in the positive $x$ direction and quickly reaches steady shape and velocity, which it maintains for a long time (Fig. \ref{tr4e}). The transition from the annular stationary state, to the polarized, crescent shaped, locomoting state is remarkably similar to the sequence of observations reported in (\cite{polarvel} Fig.2a); an example is reproduced here in Fig. \ref{tr5}.

We find that the  initial transition from static annulus to locomoting crescent is  not strongly dependent of parameters,  because the centroid  velocity is small, hence the polarization term does not play an important role.  In the absence  of polarization ($e=0$) the actin velocity Eq.~\eqref{vs1} is  radially symmetric, so it  it is the \lam that breaks radial  symmetry during the transition. 
In the presence of polarization ($e>0$)  radial symmetry of the velocity field  is broken as well when the centroid moves; thus the presence of polarization affects the long term locomoting shape of the \lamp

\textcolor{black}{ 
 \subsubsection*{Steady Motion and Parameter Dependence}
Consistent with the observations of \cite{polarvel}, our model predicts that following symmetry breaking, topological change, and flattening of the broken annulus into a crescent, the cell settles into steady motion at essentially constant shape and velocity in the low polarization regime $ e <2$,  $K<15$.  An example of full transition from static annulus to fully developed steady state can be seen in Supplemental Video  SV1K3E1p5 (standard parameters except for $e=1.5$).  The long time fully developed crescent shape depends on the parameters $K$ and $e$.  Fig. \ref{123} shows the fully developed crescent shape for various combinations of $K$ and $e$  in the low polarization regime. In particular, for fixed $K$, the aspect ratio decreases with increasing $e$, while for fixed $e$, raising  $K$ increases the length of the trailing horns and the overall diameter slightly, but decreases the aspect ratio. See \cite{wrinkle2,shapemogli,mushape} for various examples of steady shapes of different aspect ratios but similar overall form. The crescent-shaped \lam and persistent, steady motion are well known characteristics of crawling keratocytes \cite{polarvel,actinspeed1}, not only whole cells, but also separated fragments of the lamellipodium \cite{needle2,actinspeed2} without the nucleus. See the section on Tensotaxis below for further observations on fragment behavior.}
 
\textcolor{black}{ 
\subsubsection*{Bipedal Oscillations and Lamellipodial Traveling waves}
Henceforth we fix $K=3$ and focus on the effect of varying the polarization coefficient $e$. We find that there are roughly three regimes of locomotion, depending on its value.  For low polarization, approximately $0\le e <2$, following the transition from annular stationary to locomoting crescent shape, propagation quickly becomes steady with constant velocity and no shape change, as described above.}

\textcolor{black}{In the intermediate polarization regime (roughly $2\le e\le 4$), after settling to steady motion, the cell  suddenly  switches to  oscillatory propagation. The centroid follows a roughly sinusoidal trajectory that oscillates about the $x$ axis, with the onset of oscillations at $e=2$ (see Fig. \ref{t0}), and higher amplitude as $e$ increases, e.g., $e=3$ (see Fig. \ref{t1}). The \lam orientation, and the direction of polarization, oscillate in phase with the centroid about  the $x$ direction. The substrate displacements alternate from nearly symmetric to antisymmetric with respect to the $x$ axis twice over an oscillation period; see snapshots  Fig. \ref{ttr1}-\ref{ttr4} . The \lam oscillates almost rigidly with little shape change, except at the trailing edges, which alternate from a pointed to a rounded shape out of phase with each other (Fig. \ref{ttr1}-\ref{ttr4}). Thus the cell propagates through asymmetric bipedal motion, as shown in Supplemental Video  SV2K3E3.  These qualitative characteristics occur in  keratocyte motion reported in \cite{theriotbipedal}, where it is noted that ``in persistently polarized, fan-shaped cells, retraction of the trailing edge on one side of the cell body is out of phase with retraction on the other side, resulting in periodic lateral oscillation of the cell body''.  A comparison of Supplemental Video  SV2K3E3 and [\cite{theriotbipedal} Supplemental Movie S2] shows very similar alternating trailing edge  retraction shapes (alternate rounded and pointed) but a larger wavelength in the latter.}

\textcolor{black}{Increasing $e$ decreases the oscillation frequency and the cell speed, Fig. \ref{t4}, which correlate with each other, Fig. \ref{t5}, in accordance  with  \cite{theriotbipedal}. This agreement is qualitative; the spatial wavelength of centroid oscillation in our simulations seems much smaller than the ones reported in \cite{theriotbipedal}. The overall centroid trajectory (with oscillations averaged out) becomes  curved and gradually strays away from the $x$ axis more  for higher values of $e$, Fig. \ref{snapshots}. This is also observed in locomoting keratocytes \cite{theriotbipedal}.}

\textcolor{black}{ 
The high polarization regime ($e\ge5$)  is characterized by increasingly severe, more irregular \lam shape distortions, in phase with centroid oscillations that are superposed on a trajectory curving further away from the $x$ axis for higher values of $e$, Fig. \ref{t3}, \ref{t4} and \ref{ttr5}-\ref{ttr8}. A striking feature of this regime is the formation of kinks in the anterior \lam front, which is convex for lower polarization. These kinks propagate outwards to the \lam sides in an alternating fashion (Supplemental Video  SV3K3E11 and Fig.  \ref{ttr5}-\ref{ttr8}). They form traveling waves on the anterior \lam edge of keratocytes on high adhesion strength substrates \cite{thermogliwaves1}. The amplitude of these waves is generally lower in our simulations than that reported in  \cite{thermogliwaves1}, except for high values of $e$ such as in Fig.  \ref{ttr5}-\ref{ttr8}.  The trajectories of these cells are more erratic and the centroid position oscillations are nonsmooth Fig.\ref{t3}, \ref{t4}, compared to those of the intermediate polarization regime. The centroid oscillation frequency  and speed are substantially lower than those of oscillating cells with intermediate polarization Fig.\ref{t4},  in qualitative accord with \cite{thermogliwaves1}.}

\textcolor{black}{ The shift from steady motion to oscillations, as well as the emergence of traveling \lam waves as polarization is increased,  seem to be  bifurcation phenomena. Our simulations suggest that velocity polarization in the direction of  the centroid velocity plays a central role in these nonsteady propagation modes. This may happen because a polarized actin velocity field possesses  an additional degree of freedom, namely,  the direction of  polarization;  this direction can oscillate, compared to  a radial, nonpolar velocity field.}

\textcolor{black}{\subsubsection*{Substrate Wrinkling, Displacement and Traction Prediction}
We next compare predicted actin velocity, substrate displacement,  traction and wrinkle field to experiments. The velocity field Eq.~\eqref{v2} in our model, which is prescribed for given parameters,  exhibits larger inward flow at the posterior horns  of the \lam  Fig. \ref{motsim}) and smaller retrograde flow at the front (right side). This agrees to some extent with observations of \cite{polarvel} shown here in Fig. \ref{vel}, although not quantitatively. }

\textcolor{black}{The predicted substrate displacement field,  Fig. \ref{usim} shows some qualitative similarities with measured displacements using Traction Force Microscopy \cite{displ}, Fig. \ref{uexp}, in particular, arrows curve toward the rear as the $x$ axis is approached from the trailing horns in a similar way. In our model, actin velocity is proportional to traction, so Fig. \ref{motsim} is representative of  traction vectors, while Fig. \ref{trexp}, traction inferred from discrete experimental displacement \cite{displ}, does not compare so  well with Fig. \ref{motsim}.}

If the elastic substrate is sufficiently compliant, the contractile tractions exerted by keratocytes cause it to wrinkle \cite{cellsdeformsubstrate,wrinkle2}. This was first observed with fibroblasts inducing wrinkling of thin silicone substrates as a pioneering method to measure forces exerted by cells \cite{harris}. Here we compare substrate wrinkles observed in experiments involving locomoting keratocytes \cite{cellsdeformsubstrate,wrinkle2} with a prediction based on the stress field predicted by our model.

Wrinkling in thin elastic sheets is local buckling caused by compression. The direction of a wrinkle is normal to the direction of maximum compression i.e., the eigenvector of the stress tensor with the smallest (negative) eigenvalue. When the compressive force is localized, the length of a compression wrinkle emanating from the point of application was measured to be proportional \cite{wrinkleforce} to the compressive force. We use this to make a simple prediction of wrinkles from our simulations as follows. We draw straight lines emanating from grid points on or close to (but inside) the \lam boundary.  Their direction is chosen orthogonal to the direction of maximum compression, and their length is proportional to the smallest (negative) eigenvalue of the stress tensor at the cell boundary point where the line emanates. See Supporting Material for more details. The resulting line field is shown in Fig. \ref{wa}
 for a simulated steadily locomoting keratocyte, while an experimental image is in Fig. \ref{wb}. 
 There are many qualitative similarities, not only between the computed and observed \lam {} shapes, but also between the line field just described and observed wrinkles \cite{cellsdeformsubstrate,wrinkle2}. In particular, in both observed and simulated wrinkles,
(i) the wrinkle field on the anterior, advancing side of the \lam boundary is fan shaped and roughly centripetal
(directions of wrinkles diverge); (ii) the wrinkles on the posterior, retreating side are much more aligned to the (negative) direction of motion and nearly parallel;
(iii) posterior wrinkles are substantially longer than anterior ones (though the ratio is higher in the experiment than the simulation);
(iv) the rearward facing top and bottom portions of the convex side are nearly free of wrinkles.

We note that our linear elastic substrate model does not explicitly account for wrinkling, so our wrinkle prediction algorithm is somewhat crude, nonetheless it captures many features of the actual wrinkle field. We view this as a validation of our model, \textcolor{black}{since wrinkles provide  the only relatively direct way to measure aspects of the substrate stress field in this setting.}

\subsubsection*{Response to External Stimuli and Tensotaxis}
Fibroblasts respond to external forces applied remotely on the elastic substrate by changing shape and direction of motion. When microneedles are used to induce stresses on the substrate, fibroblasts---either the entire cell or a protrusion---tend to move toward tensile stresses and away from compressive stresses \cite{demboneedle}. This is known as tensotaxis. While we are unaware of similar experiments on keratocytes, we examine whether our model predicts tensotaxis. Lamellipodial fragments that are severed from the \lamc and do not contain the nucleus or organelles, behave similar to entire cells \cite{needle2}. They are disk-shaped when stationary. When pushed by a one-sided external force, they break symmetry, become crescent shaped and start propagating steadily away from the pushing force, even after the latter is removed. 
While we cannot model the direct application of force onto the cell body, we simulate a situation similar to the experiments of \cite{demboneedle}. A force (uniform traction over a disk-shaped area) is applied onto the substrate some distance from the circular stationary \lam fragment, pointing toward it. The force is applied for a short time, then removed. In response, an indentation forms as part of the fragment boundary retreats away from the applied force. This breaks the symmetry of the fragment, which becomes crescent shaped and starts propagating away from the applied force site; Fig. \ref{Push}.

Steady propagation in crescent form continues even though the force has been removed. A similar sequence of events occurs in experiments \cite{needle2} but due to direct pushing of the fragment instead of the substrate. Instead here the applied force induces compressive stress between where it is applied and the \lam fragment, which in turn causes the boundary velocity of the cell to become negative in the location closest to the applied force site and thus the symmetry is broken, eventually leading to the crescent shape and steady propagation away from the location of the force even after the latter ceases to act.

In contrast, when the direction of the applied force is opposite (away from the \lam fragment) tensile stress is generated in front of the fragment, leading to protrusion toward the force site, symmetry breaking, and in some instances, propagation in crescent shape in the direction of the applied force even after the latter is removed; Fig. \ref{Pull} . \textcolor{black}{This occurs for ellipsoidal fragments with the long axis transversal to the pulling force. Circular fragments tend to elongate in the direction of the pull, then stop after the pulling force is removed.}  These simulations exhibit tensotaxis: either motion away from higher compressive stress or protrusion and/or motion towards greater tensile stress. This behavior has similarities with that of fibroblasts \cite{demboneedle} although it seems not to have been investigated in the case of keratocytes. 
 \textcolor{black}{More  recently \cite{keratino}, relevant behavior was observed with human epithelial keratinocytes, which are closer to fish epidermal keratocytes than fibroblasts. A needle pulls the substrate behind  a locomoting cell and away from it. The cell turns, moves away  transversally to the original direction, elongates toward the needle, similar to what happens in the case of a circular fragment, then gradually turns toward the needle. See Supplemental Video  SV4Ker for a simulation capturing various stages of this behavior qualitatively.}

 \subsubsection*{Turning Towards Stiffer Substrates and Durotaxis} On a substrate with an interface between regions of different stiffness, cells that assume a crescent morphology similar to keratocytes starting on the softer region, have been observed to follow a curved trajectory, so that they turn toward, and cross into, the stiffer portion of the substrate \cite{duroturning}. 
 
 Under zero displacement boundary conditions, the simulation domain boundary becomes equivalent to an interface with a region of infinite stiffness (rigid). We find that cells starting on the central axis of the rectangular symmetric domain typically travel straight along it. However, a cell with initial position closer to the top boundary follows a curving trajectory, while also turning almost rigidly (Fig. \ref{Turn}), so that it approaches, and eventually contacts, the top boundary; see Supplemental Video    SV5Duro1. This attraction by a rigid boundary is an instance of durotaxis, and also reproduces the observations of crescent shaped fibroblasts following a curved trajectory while turning almost rigidly with slight shape change \cite{friction}.   
 
\textcolor{black}{In contrast, traction-free boundary conditions make the boundary behave like the interface with a softer material, in the limit of zero stiffness. Repeating the previous simulation with traction free conditions makes the cell turn away from the boundary toward the centerline along the $x$ axis, repelled by the interface with a much softer
substrate. See Supplemental Video SV6Duro2.} 
 
How can a cell sense an interface with a stiffer region at a distance? Our model provides insight into the mechanism responsible for the attraction of cells by a rigid boundary. Cells exert contractile forces onto the substrate. In the vicinity of a rigid boundary, this causes tensile stresses that are highest in the ligament between the boundary and the cell. These tensile stresses are sensed by the cell, which tends to protrude in their direction in accordance with the evolution law. The closer the cell approaches the boundary, the higher this stress becomes; this causes acceleration and the result is a trajectory that curves toward the stiff boundary. This strongly suggests that keratocytes and fibroblasts exert contractile forces in order to probe their surroundings by sensing  inhomogeneities in the stress field they themselves generate. In this case the inhomogeneity is caused by the vicinity of a stiff interface.

\section*{Conclusions}

We have constructed a minimal model for the evolution of fish epidermal keratocytes based on an active mechanosensing hypothesis: we posit that these cells sense the stress field that they themselves actively generate in the substrate, and evolve accordingly, by protruding in areas of tension and contracting in areas of compression.

Most previous theoretical models concentrate on the processes inside the cell, such as actin-myosin interaction. In contrast, our model focuses on the mechanical interaction between the \lam and substrate. The model of the cell itself is minimal and consists of an actin velocity field with central symmetry inside an evolving curve representing the \lam boundary. The centripetally flowing actin exerts contractile tractions onto the elastic substrate. The resulting substrate stress depends on the shape of the \lam boundary. At the same time, this stress enters the evolution law of the \lam boundary curve.

 \textcolor{black}{ In its nondimensional form, the model involves just three independent parameters for the cell and one for the substrate.}
The model predicts multiple types of observed behavior of keratocytes on elastic substrates for the same parameter set. The well known crescent shape, characteristic of keratocytes in steady locomotion, emerges through symmetry breaking bifurcation and a topological change from the annulus-shaped \lam typical of stationary keratocytes. This simulated sequence closely resembles the observed transition from the static to the locomoting state of keratocytes as reported in \cite{polarvel}. 
 
 \textcolor{black}{When the model is generalized to include actin velocity polarization in the direction of motion, it successfully  predicts two  additional types of complex observed locomotion behavior. For high enough polarization, steady motion of the crescent bifurcates into oscillatory bipedal asymmetric locomotion. Further increases of the polarization parameter yield more irregular, slower oscillations with motion facilitated by alternating traveling waves moving along the leading \lam edge. Keratocytes are known to exhibit both kinds of phenomena \cite{theriotbipedal,thermogliwaves1}.  It thus becomes clear that the three main types of locomotion, steady, bipedal oscillatory and wavelike, are due to a series of symmetry breaking bifurcations starting from the radially symmetric stationary annular \lamp The last two of these are possible only when the actin velocity field is polarized.}
 
 Additional validation of the model is provided by the successful prediction of the substrate stress field. Compressive stresses caused by contractile tractions exerted by moving keratocytes cause sufficiently thin silicone substrates to wrinkle \cite{wrinkle2}; our model predicts the direction and relative magnitude of the wrinkles based on the computed  eigenvectors and (negative) eigenvalues of the substrate stress field.

When microneedles are used to induce stresses in the substrate, fibroblasts tend to move toward tensile stresses and away from compressive stresses \cite{demboneedle,needle2,keratino}. In our simulations, applying a localized body force onto the substrate some distance away from the cell creates either a compressive or tensile stress gradient (when pointing toward or away from the cell, respectively). The cell either moves away from a force pointing towards it, or protrudes towards a force in the opposite direction. This is an example of tensotaxis, although such experiments seem not to have been performed with fish keratocytes, \textcolor{black}{but quite recently with closer related human keratinocytes \cite{keratino};  our model captures some essential aspects of these experiments.} 

Our model exhibits a form of durotaxis, whereby simulated cells are attracted by the closest rigid boundary and curve their trajectories toward, as in observed behavior of crescent shaped cells toward interfaces with stiffer regions \cite{duroturning}. The model allows us to identify the mechanism underlying this attraction as cell-induced tensile stress which is higher in the region between the cell and the closest points of the boundary, leading to preferred protrusion in the latter.   \textcolor{black}{In contrast, but as expected,  simulated cells turn away from a traction free boundary, which is in a sense an interface with an infinitely soft substrate.}

\textcolor{black}{We believe that} the present model is the first to explain  \textcolor{black}{multiple types of} the locomoting behavior of keratocytes on deformable substates through active mechanosensing. \textcolor{black}{It is remarkable that by varying velocity polarization, it displays three distinct modes of locomotion that are actually observed:  steady, bipedal-oscillatory and traveling \lam wave. This sheds light into the role played by the actin velocity polarization in these complex transitions}.  The model  also provides insight into phenomena such as tensotaxis and durotaxis, more commonly observed with fibroblasts and other cells. To investigate the validity of the active mechanosensing hypothesis further, it would be interesting to perform experiments analogous to \cite{demboneedle,needle2,duroturning}, but with fish keratocytes instead of fibroblasts or human keratinocytes \cite{keratino}, either on substrates where remote forces are exerted by microneedle, or where the substrate stiffness varies with position, either gradually or discontinuously.  \textcolor{black}{It will be especially instructive in understanding mechanosensing, to determine in what ways the morphology and locomotion characteristics of keratocytes differ from those of fibroblasts, and other cells  known to be strongly mechanosensitive.}

\section*{Author Contributions}

ZZ wrote the numerical code and ran the simulations, PR developed the model, all authors analyzed and discussed the model  and results, contributed to further refinement of the model and the preparation/revision of the manuscript.

\section*{Acknowledgments}
The research of Z. Zhang was supported in part by the Hong Kong RGC Grants (Project 27300616 and 17300817), National Natural Science Foundation of China (Project 11601457). Z. Zhang would like to thank the support and hospitality of Professor T.Y. Hou when he was a postdoctoral scholar at Caltech.  The research of P. Rosakis was partially supported by the EU
Horizon 2020 Research and Innovation Program under the Marie Sklodowska-Curie project  
ModCompShock 
agreement No 642768  (modcompshock.eu).  The research of T.Y. Hou was supported in part by an NSF Grant DMS-1613861. G. Ravichandran acknowledges the support of the National Science Foundation (DMR No. 0520565) through the Center for Science and Engineering of Materials at the California Institute of Technology.

\bibliography{biorefs}


\section*{Supplementary Material}




\subsection*{Wrinkle Field Prediction}
 \textcolor{black}{To generate the wrinkle field of Fig. \ref{figwrinkle}(b), the direction of each wrinkle is chosen normal to the direction of maximum compression, i.e., the eigenvector of the stress tensor corresponding to  the ``most negative'' eigenvalue (the negative eigenvalue with largest absolute value). The length of each simulated wrinkle is proportional to the ``most negative''  eigenvalue of the stress tensor at the cell boundary point where the line emanates. The proportionality constant depends on the substrate thickness, which we do not specify in our model,  among other factors, so we calibrate it. We choose the length of one wrinkle only, the central posterior, rear pointing wrinkle so that its ratio to the \lam diameter is the same as in the experimental image in Fig. \ref{figwrinkle}(a).  This determines the proportionality constant and the simulated wrinkle field.}

\subsection*{Supplemental Videos}

Six supplemental videos are cited in the main text:

SV1K3E1p5.avi

SV2K3E3.avi

SV3K3E11.avi

SV4Ker.avi

SV5Duro1.avi

SV6Duro2.avi

\subsection*{Level Set Formulation}
We use the level set method \cite{Chang} which has been successfully applied to cell evolution study, e.g., \cite{levelcell,mogilnervariousmodels} to solve a regularized version of the equations of the model. The regularization allows us to extend fields defined only on the moving surface $\pot$ to the entire domain.
Let $D=[-L,L]^2\subset\RR^2$ be the region occupied by the substrate, with the cell $\Omt\subset D$.
For $\e>0$ a small parameter, let $\he$ be the (smooth) \emph{regularized step function}, so that
 \[
\he'(z)>0 \quad\text{for $|z|<\e,$} \qquad \he(z)=\begin{cases} 1, & z\ge \e,
\\
0 ,&z\le -\e.
\end{cases}
\]
Its derivative, the \emph{regularized delta function} $\de$ has support $[-z,z]$ and satisfies
\[ \de(z)=\he'(z),\qquad \int_{-\e}^\e\de(z)dz=1 \]
The \emph{level set function} $\ph(\bx,t)$ vanishes on $\pot$, is positive inside $\Omt$ and negative outside it. It evolves according to the \emph{level set equation}
\be\label{lev}\ph_t-V_n|\nabla\ph|=0 \quad \text{in } D\ee
where $V_n=V_n(\bx,t)$ is the normal velocity of the level set of $\ph$ through $\bx$ at time $t$.
The characteristic function is thus $\hot(\bx,t)=H(\ph(\bx,t))$, where $H$ is the usual Heaviside step function. In the regularized scheme $\hot$ is replaced by the \emph{regularized characteristic function}
\[\he(\ph(\bx,t))\]
for $\bx\in D$. The regularized cell centroid and corresponding velocity are thus
\be\label{xbarre} \bbx_\e(t)=\frac{\int_D \bx \he(\ph(\bx,t))d\bx}{\int_D \he(\ph(\bx,t))d\bx }, \quad \bbv_\e=\dot\bbx_\e\ee
see Eq.~\eqref{cc}. The \emph{regularized body force} is
\be\label{be2}\bb_\e(\bx,t)=-K(\bI+e \bbv_\e\otimes\bbv_\e)(\bx-\bbx_\e)\he(\ph(\bx,t)),\quad\bx\in D\ee
Accordingly, the regularized version of Eq.~\eqref{equil} is $\nabla\cdot\bS+\bb_\e=\bo$ in $D$, or
\be\label{regpde}\mu\Delta\bu+(\lambda+\mu)\nabla(\diiv \bu) +\bb_\e(\bx,t)=\bo\quad \text{in } D\ee
in view of Eqs \eqref{sss}, \eqref{equil}.
Define the unit normal field
\be\label{bn}\bn(\bx,t)=-\frac{\nabla\ph(\bx,t)}{|\nabla\ph(\bx,t)|},\quad \bx\in D\ee
and the regularized normal velocity as in Eq.~\eqref{vn} with $\bv_{s\e}=(1/\eta)\bb_\e$ in place of $\bv_s$:
\be\label{vne} V_{n\e}=-\gamma(\bI+e \bbv_\e\otimes\bbv_\e)(\bx-\bbx_\e)\cdot\bn+G(\bn\cdot\bS\bn),\quad \text{in } D\ee
where
\be\label{regf} \bn\cdot\bS\bn=\frac{1}{|\nabla\ph|^2}\nabla\ph\cdot \bS \nabla\ph.\ee
Here we have used Eqs \eqref{bn} and \eqref{sss}. The Hamilton-Jacobi equation
\be\label{leve}\ph_t-V_{n\e}|\nabla\ph|=0 \quad \text{in } D\ee
governs the evolution of the level set function.
The regularized problem is to find $(\bu,\ph)$ satisfying Eqs \eqref{regpde} and \eqref{leve} subject to initial conditions specifying the initial cell domain $\Om_0\subset D$, $\bu(\cdot,0)=\bo$, $\ph(\bx,0)=\pm \rm{dist}(\po,\bx)$ with the $+$ choice inside $\po$ and the $-$ choice outside (signed distance form $\po$), and suitable boundary conditions on $\partial D$. Then the cell boundary
$\pot$ is the zero level set of $\ph(\cdot,t)$.
with $V_{n\e}$ given by Eq.~\eqref{vne} and $\bS$ by Eq.~\eqref{sss}.
\subsection*{Finite difference discretization} \label{sec:numericalmethods}

\subsubsection*{Discretization of the displacement field.} \label{sec:num-disp}
We use finite difference method to discretize the regularized model (level-set formulation) developed in the previous section. First, we specify regularized version of the singular Dirac delta function $\delta$ and the
discontinuous Heaviside function $H$. In our numerical discretizations, we define the regularized delta function as $\delta_{\e}$ as
\begin{equation}\label{regulared_delta}
 \delta_{\e}(x)= \left\{
 \begin{aligned}
 &\frac{1}{2}(1+\cos(\pi x/\e))/\e, &|x|<\e\\
 &0, &|x|\geq\e
 \end{aligned}
 \right.
\end{equation}
and the corresponding regularized Heaviside function $H_{\e}$ is defined as
\begin{equation}\label{regulared_delta}
 H_{\e}(x)= \left\{
 \begin{aligned}
 &0, & x<-\e \\
 &(x+\e)/(2\e)+\sin(\pi x/\e))/(2\pi), &|x|<\e\\
 &1, &x > \e
 \end{aligned}
 \right.
\end{equation}
We have the relation $H_{\e}^{\prime}(x)=\delta_{\e}(x)$.

\textcolor{black}{We partition the domain $D=[-L_{x},L_{x}]\times[-L_{y},L_{y}]$ into $(N_{x}+1)\times (N_{y}+1)$ 
grids $(x_i,y_j)$ with $x_i=(i-1)h-L_{x}$, $y_j=(j-1)h-L_{y}$, $1\leq i\leq N_x+1$,  $1\leq j\leq N_y+1$ and mesh $h=\frac{2L_{x}}{N_{x}}=\frac{2L_{y}}{N_{y}}$. Recall that $\textbf{u}=(u,v)^{T}$. Denote by $u_{i,j}^{n}$ the approximation of $u(x_i,y_j,t_n)$, where $t_n=n\Delta t$, $\Delta t$ is the time step, and $n$ is a nonnegative integer. The approximations to $v(x_i,y_j,t_n)$ and $\varphi(x_i,y_j,t_n)$ can be defined in the same fashion. For the discretization in space, we use a second-order, centered-difference scheme. We introduce the finite difference operators}
\begin{align}
D_{0}^{x}f_{i,j}&=(f_{i+1,j}-f_{i-1,j})/2h, \quad \text{(central difference)}, \nonumber \\
D_{-}^{x}f_{i,j}&=(f_{i,j}-f_{i-1,j})/2h, \quad \quad \text{(backward difference)},\nonumber \\
D_{+}^{x}f_{i,j}&=(f_{i+1,j}-f_{i,j})/2h, \quad \quad \text{(forward difference)}. \nonumber
\end{align}
The operators $D_{0}^{y}$, $D_{-}^{y}$, and $D_{+}^{y}$ are defined similarly. If we write in element-wise form, the regularized PDE of the displacement field satisfies,
\begin{align}
(\lambda+2\mu)u_{xx} + (\lambda+\mu)v_{xy} + \mu u_{yy} = K H_{\e}(\varphi)( x-\bar x_{\e}), \label{RegularizedPDEDispu}\\
\mu v_{xx} + (\lambda+\mu)u_{xy} + (\lambda+2\mu) v_{yy} = K H_{\e}(\varphi) (y-\bar y_{\e}), \label{RegularizedPDEDispv}
\end{align}
where $(\bar x_{\e},\bar y_{\e})=\bar\bx_\e$ is given by Eq.~\eqref{xbarre}. Using the central difference scheme, the discretized version of   Eqs \eqref{RegularizedPDEDispu} and \eqref{RegularizedPDEDispv} thus read
\begin{align}
&(\lambda+2\mu)\frac{u_{i+1,j}-2u_{i,j}+u_{i-1,j}}{h^2} +
(\lambda+\mu)\frac{v_{i+1,j+1}-v_{i+1,j-1}-v_{i-1,j+1}+v_{i-1,j-1}}{4h^2} \nonumber \\
&+\mu \frac{u_{i,j+1}-2u_{i,j}+u_{i,j-1}}{h^2} = K H_{\e}(\varphi_{i,j}^{n})(x_i - \bar x_{\e}), \label{strd1}\\
& \mu\frac{v_{i+1,j}-2v_{i,j}+v_{i-1,j}}{h^2} +
(\lambda+\mu)\frac{u_{i+1,j+1}-u_{i+1,j-1}-u_{i-1,j+1}+u_{i-1,j-1}}{4h^2} \nonumber \\
&+(\lambda+2\mu) \frac{v_{i,j+1}-2v_{i,j}+v_{i,j-1}}{h^2} =
K H_{\e}(\varphi_{i,j}^{n})(y_j - \bar y_{\e}). \label{strd2}
\end{align}
\subsubsection*{Discretization of the evolution law.} \label{sec:num-kineticrelation}
We first recall the regularized stress $S=(S^{kl})_{2\times 2}$, $1\leq k,l\leq 2$, where the entries are given by
\begin{align}
&S^{11}(x,y)=(\lambda+2\mu)u_{x}(x,y)+\lambda v_{y}(x,y), \nonumber \\
&S^{22}(x,y)= \lambda u_{x}(x,y)+(\lambda+2\mu) v_{y}(x,y) , \nonumber \\
&S^{12}(x,y)=S^{21}(x,y)=(\lambda+\mu)(u_{y}(x,y)+v_{x}(x,y)). \nonumber
\end{align}
We employ the central difference scheme to compute $S^{kl}$, 
$1\leq k,l\leq 2$. For instance, let $S^{11}_{ij}$ be the numerical approximation to $S^{11}(x_i,y_j)$.  
Away from the boundaries,  we use the central difference to discretize $u_{x}$ and $v_{y}$ and get, 
\begin{align}
S^{11}_{ij}= (\lambda+2\mu)\frac{u_{i+1,j}-u_{i-1,j}}{2h}
+ \lambda \frac{v_{i,j+1}-v_{i,j-1}}{2h} .\label{NumDisc-S11}
\end{align}
$S^{12}(x,y)$ and $S^{22}(x,y)$ can be discretized in the same way. 

At boundaries, to compute $S^{kl}$, $1\leq k,l\leq 2$, we need to impose boundary conditions of $u$ and $v$. 
If the Dirichlet boundary conditions are imposed for $u$ and $v$, we simply use an one-sided finite difference scheme to discretize $u_{x}$, $u_{y}$, $v_{x}$ and $v_{y}$ and compute $S^{kl}$, since only the stress on the interior domain has contribution to the kinetic relation. When the mixed displacement and traction free conditions are imposed, 
i.e.,  $S^{22}=S^{12}=0$ on the north and south boundary sides (traction free) and $u=v=0$ on the east and west sides, we discretize $u_{x}$, $u_{y}$, $v_{x}$ and $v_{y}$ using the central difference scheme and eliminate the ghost points (caused by $u_{y}$ and $v_{y}$) through Eqs  \eqref{strd1}, \eqref{strd2}. The corner points are discretized using a first order scheme. Once we get the stress $S=(S^{kl})_{2\times 2}$, 
we can use Eq.~\eqref{vne} to compute $V_n$.

sub\subsection*{Discretization of the level-set function.} \label{sec:num-levelset}
We employ a second-order ENO scheme to discretize  Eq.~\eqref{leve}, which describes the evolution of the level-set function $\varphi(x,y)$. Since we are interested in the accurately computing the convection of interface position, we use the nonconservative form of the ENO scheme \cite{shu1998essentially}. Define a minmod function as
\begin{equation}\label{minmod_func}
 \text{minmod}(s,t)= \left\{
 \begin{aligned}
 &sgn (s)\min(|s|,|t|), & st>0\\
 &0, & \text{otherwise}
 \end{aligned}
 \right.
\end{equation}
Here $sgn$ means the signum function.  Eq.~\eqref{leve} satisfied by the level-set function $\varphi(x,y)$ is a specialized version of the Hamilton-Jacobi equation $\varphi_{t}-V|\nabla \varphi|=0$. Given the normal velocity $V=V(x,y,t)$ of the level sets of $\varphi$, the second-order ENO discretization of the Hamilton-Jacobi equation is
\begin{equation}\label{SecondENO}
 \varphi_{i,j}^{n+1}= \left\{
 \begin{aligned}
 &\varphi_{i,j}^{n} - \Delta t V_{i,j}^{n}P_{+}, & \text{for} V_{i,j} > 0, \\
 &\varphi_{i,j}^{n} - \Delta t V_{i,j}^{n}P_{-}, & \text{for} V_{i,j}\leq 0,
 \end{aligned}
 \right.
\end{equation}
Here we have,
\begin{align}
& P_{+} = \sqrt{(\max(p_{-}^{x},0)^2 + \min(p_{+}^{x},0)^2)+ (\max(p_{-}^{y},0)^2 + \min(p_{+}^{y},0)^2)}, \nonumber \\
& P_{-} = \sqrt{(\min(p_{-}^{x},0)^2 + \max(p_{+}^{x},0)^2)+ (\min(p_{-}^{y},0)^2 + \max(p_{+}^{y},0)^2)}, \nonumber \\
& p_{-}^{x} = D_{-}^{x}\varphi_{i,j}^{n}+ 0.5h~\text{minmod}(D_{-}^{x}D_{+}^{x}\varphi_{i,j}^{n},D_{-}^{x}D_{+}^{x}\varphi_{i-1,j}^{n}), \nonumber \\
& p_{+}^{x} = D_{-}^{x}\varphi_{i+1,j}^{n}- 0.5h~\text{minmod}(D_{-}^{x}D_{+}^{x}\varphi_{i+1,j}^{n},D_{-}^{x}D_{+}^{x}\varphi_{i,j}^{n}), \nonumber \\
& p_{-}^{y} = D_{-}^{y}\varphi_{i,j}^{n}+ 0.5h~\text{minmod}(D_{-}^{y}D_{+}^{y}\varphi_{i,j}^{n},D_{-}^{x}D_{+}^{x}\varphi_{i,j-1}^{n}), \nonumber \\
& p_{+}^{y} = D_{-}^{y}\varphi_{i,j+1}^{n}- 0.5h~\text{minmod}(D_{-}^{y}D_{+}^{y}\varphi_{i,j+1}^{n},D_{-}^{y}D_{+}^{y}\varphi_{i,j}^{n}). \nonumber
\end{align}
In practice, even if we prescribe the initial value of the level-set function $\varphi$ to be a signed distance from the interface, it will not remain so at later times. For large time computations it is desirable to keep $\varphi$ as a distance function. This will ensure that the interface has a finite thickness of order $\e$ for all time. In \cite{sussman1994level}, an iterative procedure was proposed to re-initialize $\varphi$ at each time step, so that it remains a signed distance function from the evolving interface. To be specific, given a level-set function $\varphi^{n+1}(x,y)=\varphi(x,y,t_{n+1})$ at time $t=t_{n+1}$, we compute the solution of the initial-value problem as follows,
\begin{align}
\Phi_{t}&=sgn(\varphi^{n+1}(x,y))(1-|\nabla \Phi|), \quad (x,y)\in D, \label{re-initialize-levelset-eq}\\
\Phi(x,y,0)&=\varphi^{n+1}(x,y), \quad (x,y)\in D. \label{re-initialize-levelset-iv}
\end{align}
The solution converges rapidly in time to a function that has the same sign and the same zero level set as
$\varphi^{n+1}(x,y)$ and also satisfies $|\nabla \Phi|=1$, so that it equals the signed distance from the interface. After $\varphi$ evolves at each time step according to Eq.~\eqref{SecondENO}, it is re-initialized by solving Eqs \eqref{re-initialize-levelset-eq} and \eqref{re-initialize-levelset-iv}; this suffices due to rapid convergence. This procedure is crucial for our formulation, since the extension of the normal velocity $V$ in our case is not continuous across the phase boundary in the sharp-interface $\e$ limit. This makes computations more difficult than in the fluid interface problem considered in \cite{Chang,sussman1994level}, where the normal velocity is continuous across the interface.

In our calculations, we use a one-sided finite difference scheme to discretize $\varphi_x$ and $\varphi_y$ at the boundary. For example at boundaries $x=-L_x$ and $x=L_x$, $\varphi_x(-L_x,y_j,t_n)$ is approximated by $D_{+}^{x}\varphi^{n}_{0,j}=(\varphi^{n}_{1,j}-\varphi^{n}_{0,j})/h$ and $\varphi_x(L,y_j,t_n)$ is approximated by $D_{-}^{x}\varphi^{n}_{N,j}=(\varphi^{n}_{N,j}-\varphi^{n}_{N-1,j})/h$.

\end{document}